\documentclass[preprint, prX]{revtex4}

\usepackage[utf8]{inputenc}

\usepackage{xcolor}
\usepackage{algorithm}
\usepackage{algpseudocode}
\usepackage{mathtools}
\usepackage{calligra}
\usepackage{amsmath}
\usepackage{amsfonts}
\usepackage{mathrsfs}
\usepackage{graphicx}
\usepackage{relsize}
\usepackage{float}
\usepackage{svg}
\usepackage{pdflscape}
\usepackage{subfigure}
\usepackage{pdfpages}
\usepackage{comment}
\usepackage{ulem}

\DeclareMathAlphabet{\mathcalligra}{T1}{calligra}{m}{n}
\DeclareFontShape{T1}{calligra}{m}{n}{<->s*[2.2]callig15}{}
\definecolor{green}{rgb}{0,0.4,0}
\definecolor{cyan}{rgb}{0,0.7,0.7}
\definecolor{red}{rgb}{1,0,0}
\definecolor{blue}{rgb}{0,0,1}
\newcommand{\bs}[1]{\boldsymbol{#1}}

\newcommand{\RFvar}{\widetilde{\bs{u}}_m}
\newcommand{\iRFvar}{\widetilde{\bs{u}}^i_m}

\newcommand{\sOneRFvar}{\widetilde{\bs{u}}^{*1}_m}
\newcommand{\sTwoRFvar}{\widetilde{\bs{u}}^{*2}_m}
\newcommand{\sNumRFvar}{\widetilde{\bs{u}}^{*N}_m}

\newcommand{\newadd}[2]{#2}
\newcommand{\newdel}[1]{\ignorespaces}

\begin{document}

\title{\thispagestyle{empty}Generalizable Physics-constrained Modeling using Learning and Inference assisted by Feature-space Engineering}

\author{Vishal Srivastava}
\email{vsriv@umich.edu}
\affiliation{Ph.D. Candidate, Department of Aerospace Engineering, University of Michigan, Ann Arbor, 48105}

\author{Karthik Duraisamy}
\email{kdur@umich.edu}
\affiliation{Associate Professor, Department of Aerospace Engineering, University of Michigan, Ann Arbor, 48105}

\date{\today}

\begin{abstract} \noindent
  This work presents a formalism to improve the predictive accuracy of physical models by learning \textit{generalizable} augmentations from sparse data.
  Building on recent advances in data-driven turbulence modeling, the present approach, referred to as Learning and Inference assisted by Feature-space Engineering (LIFE), is based on the hypothesis that robustness and generalizability demand a meticulously-designed feature space that is informed by the underlying physics, and a carefully constructed features-to-augmentation map.
  The critical components of this approach are:
    \newdel{\sout{(1) Identification of relevant physics-informed features in appropriate functional forms to enable significant overlap in feature space for  a wide variety of cases to promote generalizability;
    (2) Explicit control over feature space to locally infer the augmentation without affecting other feature space regions, especially when limited data is available;
    (3) Maintaining consistency across the learning and prediction environments to make the augmentation case-agnostic;
    (4) Tightly-coupled inference and learning by constraining the augmentation to be learnable throughout the inference process to avoid significant loss of information (and hence accuracy).}}
    \newadd{green}{(1) Maintaining consistency across the learning and prediction environments to make the augmentation case-agnostic;
    (2) Tightly-coupled inference and learning by constraining the augmentation to be learnable throughout the inference process to avoid loss of inferred information (and hence accuracy);
    (3) Identification of relevant physics-informed features in appropriate functional forms to enable significant overlap in feature space for a wide variety of cases to promote generalizability;
    (4) Localized learning, i.e. maintaining explicit control over feature space to change the augmentation function behavior {\em{only}} in the vicinity of available datapoints.}
  To demonstrate the viability of this approach, it is used in the modeling of bypass transition.
  The augmentation is developed on skin friction data from {\em{two}} flat plate cases from the ERCOFTAC dataset.
  \newadd{green}{Piecewise linear interpolation on a structured grid in feature-space is used as a {\em{sample}} functional form for the augmentation to demonstrate the capability of localized learning.
  The impact of using a different function class (neural network) is also assessed.}
  The augmented model is then applied to a variety of flat plate cases which are characterized by different freestream turbulence intensities, pressure gradients, and Reynolds numbers.
  The predictive capability of the augmented model is also tested on single-stage high-pressure-turbine cascade cases, and the model performance is analyzed from the perspective of information contained in the feature space.
  The results show consistent improvements across these cases, as long as the physical phenomena in question are well-represented in the training. 
\end{abstract}

\maketitle\thispagestyle{empty}

\clearpage
\setcounter{page}{1}

\section{Introduction}
  
  Traditional modeling of transition and turbulence has generally relied on theory and expert knowledge to determine the model-form, and on data for the calibration of resulting parameters.
  Recent advances \cite{duraisamy2019annrev, xiao2019quantification, duraisamy2020} have explored the possibility of using data to introduce complex functional forms within existing models to improve their accuracy and applicability.
  In the context of turbulence modeling, Oliver and Moser \cite{OliverMoser2011} were among the first to quantify model-form uncertainties by introducing discrepancies in the Reynolds stresses as spatially-dependent Gaussian random fields.
  Dow and Wang \cite{DowWang2011} applied a similar approach to introduce discrepancies in the eddy viscosity field, and  assimilated data from multiple flow fields.
  These precursors to the current machine learning approaches in RANS modeling are relevant as they treat model inadequacies as fields instead of parameters.
    
  Tracey et al., \cite{Tracey2013} in 2013, proposed the idea to transform such inadequacies from a field variable (a function of space) to a function of features (local functions of states and modeled quantities).
  This was applied to learn the discrepancies in the eigenvalues of the Reynolds anisotropy tensor between RANS and DNS results.
  A kernel regression ML model used the following features - eigenvalues of the anisotropy tensor, ratio of production-to-dissiptaion rate of turbulent kinetic energy and a marker function to mask the free shear layer regions in the flow.
  Xiao and co-workers \cite{Xiao2016, xiao2017physics, wang2017physics, wu2018physics} modeled the discrepancy in not just the eigenvalues of the anisotropy tensor but also its eigenvectors and turbulent kinetic energy while using a broader feature set.
  Since the values of features and the inadequacy term used to train the ML models are extracted from the DNS data, this approach is inherently inconsistent and unsuitable for predictive use via the baseline RANS models.
  In addition to this, the requirement of full field DNS data to create such ML models takes away from the robustness and versatility of this method.
    
  Ling and Templeton \cite{Ling2016} introduced Tensor Basis Neural Networks (TBNNs) to learn the coefficients of a Tensor Basis expansion (w.r.t. mean strain rate and vorticity tensors) intended to approximate the Reynolds stress term.
  The features used for this work included five invariants based on the mean strain rate and vorticity.
  TBNNs have, since, been used to model turbulent heat and scalar fluxes \cite{milani2020generalization, milani2021turbulent}. 
    
  Another class of approaches, which attempts to optimize the \textit{functional form} of the feature-to-augmentation map, is symbolic regression.
  Weatheritt and Sandberg \cite{Weatheritt2016} employed genetic programming to construct models for Reynolds stress anisotropy in terms of invariants of the velocity gradient tensor.
  Schmelzer et al. \cite{Schmelzer2020} used sparse regression over a library of candidate functions to obtain an algebraic Reynolds stress model.
  An advantage of such approaches is that the resulting model form is simple, and presents the added benefit of interpretability.
    
  Techniques to extract model inadequacy mentioned above do not explicitly enforce model consistency, i.e. the inadequacy information embedded in the modified model is extracted from higher-fidelity data \textit{directly} and can hence be inconsistent with the lower-fidelity model.
  The importance of model consistency is discussed in a recent review~\cite{duraisamy2020}.
  To address model consistency, Duraisamy and co-workers \cite{Duraisamy2015, ParishDuraisamy2016, FIML2017a, FIML2017c} introduced the field inversion and machine learning framework which aims to learn the inadequacy in the same environment as used for predictive purposes.
  To achieve this, one needs to solve an additional inverse problem prior to the machine learning step to obtain a model-consistent inadequacy field which is optimal in the sense that the corresponding outputs from the model match the target high-fidelity data as accurately as possible.
  While this framework mitigates inconsistency, it also facilitates the use of sparse high-fidelity data from experiments in contrast to full fields of direct numerical simulation data.
  To solve this so-called \textit{field inversion} problem, an adjoint-driven gradient-based optimization technique is used.
  The following subsection focuses on the developments and applications that this framework has witnessed since its inception.
  Other techniques and frameworks introduced later (e.g. \cite{Zhao2020, Liu2019, Girimaji2020}) also suggest the use of model-consistent training, and present opportunities of bypassing the requirement for adjoint-driven gradient computations by pursuing weakly-coupled inference.
  
% \subsection{Field Inversion and Machine Learning (FIML)}
  
  \newdel{\sout{The field inversion and machine learning framework, as originally introduced in \cite{ParishDuraisamy2016}, extracts functional relationships between features and augmentation from high-fidelity data in two steps.
  An inverse problem is solved using gradient-based optimization to obtain an optimal field of augmentation values which, when injected in the augmented formulation of a low-fidelity model, results in a minimum discrepancy between the calculated observables and high-fidelity data.
  Once this field is obtained, the features are calculated at all locations in the computational domain. 
  These features along with the augmentation values from different field inversion solutions which are all obtained using the same augmented formulation of the low-fidelity model are then used to obtain optimal ML-model parameters to establish a functional relationship between the features and the augmentation term.}}
  The original formulation of FIML, hereafter referred to as the classic FIML, has been used by several research groups, with applications including, but not limited to, predictive modeling of adverse pressure gradients flows \cite{FIML2017b,Matai2019}, separated flows \cite{FIML2017c,felix2020,FIMLC2019a,FIMLC2019b}, bypass transition modeling \cite{DuraisamyDurbin2014}, natural transition modeling \cite{YangXiao2020}, hypersonic aerothermal prediction for aerothermoelastic analysis \cite{VenegasHuang2020}, turbomachinery flows \cite{Larocca2020}, shock-turbulent boundary layer interactions \cite{STBLI2017}, etc.
  Matai et al. \cite{Matai2019} proposed a zonal version of FIML, where the augmentation field obtained from field inversion was quantized into a set number of clusters, following which a decision tree based architecture was used to classify corresponding features into appropriate clusters.
  The quantization reduced the number of classes that needed to be accounted for by the learning architecture and hence improved the training performance across a wide range of flow problems. 
    
  A significant evolution in the framework was proposed  by Holland et al. \cite{FIMLC2019a, FIMLC2019b} in the context of RANS, and \newdel{\sout{MacArt et al.}} \newadd{red}{Sirignano et al.} \cite{Sirignano2020} in the context of LES.
  These techniques integrate the learning step in the inference process.
  This is an improvement over classic FIML for two reasons - (1) the augmentation fields obtained using classic FIML may be non-unique which means that two near-optimal augmentation fields might have completely different functional representations in the feature space which might result in convergence problems when learning augmentations from several field inversion problems simultaneously; and (2) significant information loss might occur when learning an augmentation field (relatively high-dimensional vector) to obtain ML-model parameters (relatively low-dimensional vector) which might result in a considerable loss in predictive accuracy.
  This tight coupling between the two inverse problems, however, requires additional infrastructure for gradient evaluation, and presents convergence challenges due to the additional non-linearity of the learning algorithm.
  This framework is, hereafter, referred to as integrated inference and learning.
    
  In this work, we introduce another step in this evolution, viz. Learning and Inference assisted by Feature-space Engineering (LIFE).
  The approach is developed on the hypothesis that robustness and generalizability demand a low-dimensional feature space that is informed by the underlying physics, and that the features-to-augmentation map has to be carefully constructed.
  LIFE uses tightly coupled inference and learning to use data in order to locally update regions of feature space and can potentially make the augmented model more generalizable, robust and reliable by inherently suppressing spurious behavior and offering the modeler more control over the feature space.
  \newadd{red}{Note that, spurious behaviour, in the context of this work, refers to the augmented model predicting non-baseline behaviour by virtue of extrapolation to physical conditions which have not been encountered during training.}
  In addition, this work also introduces new constructs \newadd{green}{(guiding principles and techniques)} across different stages in the inference process that contribute towards boosting the efficacy and generalizability of the resulting augmentation while underlining the importance of human intuition and expert knowledge in the process.

  \subsection{RANS models for bypass transition}
  
    To demonstrate the capability of LIFE, we choose the problem of developing a data-driven bypass transition model.
    Transition modeling has been studied over the past two decades, and several models are in existence.
    In this work, however, the aim is not to create a universally accurate transition model.
    Rather, the goal is to develop and demonstrate LIFE and examine the possibility of generalization in the low data limit using bypass transition as an example. 
    
    The phenomenon of laminar-to-turbulent transition is of vital importance in the design of aerospace vehicles and energy systems.
    Transition can occur either by the amplification and non-linear interactions of 2-D instability waves (also known as Tollmien-Schlichting waves) or via the perturbation of an otherwise laminar boundary layer by freestream turbulence, surface roughness, impinging wakes, etc.
    The latter is referred to as the bypass mode of transition, and is typically the preferred route for transition when freestream turbulence intensities are more than 1\% \cite{Mayle1991}.
    While Reynolds-Averaged Navier-Stokes (RANS) simulations have remained the workhorse in the industry for design and optimization and a subject of active research for decades, bypass transition models for RANS turbulence closures are relatively in their infancy. 
    
    Two major categories of approaches exist in the literature to model bypass transition induced by freestream turbulence, viz. (1) Data correlation based models, where the model predicts the transition onset location based on some criteria (usually a correlation between momentum thickness Reynolds number $Re_{\theta t}$ and freestream turbulence intensity $Tu_\infty$) and switches from laminar to turbulent computation \cite{Mayle1991, DhawanNarasimha1958, GhannamShaw1980, PraisnerClark2004}; and (2) transport equation based models, where the turbulence model is modified by introducing a new transport scalar, which is solved for using an additional equation.
    Further, there are two major approaches within transport equation-based models - laminar fluctuation models and intermittency function models:
    a) Laminar fluctuation models are based on the modeling of the so-called laminar kinetic energy, $k_L$, i.e. the energy of fluctuations in laminar boundary layers corresponding to Klebanoff modes; and the transfer of energy between the laminar disturbances and turbulent eddies by introducing a transfer term in the transport equations of $k_L$ and $k$ (modeled turbulent kinetic energy) \cite{MayleSchulz1997, Lardeau2009, WaltersCokljat2008};
    b) Intermittency function models, on the other hand, model the intermittency $\gamma$ (which is a statistical quantity referring to the fraction of the time the flow is turbulent and varies from 0 in laminar flow to 1 in fully turbulent flow) which is usually multiplied to the production term in the transport equation for $k$ or directly to the eddy viscosity, thus, attenuating the production of turbulence in the regions supposed to be laminar or transitional \cite{ChoChung1992, SteelantDick1996, SuzenHuang2000, Suzen2003, Menter2006, LangtryMenter2009, Durbin2012, Ge2014}.

    In this work, we utilize the idea of LIFE to create a data-driven intermittency-based bypass transition model using data from only two flat plate cases, and predict the bypass transition triggered by freestream turbulence across different flow conditions and geometries.
    
    The outline of this paper is as follows: The formal problem description of model augmentation, and a background of the Field Inversion and Machine Learning Framework is laid out in section \ref{sec:Background}.
    In section \ref{sec:Methodology_NewFramework}, the methodology of LIFE, which includes its philosophy, the required infrastructure and related algorithms is presented.
    Section \ref{sec:TransitionModel} begins with relevant background of the bypass transition modeling problem necessary for, and followed by, the description of baseline model-form used in this work, how the augmentation is introduced and what features are used to express the functional form of the augmentation.
    In section \ref{sec:Results}, the training and testing results of the transition model augmentation is shown and analyzed.
    Section \ref{sec:Conclusions} summarizes the work.
    \newadd{red}{The appendices details the impact of modeling choices on the results of the modeling augmentation.}
\section{Problem Statement \& Background} \label{sec:Background}

  \subsection{Premise} \label{ssec:Methodology_Premise}
  
    \subsubsection{The closure problem} \label{sssec:Methodology_Premise_Closure}
  
      Consider a physical system described on a spatial domain $\Omega$, the governing equations for which can be represented with appropriate boundary conditions as follows
      \begin{equation} \frac{\partial \bs{q}}{\partial t}+\mathscr{R}(\bs{q}) = 0 \quad\forall\quad \bs{x}\in\Omega \end{equation}
      where $\bs{q}(\bs{x},t)$ represents the highly resolved spatio-temporal field of state variables.
      For practical problems, like flow over an aircraft wing or inside a gas turbine combustor, direct simulations of the underlying physics with the required spatio-temporal resolution are presently infeasible.
      \newdel{\sout{This gives rise to the need for reduced-fidelity models which seek to lower these resolution requirements by trying to solve for coarse-grained states $\widetilde{\bs{q}}$.
      For instance, in the simulation of fluid flows, two broad classes of such models are large eddy simulations (LES) and Reynolds-averaged Navier-Stokes (RANS) simulations.
      LES achieves this coarse-graining via implicit or explicit low-pass filtering while RANS makes use of the idea of ensemble-averaging to do so.}}
      In turbulence modeling, the state variables are decomposed into  coarse-grained and unresolved parts, represented by $\widetilde{\bs{q}}$ and $\widehat{\bs{q}}$, respectively.
      Performing the coarse-graining operation on the equations representing the high-fidelity system results in an unknown closure term $\mathscr{N}(\bs{q})$, arising from any non-linearities in the operator $\mathscr{R}$ and the unaccounted contribution from $\widehat{\bs{q}}$.
      Note that the following representation is not an approximation.
      \begin{equation} \frac{\partial\widetilde{\bs{q}}}{\partial t} + \widetilde{\mathscr{R}(\bs{q})} = 0 \qquad\Rightarrow\qquad \frac{\partial\widetilde{\bs{q}}}{\partial t} + \mathscr{R}(\widetilde{\bs{q}}) + \mathscr{N}(\bs{q}) = 0 \end{equation}
      \newdel{\sout{In the context of fluid mechanics, the task of modeling the closure term using information from the coarse-grained states is referred to as turbulence modeling.}}
      In a generic sense, a closure model can be written as $\mathscr{N}_m(\widetilde{\bs{q}}_m, \widetilde{\bs{s}}_m)$, where $\widetilde{\bs{q}}_m$ refers to the state variables for the reduced-fidelity model which will be different from the coarse-grained variables $\widetilde{\bs{q}}$ because of a \textit{modeled} closure term.
      $\widetilde{\bs{s}}_m$ represents the vector of any secondary variables (e.g. turbulent kinetic energy and other scale providing variables) which might need to be solved for by using an additional system of equations as shown below.
      \begin{equation} \frac{\partial \widetilde{\bs{s}}_m}{\partial t} + \mathscr{G}_m(\widetilde{\bs{s}}_m, \widetilde{\bs{q}}_m) = 0 \quad\forall\quad \bs{x}\in\widetilde{\Omega} \end{equation}
      where $\widetilde{\Omega}$ represents the appropriately discretized version of the original domain $\Omega$.
      \newdel{\sout{The construction of operators $\mathscr{N}_m$ and $\mathscr{G}_m$ is a tedious and meticulous process which has evolved over decades through a combination of physical insight, mathematics and empiricism.}}
      In this work, the focus is restricted to steady-state reduced-fidelity models, which, using the terminology described above, can be represented as
      \begin{equation} \mathscr{R}(\widetilde{\bs{q}}_m) + \mathscr{N}_m(\widetilde{\bs{q}}_m, \widetilde{\bs{s}}_m)= 0 \quad ; \quad \mathscr{G}_m(\widetilde{\bs{s}}_m, \widetilde{\bs{q}}_m) = 0 \end{equation}
      For ease of notation, we shall refer to this system of equations in a compact manner $\mathscr{R}_m(\RFvar; \bs{\xi}) = 0$, with the state variables and secondary variables combined into a single vector of model variables $\RFvar = \left[ \widetilde{\bs{q}}_m^T \quad \widetilde{\bs{s}}_m^T \right]^T$, and the inputs to the model (discretized domain, boundary conditions, etc.) embedded into the notation via $\bs{\xi}$.
    
    \subsubsection{A surrogate measure of model inadequacy} \label{sssec:Premise_AugField}
    
      Reduced-fidelity models of turbulence and transition can be insufficiently accurate for use in predictive simulations that aid design and optimization owing to model-form inadequacies in the representation of the operators $\mathscr{N}_m$ and/or $\mathscr{G}_m$.
      To address these inadequacies, one can ``augment'' the model which involves appropriately introducing an inadequacy field, $\beta(\bs{x})$ in the model-form of $\mathscr{R}_m$, \newadd{red}{giving rise to the following augmented model.}
      \newdel{\sout{This can be done by adding or multiplying some quantity in the model formulation with $\beta$.
      This results in an augmented model, the generic mathematical representation of which can be given as}}
      \begin{equation} \mathscr{R}_m(\RFvar; \beta(\bs{x}), \bs{\xi})= 0 \end{equation}
      Examples of how such inadequacy terms have been introduced into turbulence models in the available literature include multiplication with the production term of the Spalart-Allmaras model by Singh et al. in \cite{FIML2017c}, addition to the eigenvalues of the Reynolds anisotropy tensor by Xiao et al. \cite{Tracey2013, Xiao2016}. Indeed, $\beta$ can represent the vector of coefficients in the tensor basis expansion for the Reynolds anisotropy tensor as proposed by Ling et al. \cite{Ling2016}.
      \newdel{\sout{
      The following aspects could be considered to decide how to introduce an augmentation in the baseline model:}
      \begin{itemize}
        \item \sout{\textbf{Efficacy:} The augmentation should be capable of effectively addressing the inadequacy in consideration. For instance, augmenting the coefficient of an isotropic diffusion term in a model cannot correct for an inadequacy caused as a consequence of the anisotropically diffusive behaviour in the corresponding true system.}
        \item \sout{\textbf{Spurious behavior:} The augmentation should be introduced such that it does not corrupt model behavior in regions where the inadequacy under consideration is not a concern. For instance, a RANS model augmented to predict transition should not change the model behavior outside the boundary layer or in regions inside the boundary layer in which the flow is fully turbulent.}
        \item \sout{\textbf{Sensitivity:} In practice, the predicted augmentation values might have small errors which can have significant adverse effects on the stability and accuracy of the model if it is highly sensitive to the augmentation. Hence, the augmentation should be introduced in a manner that mitigates this as much as possible.}
      \end{itemize}
      }

    \subsubsection{Model consistency and data}

      The task of quantifying the model inadequacy now reduces to inferring some ideal $\beta(\bs{x})$ field, that when used in this augmented model, minimizes \newdel{\sout{some estimate of prediction error in quantities of interest}} \newadd{red}{a chosen measure of discrepancy (e.g. L2
      norm) between the data and corresponding predictions for quantities of interest}.
      Note that, even if the full high-fidelity field of state variables is available, data alone should not be used to obtain $\beta$ values by simply plugging values into the reduced-fidelity model at different locations, as this could result in a field which is inconsistent with the model and in a general case may lead to prediction inaccuracies \cite{duraisamy2020}.
      In other words, the inferred $\beta(\bs{x})$ field is only a surrogate measure of the inadequacy, and that the true model inadequacy \textit{might} be slightly different than this field, as $\widetilde{\bs q}$ and particularly $\widetilde{\bs{s}}$ may be different from $\widetilde{\bs q}_m$ and $\widetilde{\bs{s}}_m$, respectively. 
      
      \newdel{\sout{The next step is to formulate the estimate of prediction error, which will be referred to as the cost function hereafter, in terms of available data and model predictions. The data acquired from experiments or higher-fidelity simulations correspond to certain statistical or physical quantities.
      These quantities are referred to as observables, and can be local (i.e. dependent on quantities at a single point in the domain, e.g. pressure, velocity, Reynolds stresses etc.) or integral (i.e. dependent on model quantities at multiple points in the domain, e.g. aerodynamic forces and moments).}}
      For a dataset (a dataset in the present context refers to a unique combination of domain geometry and flow conditions) referred to by the index $i$, a discrepancy between the vector of high-fidelity data for these observables, $\bs{y}^i$, and the corresponding vector of predictions using the augmented model, $\bs{y}^i_m(\iRFvar)$, can be used to formulate the cost function for this case as $\mathcal{C}^i(\bs{y}^i, \bs{y}^i_m(\iRFvar))$.
      Since the field $\iRFvar$ necessarily satisfies $\mathscr{R}_m(\iRFvar; \beta^i, \bs{\xi}^i)=0$, it implicitly depends on the augmentation field, $\beta^i(\bs{x})$, and so, the cost function is ultimately a function of $\beta^i(\bs{x})$.
      Thus, a gradient-based optimization can, potentially, be used to infer $\beta^i(\bs{x})$.
      A rudimentary candidate for the cost function can be the square of $L_2$-norm of the difference between the two vectors.
      This process of inference will be explained in greater detail in the section \ref{ssec:Methodology_FIML}.

    \subsubsection{Representing inadequacies via augmentation functions} \label{sssec:Premise_AugFn}
    
      While the augmentation field, $\beta(\bs{x})$, provides information about the local inadequacy in the model, it cannot be used for predictive improvements on its own.
      However, if a consistent functional relationship could be extracted between $\beta(\bs{x})$ and some carefully-chosen features, $\bs{\eta}(\RFvar,\bs{\zeta})$, this function could replace the $\beta(\bs{x})$ field in the reduced-fidelity model equations for predictive use.
      Here $\bs{\zeta}$ denotes local quantities independent of the state or secondary variables which are used to design features.
      Note that, for the augmentation to be usable in a predictive setting with a wide range of applicability, it should ideally have no dependence on the domain geometries or boundary conditions used for inference.
      \newdel{\sout{It is also essential that the mapping between features and augmentation is strictly local.}}
      \newadd{red}{The augmentation is assumed to be dependent only on local quantities in the present work as including non-local information in the features requires a more sophisticated, and potentially less general treatment.}
      The features must be invariant \cite{wu2018physics,duraisamy2020} to appropriate transformations (e.g. rotation, as they must not depend on the orientation of the coordinate axes.) Given the features-to-augmentation map, the model can be written as
      \begin{equation} \mathscr{R}_m(\RFvar, \beta(\bs{\eta}(\RFvar,\bs{\zeta}); \bs{w}); \bs{\xi}) = 0. \end{equation}
      It should be mentioned here that, once the model class of the augmentation (e.g. a neural network) is chosen, the goal is to infer the parameters $\bs{w}$.
      The section \ref{ssec:Methodology_FIML} focuses on  variants of the FIML framework to obtain the parameters $\bs{w}$ from sparse data.
  
  \subsection{Field Inversion and Machine Learning (FIML)} \label{ssec:Methodology_FIML}
  
    The field inversion and machine learning framework, originally proposed by Duraisamy and co-workers, \cite{Duraisamy2015,ParishDuraisamy2016}, was formulated to reduce inadequacies in a given reduced-fidelity model by inferring optimal augmentations such that the predictive accuracy of the model is improved.
    Point-wise and integrable data of variable sparsity across multiple cases can be used from experiments and higher-fidelity simulations.
    The two main variants of this framework are been discussed in the following sections.
    
    \subsubsection{Classic FIML} \label{sssec:Methodology_FIML_Classic}
    
      FIML seeks the optimal augmentation into two steps: Firstly, a ``field inversion'' problem is solved separately over each dataset to infer optimal augmentation fields, $\beta(\bs{x})$, in the respective discretized domains.
      Then, the step of machine learning uses these inferred augmentation values and respective features for all spatial locations across all datasets and infers the optimal augmentation function parameters $\bs{w}$. These two steps are described below.
      
      \paragraph{Field Inversion}
        \begin{equation} \label{eq:FI} \begin{split} \beta^{*i}(\bs{x}) = \text{arg}\min_{\beta^i(\bs{x})} \left\lbrace \mathcal{C}^i(\bs{y}^i,  \bs{y}^i_m(\iRFvar)) + \lambda_\beta^i \mathcal{T}^i_\beta(\beta^i(\bs{x})) \right\rbrace\\
        \text{s.t.} \quad \mathscr{R}_m(\iRFvar; \beta^i(\bs{x}), \bs{\xi}^i) = 0 \quad\forall\quad  i=1,2,\hdots,N \end{split} \end{equation}
        Equation \ref{eq:FI} represents the field inversion problem which is solved individually over each of the $N$ training cases (cases for which available high-fidelity data is to be used to infer the augmentation function parameters) to obtain optimal augmentation fields $\beta^{*i}(\bs{x})$ in the respective domains, where $i$ is the case index.
        The augmentation field is optimal in the sense that it minimizes an objective function which consists of a cost function, $\mathcal{C}^i$, and a regularization term, $\mathcal{T}^i_\beta$, with $\lambda_\beta^i$ as the regularization constant.
        \newdel{\sout{The cost function, as described previously, quantifies the discrepancy between the available data and the corresponding model predictions for observables belonging to a given case.
        Since, this inverse problem, in general, will be ill-posed, a regularization on the augmentation field can be used as a part of the objective function to address this issue.
        A number of regularization candidates can be used, including those resulting from Bayesian inference.
        One of the simplest combinations of cost function and regularization is to use an $L_2$ norm to estimate both these quantities as shown below.}}
        \begin{equation} \mathcal{C}^i = \lVert \bs{y}^i - \bs{y}^i_m(\iRFvar) \rVert_2^2 \qquad ; \qquad \mathcal{T}^i = \lVert \beta^i(\bs{x}) - \beta_0 \rVert_2^2 \end{equation}
        \newdel{\sout{where $\beta_0$ is the baseline value of the model, i.e. $\beta_0=0$ if $\beta$ was added to some term in the model and $\beta_0=1$ if it was multiplied to some term in the model.
        Here, it is also worth noting that it is preferred to multiply the augmentation to some term in the model as then the augmentation is a dimensionless quantity which could, arguably, make generalization easier.}}
        \newadd{red}{Note here that an additive inadequacy term which is non-dimensionalized with the same turbulent length and time scales as the source term, can also be viewed as a multiplicative inadequacy term.}
        
        Given the very high-dimensional nature of the augmentation fields that need to be inferred, the field inversion problem is usually solved using a gradient-based optimization technique.
        \newdel{\sout{in which the gradients are computed via discrete adjoints, the details for which can be found in Appendix \ref{app:Adjoints}.
        In case, the reduction in the objective function relative to its baseline value is not sufficient (greater than a set threshold $\varepsilon$), the procedure can be repeated by starting from the last obtained augmentation field with a reduced step size, if needed.
        This does not imply that the objective function will be minimized to a value close to zero for every inference problem, rather $\varepsilon$ could be a threshold significantly greater than zero set by the user, as deemed appropriate for a given problem and inadequacy term.}}
      
      \paragraph{Supervised Learning}~\bigskip{}\vspace{-1em}
      
        Once the field inversion problem is solved for all available datasets, the supervised learning problem can be solved using the augmentation fields and correspondingly calculated feature fields from all datasets to optimize for the function parameters $\bs{w}$ as 
        $ \bs{w} = \text{arg}\min_{\bs{w'}} \mathcal{L}(\bs{\beta}^{*},\beta(\bs{\eta}^*;\bs{w'})) $
        where, $\bs{\beta}^*$ refers to the stacked vector containing optimal augmentation fields across all datasets, and $\bs{\eta}^*$ contains the respectively stacked feature values.
        \newdel{\sout{This can be written as a vector/matrix created by stacking vectors/matrices from individual cases one below another}
        \[
          \hbox{\sout{$
          \bs{\beta}^* = \begin{bmatrix}
            \beta^{*1}(\bs{x}) \\ \beta^{*2}(\bs{x}) \\ \vdots \\ \beta^{*N}(\bs{x})
          \end{bmatrix} \qquad
          \bs{\eta}^* = \begin{bmatrix}
                \bs{\eta}(\sOneRFvar(\bs{x}), \bs{\zeta}^1(\bs{x})) \\
                \bs{\eta}(\sTwoRFvar(\bs{x}), \bs{\zeta}^2(\bs{x})) \\
                \vdots \\ \bs{\eta}(\sNumRFvar(\bs{x}), \bs{\zeta}^N(\bs{x}))
          \end{bmatrix}$
          }}
        \]
        \sout{ $\mathcal{L}(.,.)$ refers to some appropriate loss function that measures the discrepancy between the optimal augmentation field data obtained from field inversion and predictions made by the augmentation function under training with respective features as inputs.
        A simple loss function is the squared $L_2$ norm of the difference between the two, given as follows}
        \[
          \hbox{\sout{$\displaystyle
          \mathcal{L}(\bs{\beta}^{*},\beta(\bs{\eta}^*;\bs{w}')) =
          \left\lVert \bs{\beta}^{*},\beta(\bs{\eta}^*;\bs{w}') \right\rVert_2^2$
          }}
        \]
        \sout{A mini-batch gradient descent technique is well-suited to learn the function parameters in most cases when used with Adam or L-BFGS optimizer.}}
        Note that, while the inferred augmentation from the field inversion step is fully consistent with the underlying model, the augmentation field provided by the field inversion process is not necessarily learnable as a function of the chosen features~\cite{FIMLC2019b}. This can lead to a loss of information extracted in the field inversion step.
        \newadd{cyan}{
          A simple measure of learnability can be defined as follows.
            $$ \text{Learnability} = 1.0 - 
               \frac{1}{n}\sum_{i=0}^N
               \frac{\lvert\beta(\bs{\eta}_i;\bs{w}^*)-\beta^*_i\rvert}
                    {\lvert\beta(\bs{\eta}_i;\bs{w}^*)\rvert+\lvert\beta^*_i\rvert}
            $$
        where $\bs{w}^*$ is the globally optimal set of parameters that minimizes the discrepancy between the augmentation values obtained using field inversion and corresponding machine learning predictions, and $n$ is the number of datapoints.
        This measure can assume values between $0$ and $1$ such that higher value refers to higher learnability.}
        As discussed previously, the augmentation field obtained from field inversion is not unique, it is possible that the field inversion results obtained from two different datasets correspond to different features-to-augmentation mappings.
        This inconsistency can also degrade the learnability of the model.
        The more recent variant of FIML, referred to as integrated inference and learning addresses these concerns.

    \subsubsection{Integrated Inference and Learning} \label{ssec:Methodology_FIML_Integrated}
    
      The integrated inference and learning approach, proposed by Holland et al. \cite{FIMLC2019a, FIMLC2019b} and Sirignano et al. \cite{Sirignano2020} replaces the two-step inference problem introduced in classic FIML by just one inverse problem, mentioned as follows
      \begin{equation}
        \begin{split}
          \bs{w} = \text{arg}\min_{\bs{w}'} \bigsqcup_{i=1}^N
          \left(\mathcal{C}^i(\bs{y}^i, \bs{y}^i_m(\iRFvar)) +
          \lambda_\beta^i\mathcal{T}_\beta^i(\beta(\bs{\eta}^i(\iRFvar,\bs{\zeta});\bs{w}'))\right)\\
          \text{s.t. } \mathscr{R}_m(\iRFvar, \beta(\eta(\iRFvar,\bs{\zeta}^i); \bs{w}'); \bs{\xi}^i) = 0
          \quad\forall\quad i=1,2,\hdots,N,
        \end{split}
      \end{equation}
      where $\bigsqcup$ is a generic assembly operator to build a composite objective function using the individual objective functions calculated for each dataset.
      It can be as simple as a sum, if all datasets are equally important for the inference, or it could be a weighted sum, if some datasets are to be assigned more importance than others, or it could be something
      even more complex as designed/needed by the user.
      Notice here that using several cases at once adds an implicit regularization to the problem.
      Similar to the field inversion problem, the required sensitivities can be obtained using discrete adjoints as described in Appendix \ref{app:Adjoints}.
      \newdel{\sout{Here, an interesting observation can be made.
      Since the calculation of sensitivities requires $\dfrac{\partial\beta^i}{\partial{\bs{\eta}^i}}$, and since it is tractable to estimate derivatives of any differentiable functional form chosen for the augmentation function analytically, the reduced-fidelity model can be provided with only a locally linearized approximation of the augmentation w.r.t. the feature space during inference, without any loss of information.
      Thus, the adjoint framework need not be aware of the functional form of the augmentation, which makes the implementation slightly easier.}}

      \paragraph*{Benefits over classic FIML}
        \begin{itemize}
            \item \newdel{\sout{It has been observed that}}
            Classic FIML can suffer from a loss of information during the machine learning step owing to the inability of the chosen functional form to accurately approximate the augmentation inferred during the field inversion step as a function of the specified features.
            Integrated inference and training bypasses this problem as the weights are directly updated and consequently, the obtained augmentation field remains consistent with the functional form of the augmentation.
            Thinking in terms of the augmentation field, this constrains the optimization to minimize the objective function to find an augmentation field which is realizable w.r.t. its functional form.
            
            \item Since integrated inference and learning can infer the function parameters $\bs{w}$ while simultaneously assimilating data from multiple datasets, the augmentation fields for all datasets is constrained to be realizable w.r.t. the functional form of the augmentation as explained in the previous point.
            This has an added advantage that this procedure, unlike classic FIML, is by design prevented from learning augmentation fields from different datasets that behave differently in the feature space.
            In other words, consistency in the features-to-augmentation mapping across datasets is automatically enforced when using integrated inference and learning.
            
            \item Building on the previous point, this means that if we have even a handful of DNS(Direct Numerical Simulation) fields from which true augmentation values can be extracted (or are readily available), they can be used to enforce a near-physical relationship between the features and augmentation by simultaneously using a plethora of other sparse field data from experiments or higher fidelity simulations. \textbf{This is critical from the viewpoint of generalizable and physical augmentations.}
        \end{itemize}
        
      \paragraph*{Limitations}
        \begin{itemize}
            \item Unlike the classic FIML approach, the task of designing an appropriate feature space must precede the inference from data.
            This, at times, could be difficult for the modeler and demonstrates the preliminary need of an independent field inversion step which can lend crucial information about the quantities correlated to the augmentation field that may then be used to formulate features.
            
            \item In its original form, integrated inference and learning does not consider the significance of localized learning.
            For practical problems, the data being used might not populate the entire feature space, which means that if the optimization problem is not constrained to change the augmentation only in the vicinity of the feature space locations for which data is available, it might lead to spurious predictions in other regions, which might not only result in worse accuracy compared to the baseline model but also severely affect the stability of the numerical solver.
            
            \item When using complex learning algorithms such as neural networks or decision trees, the augmented model inherits non-linearities from  the baseline model {\bf and} the learning algorithm.
            This, combined with the previous point, can lead to a disorderly optimization behavior when solving the inference problem, in addition to the aforementioned potential deterioration in accuracy and/or numerical stability with every successive optimization iteration.
        \end{itemize}

\section{Model-Consistent Learning and Inference assisted by Feature-space Engineering (LIFE)}
\label{sec:Methodology_NewFramework}

  \newdel{\sout{The Learning and Inference assisted by Feature-space Engineering approach is described below.
  To create an augmentation which is generalizable, i.e., applicable across flows with significantly different geometries and boundary conditions such that it provides consistent improvements, the following questions should be considered:}}
  
  \newadd{green}{To alleviate the limitations in integrated inference and learning, and to make the augmentation robust and generalizable we present a set of guiding principles to choose features along with the notion of localized learning in the feature space which requires controlling how learning takes place in different parts of the feature space. Since a significant effort is put into how features are chosen and how the learning takes place in different parts of the feature space, we call this version of integrated inference and learning as ``Learning and Inference assisted by Feature-space Engineering (LIFE)''.
  
  This section is structured as follows.
  Section \ref{ssec:augmentation_design} and \ref{ssec:feature_design} propose guiding principles to take into consideration when designing the augmentation and features.
  Section \ref{ssec:localized_learning} deals with the notion of localized learning and provides three different ideas on how to make it work.
  Finally, \ref{ssec:implementation} deals with the algorithm and practical concerns of implementing the LIFE framework.
  
  \subsection{How should a model augmentation be designed?} \label{ssec:augmentation_design}

    The following aspects could be considered to decide how to introduce an augmentation in the baseline model:
    
    {\bf 1. Efficacy:}
    The augmentation should be capable of effectively addressing the inadequacy in consideration.
    For instance, augmenting the coefficient of an isotropic diffusion term in a model cannot correct for an inadequacy caused as a consequence of the anisotropically diffusive behaviour in the corresponding true system.
    
    {\bf 2. Spurious behavior:}
    The augmentation should be introduced such that it does not corrupt model behavior in regions where the inadequacy under consideration is not a concern.
    For instance, a RANS model augmented to predict transition should not change the model behavior outside the boundary layer or in regions inside the boundary layer in which the flow is fully turbulent.
    
    {\bf 3. Sensitivity:}
    In practice, the predicted augmentation values might have small errors which can have significant adverse effects on the stability and accuracy of the model if it is highly sensitive to the augmentation.
    Hence, the augmentation should be introduced in a manner that mitigates this as much as possible.}

  \subsection{What features should the augmentation depend on?} \label{ssec:feature_design}

    \newadd{green}{To decide the features that the augmentation function would depend on, one should use the following guiding principles.
    
    {\bf 1. Choice of features using expert knowledge}
    
    The choice of features has a major role to play in how effective the augmentation function can be at addressing the model inadequacy in consideration.
    While automated feature selection techniques exist in literature, they heavily (if not prohibitively) depend on data to find the best features from amongst hundreds/thousands of possible candidate functions of local quantities for a complex problem like transition or turbulence.
    The authors, hence, believe that human intuition and expert knowledge can lead the way in feature selection as it has proven effective for traditional modeling.
    This is one of the major reasons why the LIFE framework is aimed towards use by expert modelers. 
    While choosing features, it is desired that there exists a causal relationship between the chosen features and the inadequacy targeted by the augmentation function.
    However, for steady-state models, the augmentation in turn influences the features by virtue of feedback, and hence quantities which do not share a causal relationship with the inadequacy and, rather are only correlated with it, can also serve as features.
    
    {\bf 2. Physics-based non-dimensionalization}
    
    Once the quantities to be used as features are chosen, they must be non-dimensionalized in order to be generalizably used for prediction.
    This is because similar physical phenomena can occur for significantly different magnitudes of dimensional model quantities and, since the LIFE framework aims at using as little data as possible to discover generalizable augmentation functions, properly non-dimensionalization of the features becomes imperative.
    While statistics from the training datasets can be used to non-dimensionalize features, the available data might not be sufficiently representative of the complete range of values that could be encountered during prediction on an unseen case.
    Thus, it is better to make use of model quantities to non-dimensionalize features. 
      
    A simple example can be used to illustrate this point.
    Let us consider that the eddy viscosity $\nu_t$ is intended for use as a feature.
    Technically speaking, $0\leq\nu_t<\infty$.
    With limited data, it seems hopeless that the value of $\nu_t$ can be used as a feature in a generic predictive setting.
    However, simply using the turbulent Reynolds number $\nu_t/\nu$ can immediately improve the quality of results.
    This is because we take some burden of discerning physical conditions for a given combination of features off of the learning algorithm and make the integrated inference and learning problem better-posed.
    Note that one can also non-dimensionalize $\nu_t$ as $\nu_t\lvert\bs{\Omega}\rvert/\lvert\bs{U}\rvert^2$, but the physical conditions represented here will be different than the ones represented by $\nu_t/\nu$. 
    Which non-dimensionalization is chosen depends on its relevance w.r.t. the inadequacy to be alleviated. 
    While the example shown here seems trivial, physics-based non-dimensionalization can prove to be a challenge in some scenarios which might require domain-specific theoretical knowledge and/or experimentally obtained correlations to resolve.
    
    {\bf 3. Effectively bounded feature-space}
    
    While non-dimensionalization is critical, it should be remembered that a non-dimensionalized feature can still be unbounded (e.g. $\nu_t/\nu$).
    Since the data used to learn the augmentation is limited, this can lead to extrapolated predictions by the augmentation function for features outside the range of values available from the training data.
    To circumvent this, the functional form of the non-dimensionalized feature must be chosen such that either the feature is mathematically bounded or it takes non-baseline values (baseline value is $0$ for an additive augmentation and $1$ for a multiplicative augmentation) only in a bounded part of the feature space.
    We shall call such a feature as ``effectively'' bounded, hereafter.
    Effective boundedness minimizes extrapolation errors and makes the augmentation more robust and generalizable.
    
    {\bf 4. Appropriate functional forms for features}
    
    Several functional forms of a non-dimensionalized feature can offer effective boundedness.
    But, only a few of these forms might address the inadequacy in a major part of the feature space.
    For instance, the part of the feature space covered by the feature $\nu/(\nu_t+\nu)$ between the values of $0$ and $0.3$ reduces to being covered between the values $0$ and $0.000729$ for a different form of the same feature, viz. $(\nu/(\nu_t+\nu)^6)$.
    Note that, both features have the same mathematical range between $0$ and $1$.
    However, by the virtue of different functional forms, the regions denoting the same physical conditions span very differently in the feature space. For an augmentation which is predominantly affected within the said range, the feature $\nu/(\nu_t+\nu)$ offers a better-conditioned learning problem.
    Hence, the choice of which among different effectively bounded non-dimensional functional forms of a feature to use can play a significant role in setting up a well-conditioned learning problem.
      
    {\bf 5. Parsimonious combination of features vs. one-to-one features-to-augmentation mapping}
    
    Finally, the augmentation should be a function of as small a number of features as possible to maintain simplicity, as a simpler augmentation could mean better generalizability when training on limited data.
    On the other hand, the features should be chosen such that, there exists a one-to-one feature-to-augmentation mapping.
    While a perfect one-to-one mapping might be exceedingly difficult to achieve in the entire feature space, the property should be virtually preserved (i.e. if more than one, all possible augmentation values should be close to each other) in as large a region of the effectively bounded feature space as possible.
    More features imply a better chance of ensuring such a mapping.
    Note that ``mapping'' here refers to the true relation between features and corresponding optimal augmentation values for all locations in the feature space.
    It does \textbf{not} refer to the augmentation function (which is one-to-one by definition and tries to approximate this mapping).
    If the mapping is not one-to-one, then during the inference and learning process, sensitivities from two different datapoints might try to pull the augmentation value at some location in the feature space in opposite directions and the so-obtained optimal augmentation function would predict a compromise between two significantly different values which are optimal w.r.t. each of the two datapoints.
    Physically, this translates to the feature space being inadequate to uniquely represent distinct physical phenomena corresponding to significantly different augmentation values.
    Thus, a balance has to be attained by choosing just enough features to ensure a virtually one-to-one mapping in most of the effectively bounded feature space.}
      
  \subsection{Which class of functions should be used to model this augmentation?} \label{ssec:localized_learning}
  
    Depending on the complexity of the behavior of the augmentation in feature space, there exist different alternatives to choose the functional form of the augmentation prior to the inference and learning procedure.
    If the behavior is very simple and the feature space is very low-dimensional, one can hand-fit a function on the inferred set of augmentation values.
    If the behavior is not intractably complex, a user-defined functional form can be used.
    Lastly, if the behavior is highly non-linear, a class of functions available in the machine learning literature (decision trees, neural networks, etc.) could be used\newdel{\sout{, with the associated convergence difficulties in model-consistent inference}}.
          
    \newdel{\sout{In this work, we use a class of functions that relies on feature space discretization which aids in a greater control over augmentation behavior in different parts of the feature space. This is described in detail in section III C.}}
    
    \newadd{green}{This choice becomes even more important when the available data does not span/represent all parts of the effectively bounded feature space.
    In such a scenario the augmentation function might extrapolate to predict values for feature space regions corresponding to which no data was available.
    This could lead to spurious model behavior and could make the prediction capabilities of the augmented model worse than its baseline counterpart.
    To preempt this complication the idea of localized learning can be used.
    
    {\bf Localized Learning}, here, refers to the class of learning techniques that modifies the augmentation function behavior only in the vicinity of available data without perturbing augmentation values in regions far from available datapoints. Thus, the augmentation value at any point in the feature space could be either ``influenced'' by one or more datapoints, or remain ``uninfluenced''.
    Three different ways to perform localized learning are mentioned as follows:
    \begin{enumerate}
        \item {\bf Discretizing the feature space} into subdomains and representing the augmentation function as a piecewise combination of local augmentation functions corresponding to each of these subdomains.
        Here, a datapoint influences only those points in the feature space that lie in its own subdomain.
        \item {\bf Introducing artificial datapoints} in the learning process by sampling from the existing augmentation field in regions where no datapoints are available.
        Here, the sampling strategy determine the influenced regions in the feature space.
        \item {\bf Superposing finite support functions} (e.g., truncated Gaussians) centered at available datapoints can be added on top of the existing augmentation to update it.
        Here the support width of these individual functions characterize the influence of the datapoints in the feature space.
    \end{enumerate}
    
    Choosing how fine/coarse the discretization is (in case of method 1), or the strategy used to sample points from existing augmentation field (in case of method 2), or the support width (in case of method 3) in order to control the influence of datapoints in the feature space requires the user to make a trade-off. If, for a given datapoint, the region of influence is very small, the augmentation would behave as an over-fitted function which, while performing accurately on the training cases, will not be able to generalize well. On the other hand, if the region of influence is too large, the augmentation will be over-regularized and the predictive accuracy will suffer, again resulting in a loss of generalizability.
    
    Localized learning forces the augmentation function to retain the baseline behavior for feature space regions which remain uninfluenced throughout the LIFE process.
    The certainty that the augmented model reverts back to its baseline behavior when faced with feature combinations (physical conditions) not encountered during the LIFE process means that the performance of the augmented model would always be better than or equal to the baseline version.
    
    Also note that, for steady-state models like RANS, the feature values obtained in one solver iteration influence the feature values obtained in the next iteration.
    The inference and learning process, however, works with the feature values obtained at solver convergence only.
    Hence, uninfluenced feature space regions (according to the converged solutions) could be accessed by the solver when the residuals are not converged.
    So, if the augmentation function is modified in these regions, it could consequently affect the converged result, which is undesirable.
    Thus, it is even more important that in uninfluenced feature space regions, the augmentation function does not learn from data, and rather predict from the unaltered existing augmentation behavior.
    
    Clearly, there exists ample scope to explore different localized learning techniques (along with strategies to optimally choose regions of influence).
    However, our main motive in this work w.r.t. localized learning is to introduce it in the LIFE framework and emphasize on its importance in creating robust and generalizable augmentations, and so we choose to deal with these explorations in future works.}
    
    \newdel{\sout{Thus, the use of domain expertise on how to augment the model, an informed choice of features, and using an appropriate function class with control over the feature space is instrumental in creating a generalizable and robust augmentation.}}
    
    \subsection{Implementation details} \label{ssec:implementation}
      
      \begin{algorithm}[h!]
      \small
      \caption{Simultaneous Integrated Inference and Learning}
      \label{algo:LIFE}
      \begin{algorithmic}[1]
      %\vspace{0.25em}
      \Require $\bs{w}^{0}$, $\bs{y}^1$, \dots, $\bs{y}^N$, $n$, $\alpha$
      \vspace{-0.5em}\Ensure $\bs{w}^{n}$
      
      \For{$i=1:N$} \hspace*{\fill} \texttt{[$N$ is number of datasets]}
        
        \State Solve for $\iRFvar$ s.t. $\displaystyle{\mathscr{R}_m(\iRFvar; \beta(\bs{\eta}(\iRFvar,\bs{\zeta}^i);\bs{w}^{0}), \bs{\xi}^i) = 0}$ \hspace*{\fill} \texttt{[Forward problem]}
      
        \State $\mathcal{J}^i \coloneqq \mathcal{C}^i(\bs{y}^i, \bs{y}^i_m(\iRFvar)) + \lambda_\beta^i\mathcal{T}_\beta^i(\beta(\bs{\eta}(\iRFvar,\bs{\zeta}^i);\bs{w}^{0}))$ \hspace*{\fill} \texttt{[Objective evaluation]}
      
      \EndFor \vspace{0.5em}
      
      \State $\displaystyle{\mathcal{J} \coloneqq \bigsqcup_{i=1}^N \mathcal{J}^i}$ \hspace*{\fill} \texttt{[Assembly operator (e.g., weighted sum)]}
        
      \vspace{0.5em}\For{$k=1:n$} \hspace*{\fill} \texttt{[$n$ is number of LIFE iterations]}
      
        \For{$i=1:N$}
        
            \vspace{0.25em}\State $\displaystyle{\bs{\psi}^T \coloneqq \frac{d\mathcal{C}^i}{d\iRFvar} \left[\left.\frac{d\bs{\mathscr{R}}_m}{d\RFvar}\right|_{\bs{\xi}^i}\right]^{-1}}$ where\text{ } $\displaystyle{\frac{d}{d\iRFvar}=\left.\frac{\partial}{\partial\iRFvar}\right|_{\beta}+ \frac{\partial}{\partial\beta}\frac{\partial\beta}{\partial\bs{\eta}}\frac{\partial\bs{\eta}}{\partial\iRFvar}}$ \hspace*{\fill} \texttt{[Adjoints]}
            
            \State $\displaystyle{\frac{d\mathcal{J}^i}{d\bs{w}^{k-1}} \coloneqq \bs{0}}$  \hspace*{\fill} \texttt{[Set sensitivities to zero]}
            
            \vspace{0.25em} \For{all $j$ s.t. $x_j\in\widetilde{\Omega}^{\hspace{0.1em}i}$}
        
                \vspace{0.25em}\State $\displaystyle{ \frac{d\mathcal{J}^i}{d\bs{w}^{k-1}} = \frac{d\mathcal{J}^i}{d\bs{w}^{k-1}} + \lambda_\beta^i\frac{\partial\mathcal{T}_\beta^i}{\partial\beta}\left[\frac{\partial\beta}{\partial\bs{w}^{k-1}}\right]^i_j - \bs{\psi}^T\left[\frac{\partial\mathscr{R}_m}{\partial\beta}\right]^i}\left[\frac{\partial\beta}{\partial\bs{w}^{k-1}}\right]^i_j$
            
            \vspace{0.25em}\EndFor
        
        \EndFor
        
        \vspace{0.25em}\State $\displaystyle{
        \frac{d\mathcal{J}}{d\bs{w}^{k-1}} \coloneqq \sum_{i=1}^N\frac{\partial\mathcal{J}}{\partial\mathcal{J}^i}\frac{d\mathcal{J}^i}{d\bs{w}^{k-1}}}$ \hspace*{\fill} \texttt{[Sensitivity assimilation from all cases]}
        
        \vspace{0.25em}\State $\displaystyle{
        \bs{w}^{k} = \bs{w}^{k-1} - \alpha\frac{d\mathcal{J}}{d\bs{w}^{k-1}}
        \left\lVert\frac{d\mathcal{J}}{d\bs{w}^{k-1}}\right\rVert_\infty^{-1}
        }$ \hspace*{\fill} \texttt{[Steepest gradient descent update]}
        
        \vspace{0.25em}\For{$i=1:N$}
        \vspace{0.25em}\State Solve for $\iRFvar$ s.t. $\displaystyle{\mathscr{R}_m(\iRFvar,   \beta(\bs{\eta}(\iRFvar,\bs{\zeta}^i;\bs{w}^{k})); \bs{\xi}^i) = 0}$ \hspace*{\fill} \texttt{[Forward problem]}
        \vspace{0.25em}\State $\mathcal{J}^i=\mathcal{C}^i(\bs{y}^i, \bs{y}^i_m(\iRFvar)) +   \lambda_\beta^i\mathcal{T}_\beta^i(\beta(\bs{\eta}^i(\iRFvar,\bs{\zeta});\bs{w}^{k}))$ \hspace*{\fill} \texttt{[Objective evaluation]}
        \vspace{0.25em}\EndFor
        
        \vspace{0.5em}\State $\displaystyle{ \mathcal{J} = \bigsqcup_{i=1}^N \mathcal{J}^i }$
      
      \vspace{0.25em}\EndFor
      
      \end{algorithmic}
      \end{algorithm}
      
      \newadd{red}{To be able to evaluate different functional forms for the augmentation, the solver has to be made agnostic of how the augmentation is implemented.
      To do so, the solver must interact with only a linearized version of the augmentation function, $\beta_\ell$.
      In other words, given feature values $\bs{\eta}_0$, an implementation of the augmentation function should internally compute the numerical values $\beta_0 = \beta(\bs{\eta}_0;\bs{w})$ and $\bs{g}_\beta = \left.\dfrac{\partial\beta}{\partial\bs{\eta}}\right|_{\bs{\eta}_0}$.
      These numerical values should then be passed to the solver which should construct the linearized augmentation using AD variables $\bs{\eta}$ and non-AD type numerical values for $\bs{\eta}_0$ as $\beta_\ell(\bs{\eta}) = \beta_0 + \bs{g}_\beta(\bs{\eta}-\bs{\eta}_0)$.
      Notice here, that the value of $\beta_\ell$ at $\bs{\eta}=\bs{\eta}_0$ (which is the value the direct solver uses in the model) remains $\beta_0$ and the second term involving $\bs{g}_\beta$ is present only to record $\left.\dfrac{\partial\beta}{\partial\bs{\eta}}\right|_{\bs{\eta}_0}$ in the AD tape.
      Notice that this also presents the advantage of making the AD tape shorter.
      
      The detailed algorithm for the LIFE formulation is presented in Algorithm \ref{algo:LIFE}.
      The jacobians presented in the algorithm can be calculated using hard-coded analytic derivatives or using an automatic differentiation (AD) package.
      In the transition modeling demonstration presented in this work (section \ref{sec:TransitionModel}), the in-house solver makes use of the open-source automatic differentiation package ADOL-C to calculate these jacobians.}
    
\section{Augmenting a bypass transition model} \label{sec:TransitionModel}

  \subsection{Relevant ideas from bypass transition modeling literature}
  
    \subsubsection{The Praisner-Clark estimate of transition momentum-thickness Reynolds number} \label{sssec:DataCorr}
      
      Praisner and Clark \cite{PraisnerClark2004}, in 2004, presented a correlation that can approximate the transition momentum thickness Reynolds number over a wide range of pressure gradients, freestream turbulence intensities and Mach numbers. This correlation is given as follows.
      \begin{equation} Re_{\theta,t} = A\left(Tu_\infty\frac{\theta_t}{\lambda_\infty}\right)^B \end{equation}
      where $A=8.52$ and $B=-0.956$. Approximating $B$ to be -1, recasting this correlation and assuming $C_\mu\omega=u'/\lambda$ for the Wilcox $k-\omega$ model one can write this correlation as
      \begin{equation} \theta_t^2 = A_1\frac{\nu\lambda_\infty}{u'_\infty} =     \frac{A_1}{C_\mu}\frac{\nu}{\omega_\infty} \end{equation}
      where $A_1=0.07 \pm 0.011$.
      The universality of this estimate across such a varied set of flow conditions, even though the dataset comes predominantly from turbomachinery setups, is adopted in the model presented in this work.
      
      \subsubsection{A brief review of recent intermittency-based models} \label{sssec:IntroToGamma}
      
      Langtry and Menter \cite{LangtryMenter2009} proposed one of the most widely used intermittency-based transition models with transport equations for intermittency ($\gamma$) and an estimate of transition momentum thickness Reynolds number ($\widetilde{Re_{\theta,t}}$) to supplement the SST turbulence model equations.
      \newadd{red}{The additional transport equation for $\widetilde{Re_{\theta,t}}$ is used to diffuse the values calculated using algebraic correlations from outside the boundary layer to regions inside the boundary layer.}
      This quantity is then used to evaluate the critical momentum thickness Reynolds number, $Re_{\theta_c}$, which when compared with the vorticity Reynolds number, $d^2\Omega/\nu$, can be used to predict the transition onset and hence trigger intermittency production.
      The intermittency takes effect in the solver by scaling the production term of the TKE transport equation.
      
      In 2012, Durbin \cite{Durbin2012} proposed a simple transition model with a single production term which is directly proportional to the local vorticity, $\Omega$ and  $(\gamma_\text{max} - \gamma)\sqrt{\gamma}$, where $\gamma_\text{max}=1.1$, to supplement the Wilcox's 1988 $k-\omega$ turbulence model. The $(\gamma_\text{max}-\gamma)$ term would give a trivial solution of $\gamma = \gamma_\text{max}$ in the entire field if the model is used as it is.
      For this reason, the author proposed to set the intermittency to zero for all mesh locations where the eddy viscosity was very small compared to the local molecular viscosity and the vorticity Reynolds number is less than a pre-defined threshold, after every solver iteration.
      After every solver iteration, the values of $\gamma$ is set as $\min(\gamma, 1.0)$ as $\gamma_\text{max} > 1$ helps the intermittency to rapidly increase, but at the same time, results in a non-physical growth in turbulent production.
      As in the Langtry-Menter model, the intermittency is multiplied to the production term in the transport equation for $k$.
      
      Ge and Durbin \cite{Ge2014} proposed another transition model based on the structure of the previous model in 2014 which attempts to predict bypass transition due to freestream turbulence and flow separation.
      Separation effects were accounted for by multiplying the production term in the transport equation for $k$ with a separation modification.
      It is noteworthy, here, that the presence of a dissipation term in this equation removes the requirement of setting the intermittency to zero, externally.
      
      In the models mentioned above, one can see a common theme.
      All of these models attenuate the production of turbulent kinetic energy using the intermittency function rather than attenuating the eddy viscosity directly.
      This is justified as turbulent kinetic energy should be very low in the laminar regions of the flow, something which is hard to implement by just altering the eddy viscosity term.
      Secondly, one of the most important terms in all these models is the vorticity Reynolds number, the maximum value of which approximates the momentum thickness Reynolds number fairly well for a zero pressure gradient boundary layer, as shown by van Driest and Blumer \cite{vanDriestBlumer1963}.
      Even in the presence of favorable/adverse pressure gradients, the approximation still provides an excellent tool for modeling the transition onset as a function of a local quantity, as opposed to the momentum thickness Reynolds number which is an integral quantity.
      These ideas were instrumental in designing the transition model presented in this work.

  \subsection{Chosen baseline model and proposed augmentation}
    
    The transport equations that constitute the chosen baseline model have been shown below.
    This model is a variation of the model proposed by Durbin in 2012 \cite{Durbin2012} which used Wilcox's 1988 $k$-$\omega$ model as the underlying turbulence model.
    The source term for the intermittency transport equation in Durbin's model is expressed as $F_\gamma\lvert\Omega\rvert(\gamma_\text{max}-\gamma)\sqrt{\gamma}$, where $\lvert\Omega\rvert$ is the local vorticity magnitude, $\gamma_\text{max}=1.1$ and $F_\gamma$ is a piecewise linear function that modulates the source term.
    Instead, we consider the baseline source term as $\rho(1.0-\gamma)\sqrt{\gamma}\lvert\Omega\rvert$ (the additional $\rho$ is attributed to the compressible formulation of the model).
    The version of Wilcox's 1988 turbulence model containing a vorticity-based production term is used for this work, in contrast to the strain-rate based production term.
    \begin{equation}\label{eqn:k_transport}
    \frac{D\rho k}{Dt} = 
    \boldsymbol\nabla\cdot\left(\left(\mu+\frac{\mu_T}{\sigma_k}\right)\boldsymbol\nabla k\right) +
    \mu_t\Omega^2 - \frac{2}{3}\rho k\nabla\cdot\bs{u} - C_\mu k\omega
    \end{equation}
    \begin{equation}   
    \frac{D\rho\omega}{Dt} = 
    \boldsymbol\nabla\cdot\left(\left(\mu+\frac{\mu_T}{\sigma_\omega}\right)\boldsymbol\nabla\omega\right) +
    C_{\omega 1}\frac{\omega}{k}\left(\mu_t\Omega^2 - \frac{2}{3}\rho k\nabla\cdot\bs{u}\right) - C_{\omega 2}\omega^2
    \end{equation}
    \begin{equation}   
    \frac{D\rho\gamma}{Dt} = 
    \boldsymbol\nabla\cdot\left(\left(\frac{\mu}{\sigma_l}+\frac{\mu_T}{\sigma_\gamma}\right)\boldsymbol\nabla\gamma\right) +
    \rho(1.0-\gamma)\sqrt{\gamma}\lvert\boldsymbol\Omega\rvert,
    \end{equation}
    with the eddy viscosity given by $\mu_t=\rho k/\omega$.
    Clearly, this model will give a fully turbulent solution, as the intermittency equation has a trivial solution with the value 1.0 throughout the domain.
    Multiplying the value $1.0$ in the intermittency transport equation with the augmentation $\beta$, we have an augmented intermittency transport equation as follows
    \begin{equation}   
    \frac{D\rho\gamma}{Dt} = 
    \boldsymbol\nabla\cdot\left(\left(\frac{\mu}{\sigma_l}+\frac{\mu_T}{\sigma_\gamma}\right)\boldsymbol\nabla\gamma\right) +
    \rho(\beta-\gamma)\sqrt{\gamma}\lvert\boldsymbol\Omega\rvert.
    \end{equation}
    This particular form of baseline model and augmentation was chosen for the following reasons:
    \begin{itemize}
      \item For most cases, the range for this augmentation is bounded between $0$ and $1$.
      \item Any discontinuities in the predicted augmentation field will be smoothed over by the transport equation and will not require special attention. 
      Indeed, this is the reason why the intermittency transport equation was retained in contrast to augmenting the production term in the $k$-equation directly.
      \item The knowledge  of correlations between the intermittency field and other flow quantities can be readily used to formulate this augmentation, as the converged intermittency field would closely resemble a smoothed augmentation field.
    \end{itemize}
    
  \subsection{Choice of features and limiters for the augmentation}
  
    \newadd{red}{
      A clear understanding of the effect of intermittency term $\gamma$ on the modeled turbulent kinetic energy $k$ played a vital role in the choice of features.
      A brief explanation of this effect is given as follows.
      
      From equation \ref{eqn:k_transport}, it can be observed that the production of $k$ is directly proportional to eddy viscosity ($\nu_t$) and vorticity ($\Omega$), and inversely proportional to the modeled specific dissipation ($\omega$).
      $\Omega$, for external flows, is usually the highest at the wall and decreases away from the wall.
      Also, the freestream eddy viscosity (which is usually the initial condition for $\nu_t$ in RANS solvers) is very small (around an order of magnitude higher than molecular viscosity).
      Hence, the production of $k$ starts in a region very close to the wall at a distance where $\omega$ has sufficiently decreased and $\Omega$ remains high enough to enable a net production of $k$ (for a fully turbulent boundary layer, this region lies within the inner layer but outside the viscous sublayer).
      By the virtue of diffusion, $k$ starts increasing slightly farther from the wall relative to where it is being produced, thus increasing local $\nu_t$ and triggering a net production of $k$.
      This process keeps spreading away from the wall until $\Omega$ drops so low that a net production is no longer possible.
      Also, the convective term in equation \ref{eqn:k_transport} ensures that a downstream production of $k$ has minimal effect on upstream locations.
      Using all this information, it can be concluded that in order to laminarize the upstream of the transition location, the production of $k$ needs to be attenuated by sufficiently reducing $\gamma$ only in a region close to the wall.
      Doing this requires - (1) feature(s) that can indicate the transition onset (the location up to which the augmentation should be effective); (2) feature(s) to uniquely identify the regions close to the wall where the augmentation should be active.
      The following sections describe the rationale behind the choice of three such features.
    }
  
    {\bf Feature 1: Ratio of Vorticity Reynolds Number and an estimate of $Re_{\theta,t}$}
    
      Traditional models have used experimentally obtained correlations for transition momentum thickness Reynolds number ($Re_{\theta,t}$) to predict transition by comparing it with the local momentum thickness Reynolds number $Re_\theta$.
      As mentioned in section \ref{sssec:IntroToGamma}, the vorticity Reynolds number, $Re_\Omega$, provides a good surrogate to use instead of $Re_\theta$, which is an integral quantity.
      The vorticity Reynolds number, as defined in \cite{Durbin2012}, is given as follows.
      \begin{equation} Re_\Omega = \frac{d^2\Omega}{2.188\nu} \end{equation}
      The transition momentum thickness Reynolds number, $Re_{\theta,t}$, is given as follows using the Praisner-Clark estimate \cite{PraisnerClark2004} of transition momentum thickness described in section \ref{sssec:DataCorr}.
      \begin{equation} Re_{\theta,t} = \frac{U_\infty\theta_t}{\nu} \approx \frac{U_\infty}{\nu}\sqrt{\frac{7\nu}{9\omega_\infty}}=\overline{Re_{\theta,t}} \end{equation}
      \newadd{cyan}{Note that this correlation does not depend on $k_\infty$ which is physically inconsistent and might render it ineffective in certain scenarios.
      However, in their study, Praisner and Clark found this correlation seems to work quite well on a range of turbomachinery configurations across different Reynolds numbers, Mach numbers, pressure gradients and freestream turbulence intensities.
      Hence, this correlation is currently being used as a physics-based non-dimensionalization for the first feature which is formulated as the ratio $Re_\Omega/\overline{Re_{\theta,t}}$.}
      \newdel{\sout{The first feature is the ratio of these two Reynolds numbers.}}
      A conservative upper-bound for this ratio is assumed to be $3$.
      Hence, the feature can be reformulated as follows to strictly bound the feature space as follows.
      \begin{equation} \eta_1 = \min\left(\frac{d^2\Omega}{2.188\nu\overline{Re_{\theta,t}}}, 3.0\right) \end{equation}
      Note here, that the augmentation resulting from this feature would be considerably sensitive to the far-field boundary condition of $\omega$, and hence the boundary conditions for $\omega$ must be set carefully to simulate the decay of the freestream turbulence intensity as accurately as possible.
      
      The remaining issue is to extract $\overline{Re_{\theta,t}}$ from the freestream.
      In the current work, this is done as follows.
      A preset distance interval $[r-\delta r, r+\delta r]$ is chosen for every case.
      For every discrete element on a wall boundary, $\partial\widetilde{\Omega}_k^w$, the volume element inside the domain located at the minimum normal distance within the preset distance interval, is chosen to be the corresponding freestream cell, $\widetilde{\Omega}_k^\infty$.
      If no such element is present, $\delta r$ needs to be increased.
      Now for every volume element, $\widetilde{\Omega}_j$ inside the domain which is at a minimum distance from the wall boundary element $\partial\widetilde{\Omega}_k^w$ among all such wall boundary elements, any freestream quantity is set as the corresponding quantity in $\widetilde{\Omega}_k^\infty$.
      While this can be inaccurate at large distances from the wall, it works fairly well inside the boundary layer with appropriate choice of the distance interval.
      \newadd{red}{Features 2 and 3 distinguish between the volume elements within the boundary layer where the augmentation needs to be significantly lower than the baseline value and volume elements elsewhere. Hence, the freestream estimate of quantities being poor far from walls is of no concern.}
      The distance interval is usually chosen to be as close to the wall and as small as possible, such that the maximum boundary layer thickness should remain within the lower bound of this interval.
      It was also noted that the augmentation is not very sensitive to small changes in the choice of this interval as demonstrated in appendix \ref{app:preset}.
      
      \newadd{red}{Note that although $\omega_\infty$ and $U_\infty$ seem to be non-local quantities at a first glance, but since the augmentation mainly affects the flow only within the boundary layer, the freestream quantities virtually remain constant across LIFE iterations.
      Thus, $\omega_\infty$ and $U_\infty$ can be thought of as frozen quantities local to each volume element in the discretized domain.}
      \newadd{cyan}{Finally, while $U_\infty$ can be used in moderate pressure gradient problems, the quantity could be ambiguous for very high pressure gradients.
      Also, note that the maximum value of $Re_\Omega$ in the wall-normal direction approximates $Re_\theta$ for zero pressure gradients and hence, $\displaystyle\max_{d_w} Re_\Omega$ would also poorly estimate $Re_\theta$ for very high pressure gradients.}
      
    {\bf Feature 2: Ratio of wall distance and turbulent length scale}
    
      \newdel{\sout{The second feature that was considered has to do with turbulent length scales. Intermittency inside the outer and buffer layers can be high only where the flow is sufficiently
      turbulent, which can be diagnosed based on the turbulent length scales. This feature is non-dimensionalized using the wall distance and formulated
      as shown below such that it remains bounded between 0 (limit of turbulent length scale approaching infinity) and 1 (limit of turbulent length scale approaching  zero).}}
      \newadd{red}{To identify laminar regions within the boundary layer, one can compare the local turbulent length scale $\sqrt{k}/\omega$ with the wall distance $d$.
      A predominantly laminar region is characterized by an overwhelmingly larger wall distance compared to the turbulent length scale.
      A modified functional form for the ratio of these two quantities which is mathematically bounded between the numerical values of $0$ and $1$ can be written as,
      \begin{equation} \eta_2 = \frac{d}{d+\sqrt{k}/\omega} \end{equation}
      The laminar regions in the flow are denoted by values of $\eta_1$ that are significantly close to $1$. Note here that, the value of this feature will be close to $1$ for regions outside the boundary layer as well.}
      
    {\bf Feature 3: Ratio of laminar viscosity and eddy viscosity}
    
      \newdel{\sout{The intermittency is very small in the region between the leading edge and the transition onset. In this nearly laminar part of the boundary layer, the viscous sub-layer would
      be much thicker compared to the one observed in the fully turbulent zones. It is, thus, important to diagnose the inner part (mainly viscous sub-layer, but also how it varies to
      buffer and outer layers) of the boundary layer. An excellent candidate to identify these zones is the following feature which varies from 1 inside the viscous sub-layer to very
      small (nearly zero) values inside the outer layer and increasing to the freestream value (significantly less than 1 for moderately high viscosity ratios) outside the
      boundary layer.}}
      \newadd{red}{Since the magnitude of $\nu_t$ in the freestream is usually greater than $\nu$, the following modified functional form for the ratio of these two quantities would be less than $0.5$ in the freestream while being close to $1$ in the laminar boundary layer and the viscous sublayer in the turbulent boundary layer.
      \begin{equation} \eta_3 = \frac{\nu}{\nu_t+\nu} \end{equation}
      Together with feature 2, this feature can distinguish regions within the boundary layer for locations upstream of the transition location.
      Note here, that the feature design in this case is inspired from the augmentation being correlated to (and not caused by) the features.}
    
    {\bf Limiter for flows without separation}
    
      As we are not considering flows with separation, where the intermittency might be allowed to attain higher
      values in order to match the physical behavior, the resulting intermittency is simply limited to values 
      between the theoretical bounds of 0 and 1.
      
  \subsection{Augmentation function as Piecewise Linear Interpolation on Uniformly Structured Grid in Feature-space} \label{ssec:PLUS}
      
    One of the simplest class of functions that can perform localized learning while approximating sensitivities w.r.t. features is linear interpolation on a uniform grid.
    While this is not the most sophisticated class of functions to approximate an augmentation, it certainly demonstrates the capability of localized learning.
    Note that the discretization need not be uniform, or structured for that matter, but these assumptions have been made to simplify calculations, implementation and embedding of the augmentation function within the numerical solver.
    
    In this method, the effectively bounded feature space is discretized into a uniformly spaced structured grid.
    An augmentation value can be attributed to the center of each grid cell.
    These values can then be linearly interpolated using second order finite difference approximations inside every cell, to obtain the augmentation for any feature space location.
    Hence, these cell-centered values serve as the augmentation function parameters, $\bs{w}$.
      
    Given a vector of feature values, $\bs{\eta}$, and a value of the augmentation function at the center of each grid cell $\beta_{c_\ell}$, a linearized approximation of the augmentation can be obtained as described below.
    Note that the vector of values $\beta_{c_\ell}$, viz. $\bs\beta_c$, here represents the function parameter vector $\bs{w}$ mentioned in the previous sections.
      
    \begin{enumerate}
        
      \item Find the grid cell $\ell$ which $\bs{\eta}$ belongs to. Let the features at the center of this grid cell be $\bs\eta_{c_\ell}$.
        
      \item Based on the augmentation value at the centers of the neighboring grid cells, calculate a gradient approximation $\bs\nabla^{FD}_{\bs\eta}\beta_\ell$ for the $\ell^\text{th}$ grid cell using a second order accurate central finite difference scheme along each feature.
        
      \item Return the following linearized approximation of the augmentation function
      \begin{equation} \beta_\ell(\bs\eta) = \beta_{c_\ell} + (\bs\eta - \bs\eta_{c_\ell})\cdot\bs\nabla^{FD}_{\bs\eta}\beta_\ell \end{equation}
        
    \end{enumerate}
      
    The sensitivity approximation of the augmentation function in a locally linear fashion is sufficient for use in method of adjoints which is used to evaluate the sensitivities $\dfrac{dJ}{d\bs{w}}$ to perform integrated inference and learning as sensitivities by definition only require the linearized functional form.
    Note here, that the chosen structure of the feature space, i.e. the uniform discretization characterized by $\Delta\eta_1$ and $\Delta\eta_2$, and the functions parameters $\bs{w}=\lbrace\beta_{00},\beta_{01},\hdots,\beta_{21},\beta_{22}\rbrace$ are sufficient to determine the linearized approximation of the augmentation as shown above, at any point in the feature space.
    The sensitivities $\dfrac{d\beta_\ell}{d\bs{\eta}}$ and $\dfrac{d\beta_\ell}{d\bs{w}}$, which are used in the algorithm given for integrated inference and learning in Algorithm \ref{algo:LIFE} (also refer Appendix \ref{app:Adjoints} for the adjoint method), are given as follows:
    \begin{equation}
      \frac{\partial\beta_\ell}{\partial\bs{\eta}} = \bs{\nabla}^{FD}_{\bs\eta}\beta_\ell
      \qquad\qquad
      \frac{\partial\beta_\ell}{\partial w_m} = \frac{\partial\beta_\ell}{\partial\beta_{c_m}} =
      \begin{cases}
        1 & \text{if } \ell=m \\
        \dfrac{\eta_j-\eta_{c_\ell,j}}{2\Delta\eta_j} & \text{if } \eta_{c_m,j} =
        \eta_{c_\ell,j} + \Delta\eta_j \\
        \dfrac{\eta_{c_\ell,j}-\eta_j}{2\Delta\eta_j} & \text{if } \eta_{c_m,j} =
        \eta_{c_\ell,j} - \Delta\eta_j \\
        0 & \text{otherwise}
      \end{cases}
    \end{equation}
      
    \begin{figure}[h]
    \centering
    \includegraphics[width=0.3\textwidth]{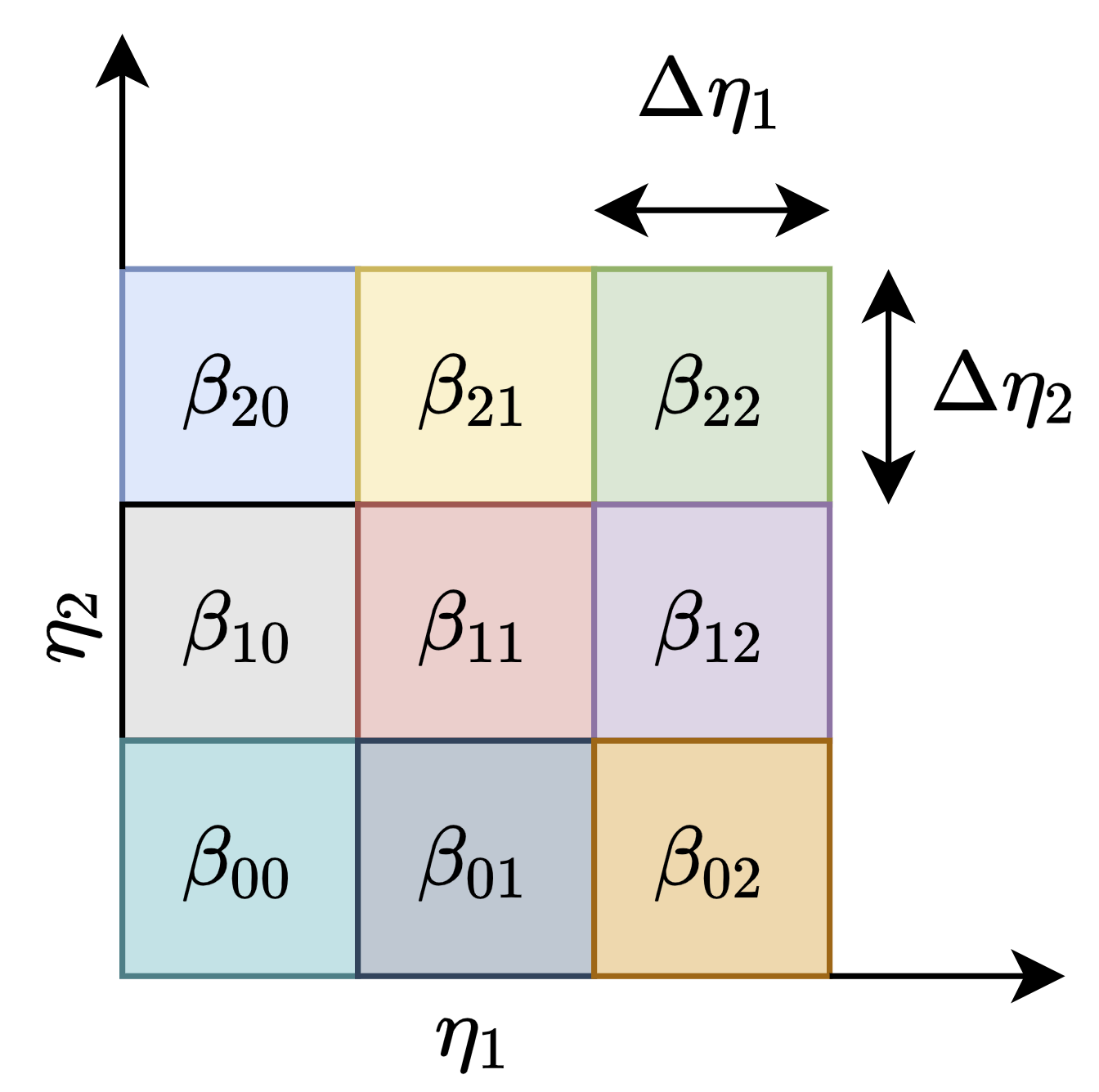}
    \caption{Schematic of feature space discretization}
    \label{fig:LIST}
    \end{figure}
      
    For instance, given a feature pair $(\eta_1, \eta_2)$ for the above schematic, such that the feature pair lies in the cell with center $(1.5\Delta\eta_1, 1.5\Delta\eta_2)$, the augmentation value can be calculated as
    \begin{equation} \beta(\eta_1,\eta_2) = \beta_{11} + \frac{\beta_{12}-\beta_{10}}{2\Delta\eta_1}\left(\eta_1-1.5\Delta\eta_1\right) + \frac{\beta_{21}-\beta_{01}}{2\Delta\eta_2}\left(\eta_2-1.5\Delta\eta_2\right) \end{equation}
    For this specific function class, a trade-off is required in deciding the feature space grid resolution while learning the augmentation.
    If the resolution is too high, the augmentation might not be learned for a majority of grid cells as they might not be used during the optimization process, which can lead to a loss of generality of the augmentation.
    If the resolution is too low, the augmentation would not have enough resolution in the feature space and might lead to sub-optimal learning of the augmentation.
    This is further complicated by the curse of dimensionality, which restricts the resolution per feature with increasing number of features.
    Thus, this particular technique of approximating local augmentations using linear interpolation cannot be practically used for a ``large'' number of features.

\section{Results} \label{sec:Results}

  We use the datasets for the T3 series of experiments conducted by ERCOFTAC \cite{ercoftac} to study bypass transition over flat plates across a range of inflow freestream turbulence intensities and pressure gradients.
  The T3A and T3B cases from this dataset have zero pressure gradient along the length of the flat plate.
  The freestream turbulence intensity at the inflow for T3B is significantly higher when compared to T3A. 
  Other cases in the dataset include T3C1, T3C2, T3C3, T3C4 and T3C5.
  Out of these, T3C4 exhibits flow separation and since we are not considering separation-induced transition in this work, this case has been ignored.
  All of the T3C cases have a favorable pressure gradient near the leading edge which gradually decreases in magnitude along the length of the plate and turns into an increasingly adverse pressure gradient.
  T3C1 and T3C5 exhibit bypass transition in the favorable pressure gradient region of the flow, while T3C2 and T3C3 exhibit transition in regions with mild and strong adverse pressure gradients respectively. 
  
  {\bf Training data:} The augmented model presented below is obtained by inferring the augmentation from the T3A and T3C1 cases only.
  Hence, the following results should be observed in the context of generality of the augmentation obtained in the low data limit.
  To examine sensitivity of the results to the training data, a model trained on just T3A is also shown in appendix \ref{app:T3A_only}. 
  
  {\bf Test cases:} To test the model on unseen data, the T3B, T3C1, T3C2 and T3C3 cases are used from the ERCOFTAC dataset.
  Further, to test the model on unseen geometries, the VKI dataset \cite{Arts1990} for single-stage high-pressure-turbine cascade is used.
  Specifically, MUR116, MUR129, MUR224 and MUR241 are used from the VKI dataset.
  
  \subsection{Training the model on T3A and T3C1 simultaneously}
  
    \subsubsection{Mesh for T3A and T3B}
    
      The length of the flat plate is $1.5$ m with the inlet being $0.04$ m upstream of the leading edge of the flat plate. The mesh resolution next to the wall is on the order of $y^+\approx1$.
      \begin{figure}[h]
      \centering
      \includegraphics[width=0.5\textwidth]{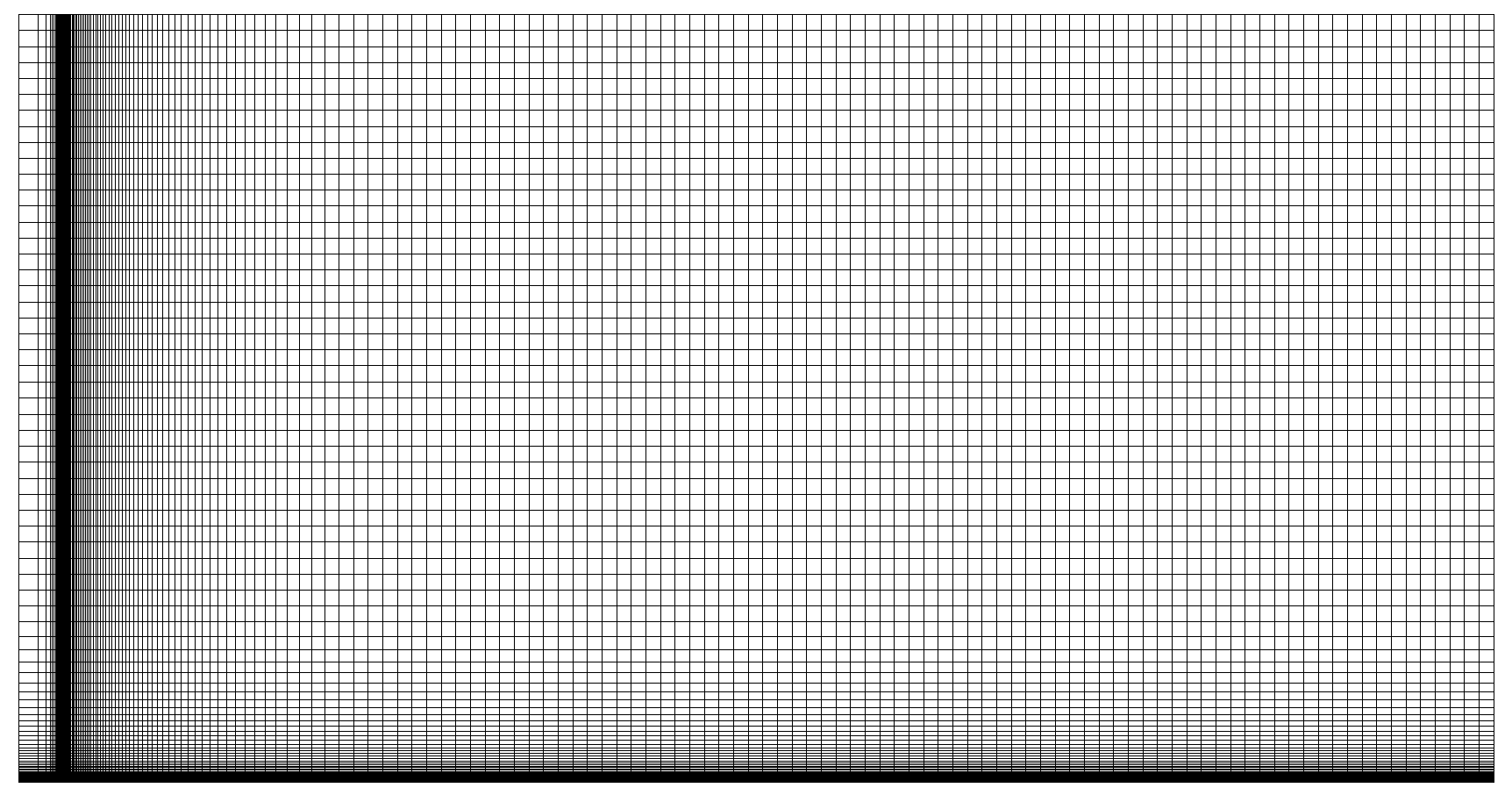}
      \caption{Mesh used for RANS simulation of T3A and T3B cases}
      \label{fig:T3AB_mesh}
      \end{figure}
      
    \subsubsection{Mesh for T3C cases}
\begin{figure}[h]
      \centering
      \includegraphics[width=0.5\textwidth]{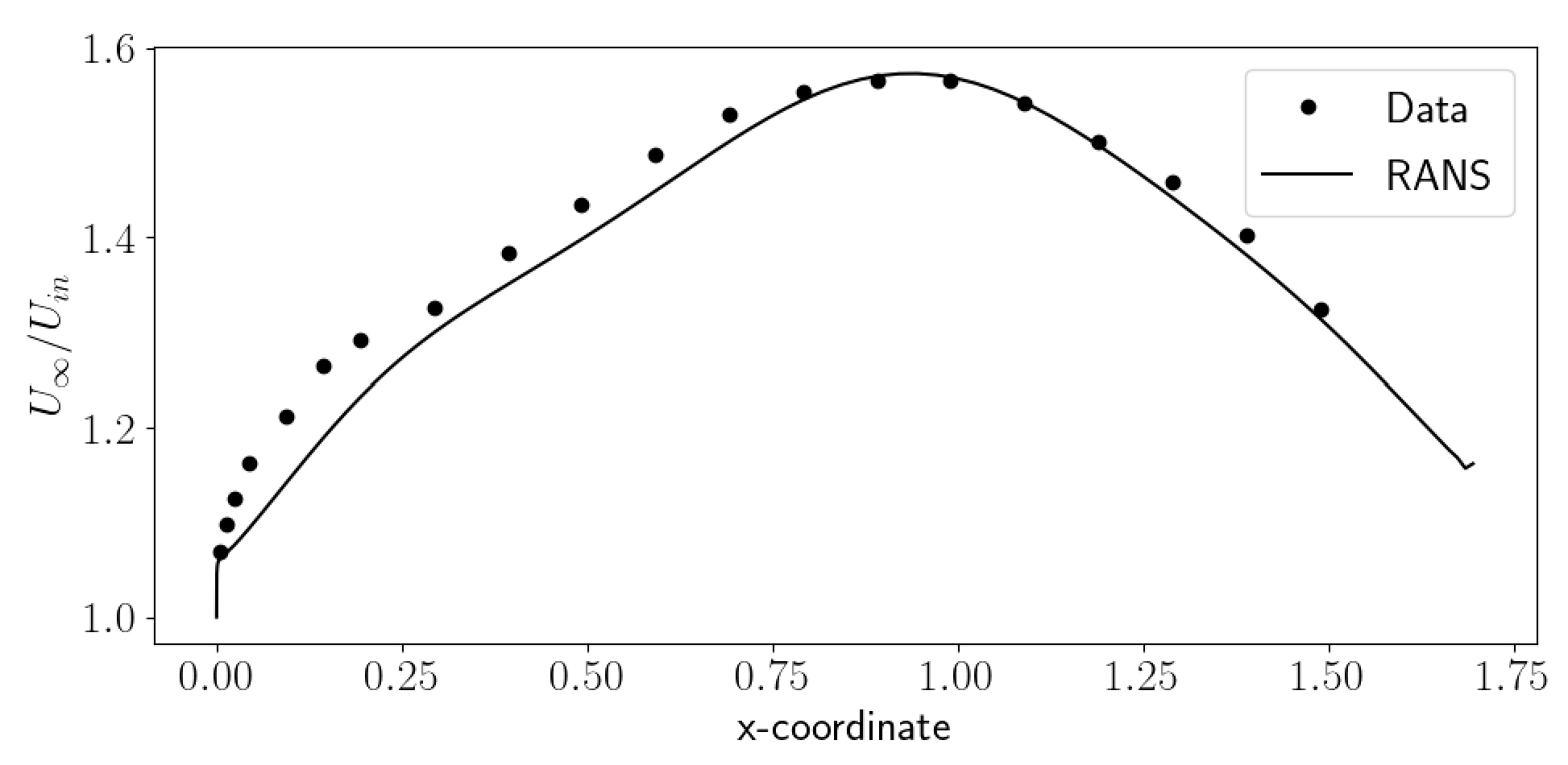}
      \caption{$U_\infty/U_{in}$ vs $x$ for T3C1}
      \label{fig:T3C_valid}
      \end{figure}

      The length of the flat plate is $1.7$ m with the inlet being $0.15$ m upstream of the leading edge of the flat plate.
      The contouring of the upper wall follows the correlations given in the work by Suluksna et al \cite{Suluksna2009}, which results in the trends of $U_\infty/U_\text{in}$ along the length of the plate as shown in figure \ref{fig:T3C_valid}.
      Here $U_\infty$ is calculated using the surface pressure, $p_w$, at the flat plate wall assuming incompressible flow as
      $$U_\infty(x) = \sqrt{2 \left(p_{in} + (1/2)\rho U_{in}^2 - p_w(x)\right)/\rho}$$
      The mesh resolution next to the wall is on the order of $y^+\approx1$.
      \begin{figure}[h]
      \centering
      \includegraphics[width=0.9\textwidth]{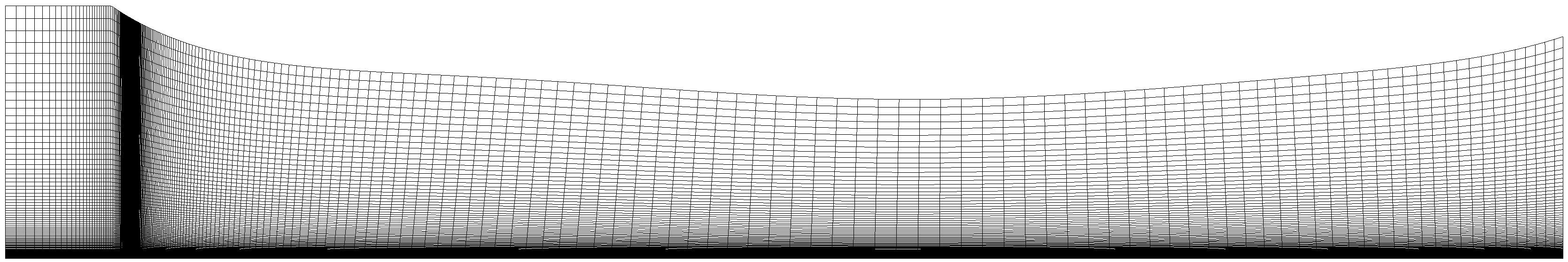}
      \caption{Mesh used for RANS simulation of T3C cases}
      \label{fig:T3C_mesh}
      \end{figure}
      
    \subsubsection{Inflow conditions}
    
\begin{figure}[h!]
      \centering
      \subfigure[T3A] { \includegraphics[width=0.3\textwidth]{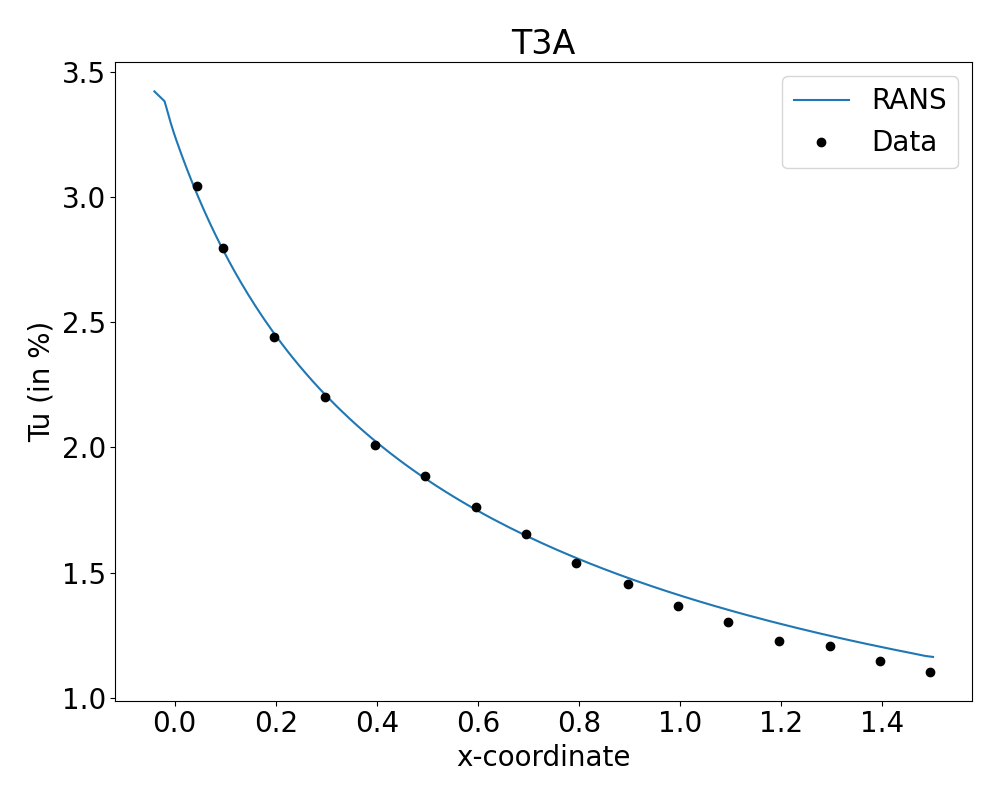} }
      \subfigure[T3C1] { \includegraphics[width=0.3\textwidth]{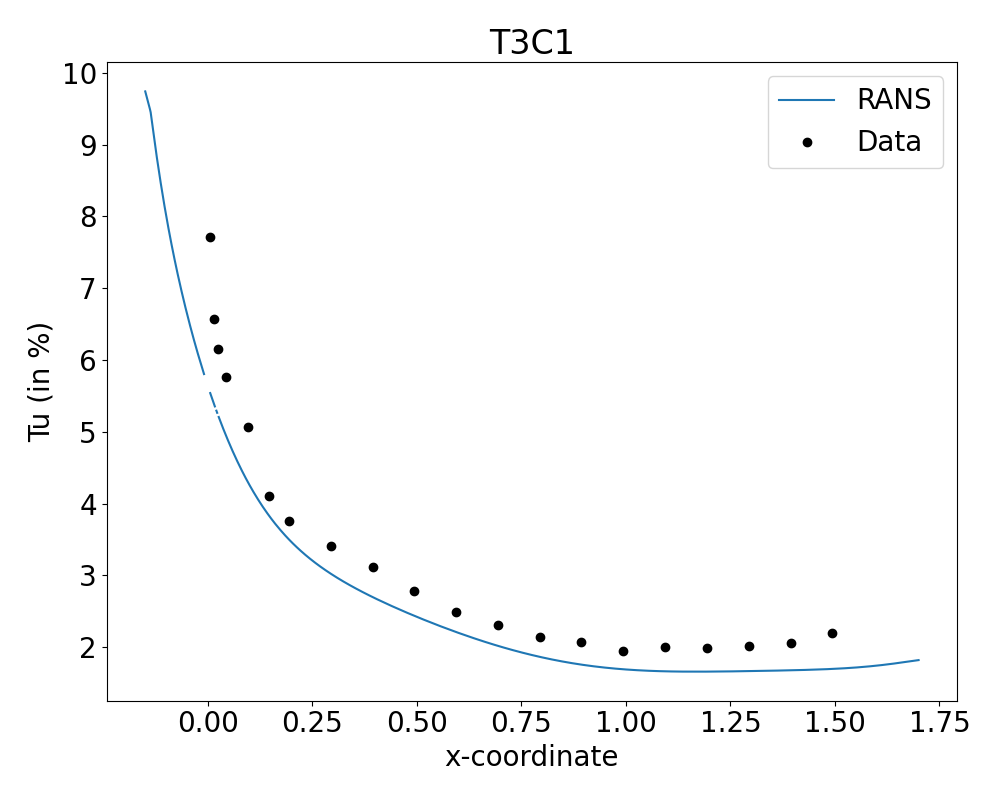} }
      \caption{Decay of freestream turbulence intensity}
      \label{fig:train_TI_decay}
      \end{figure}

      The information needed to characterize the inflow for both the training cases has been presented in table \ref{table:T3_train_inflow}.
      \newadd{red}{The turbulence intensity decay plots shown in figure \ref{fig:train_TI_decay} verify the $\omega$ boundary conditions.}
      \begin{table}[h!]
      \begin{center}
      \begin{tabular}{c|c|c}
        \hline
        \textbf{Cases}                               & \textbf{T3A}   & \textbf{T3C1} \\ \hline
        $Tu_\text{in}$                               & $0.035$        & $0.1$         \\
        $\nu_{t_\text{in}}/\nu$                      & $14.0$         & $50.0$        \\
        $L$(in $m$)                                  & $1.5$          & $1.65$        \\
        $Re_{L,\text{in}}$                           & $520000$       & $660000$      \\
        \newadd{cyan}{$\omega_\text{in}$(in $s^{-1}$)}  & $9.1$          & $24.0$        \\ \hline
      \end{tabular}
      \caption{Inflow conditions for the T3 cases used for training}
      \label{table:T3_train_inflow}
      \end{center} \vspace{-1em}
      \end{table}
      
    \subsubsection{LIFE on T3A and T3C1}
    
\begin{figure}[h]
        \centering
        \subfigure[T3A]
        {\includegraphics[width=0.48\textwidth]{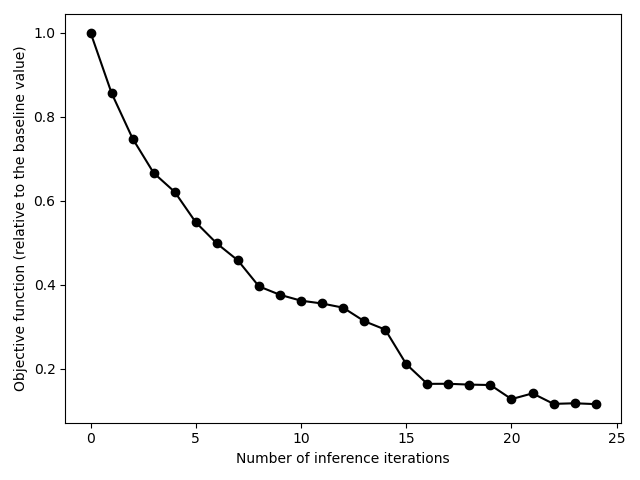} \label{fig:optim_T3A}}
        \subfigure[T3C1]
        {\includegraphics[width=0.48\textwidth]{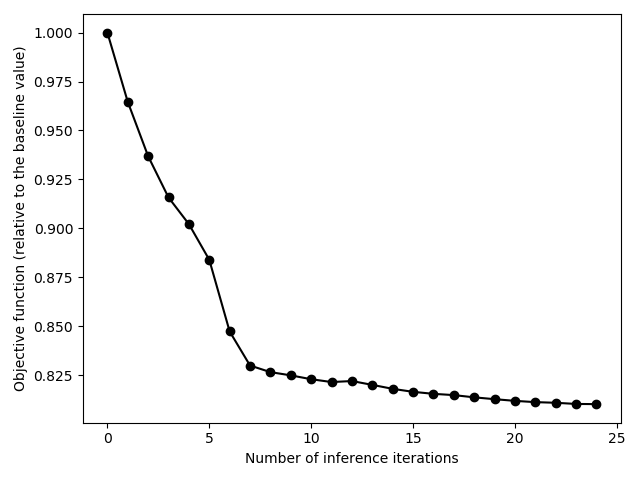} \label{fig:optim_T3C1}}
        \caption{Field Inversion: Optimization plots}
        \label{fig:optim_T3}
      \end{figure}

\begin{figure}[h]
        \centering
        \subfigure[Residual convergence for energy]
        {\includegraphics[width=0.48\textwidth]{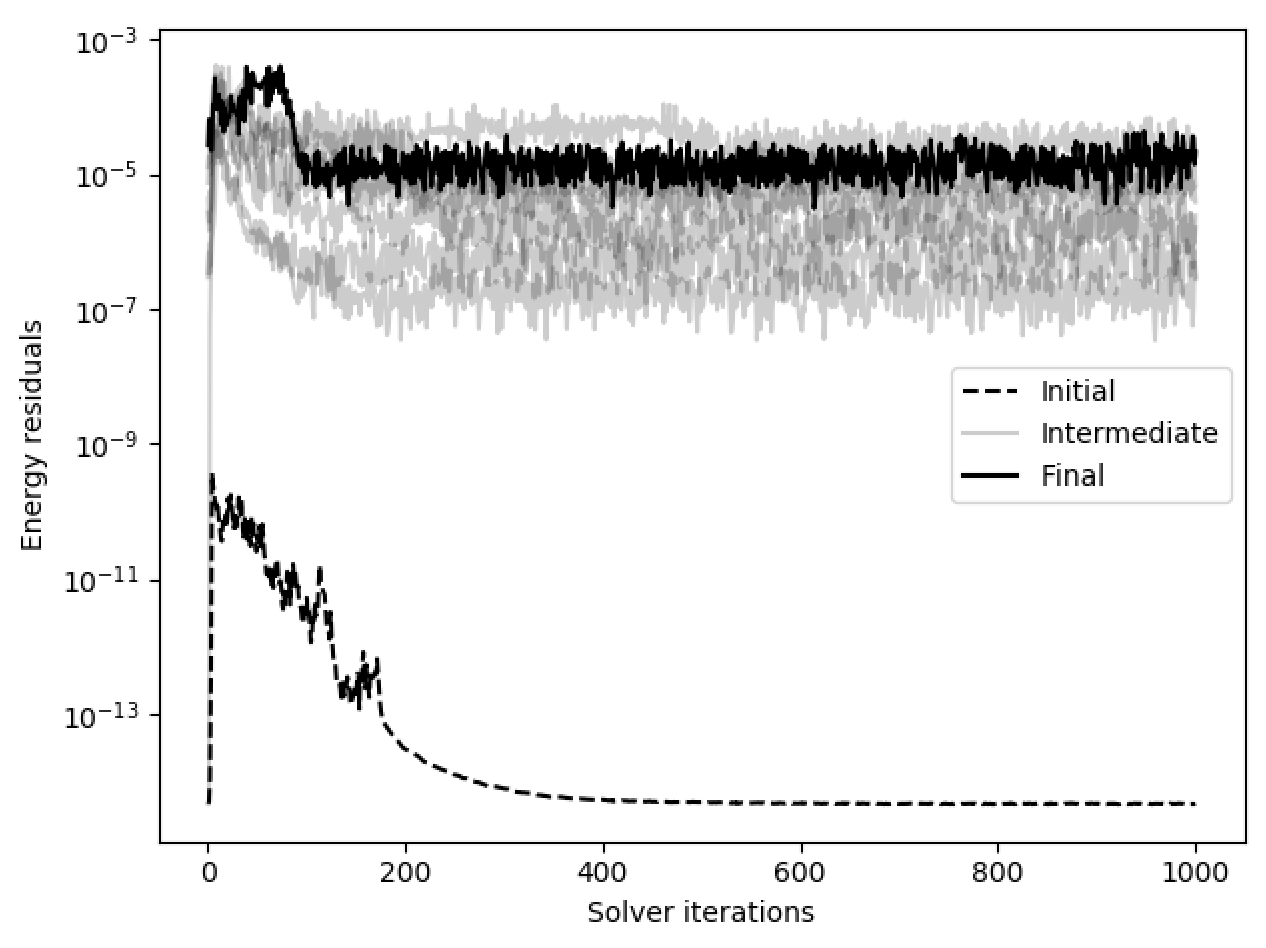} \label{fig:Direct_residual_histories}}
        \subfigure[Residual convergence for energy adjoints]
        {\includegraphics[width=0.48\textwidth]{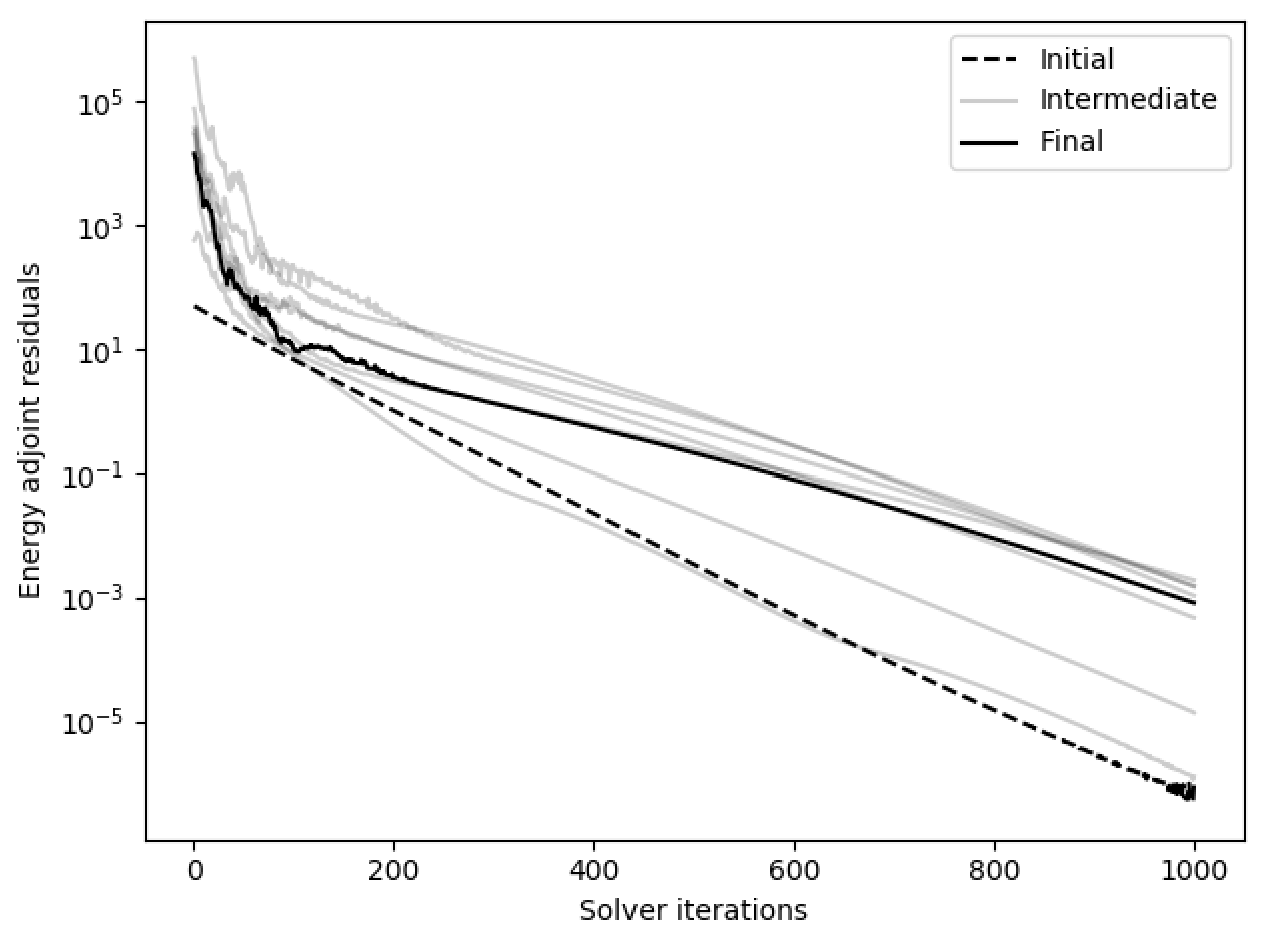} \label{fig:Adjoint_residual_histories}}
        \caption{$\rho E$ Residual histories for all optimization iterates}
        \label{fig:Residual_histories}
      \end{figure}

\begin{figure}[h]
        \centering
        \subfigure[T3A]
        {\includegraphics[width=0.48\textwidth]{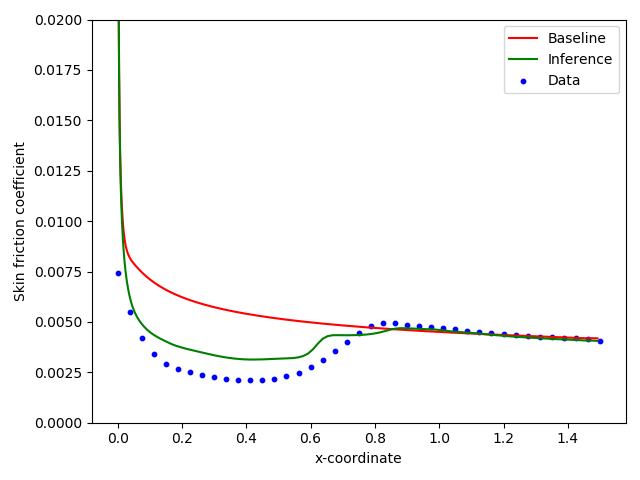} \label{fig:Cf_T3A}}
        \subfigure[T3C1]
        {\includegraphics[width=0.48\textwidth]{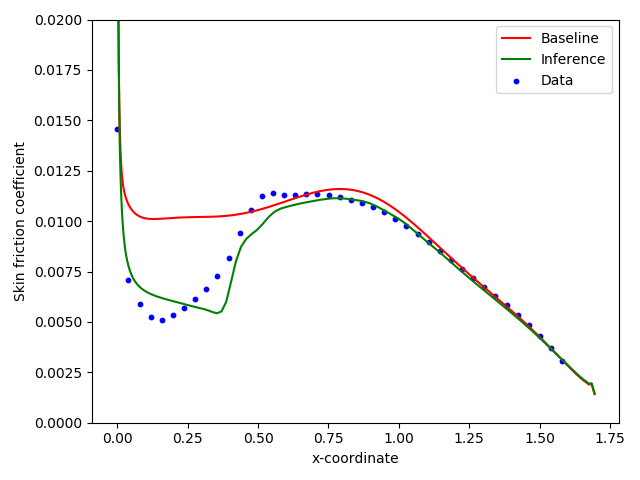} \label{fig:Cf_T3C1}}
        \caption{Skin friction coefficients along the length of the flat plate}
        \label{fig:Cf_T3}
      \end{figure}

      To obtain the augmentation, LIFE was used to infer the functional parameters (see \ref{ssec:PLUS}) in the feature space using the T3A and T3C1 cases simultaneously.
      The cost function for this inference is defined as follows.
      \begin{equation} \mathcal{C} = \sum_{i_\text{wall}=0}^{n_\text{wall}} \left\lVert C_{f,i_\text{wall}} - C_{f,\text{data},i_\text{wall}} \right\rVert_2^2 \end{equation}
      where $n_\text{wall}$ is the number of faces corresponding to the flat plate in the mesh \newadd{red}{and $C_f$ refers to the local skin friction coefficient.}
      The two different problems and the discretization of the feature space acts as an implicit regularizer.
      Figures \ref{fig:optim_T3} and \ref{fig:Residual_histories} show the optimization and residual histories.
      Figure \ref{fig:Cf_T3} shows the $C_f$ of the baseline ($\beta=1$ throughout the domain) and optimal skin friction coefficient distributions compared to data.
      The initial residual convergence for the direct solver is much better compared to the subsequent optimization iterates.
      This is due to the change in the feature-to-augmentation relationship, which affects the convergence. 
      However, the magnitude of the residual suggests sufficient convergence.
      Also, one does not see a significant drop in the residuals because each optimization iterate is restarted from the converged state of the previous iterate.
      The effectively bounded feature space was uniformly discretized into a cartesian grid with 30, 10 and 10 cells along the features $\eta_1$, $\eta_2$ and $\eta_3$, respectively.
      \newadd{red}{It should be noted that the discretization of the feature space presented here is not necessarily the most efficient one.
      The objective rather is to demonstrate the capability of localized learning.
      Further, the discretization need not be in the form of a uniform grid as shown here and could be made adaptive.
      Finally, as mentioned in section \ref{ssec:PLUS}, the discretization needs to be just enough to characterize the augmentation satisfactorily.
      We found that the discretization presented here predicts the transition location well.}
    
      As can be seen in the plots above, the transition locations are inferred with a reasonable accuracy for both the cases.
      Notice, that in the T3C1 case, the skin friction coefficient deviates from the data in the vicinity of transition.
      The same happens for T3A, though the effects there are much subtler.
      Also, it can be noticed that the laminar part of the flow does not show fully laminar friction coefficients as can be seen when compared to data.
      This can be attributed to the predicted intermittency having low values in a very narrow region in the flow field.
      Better results may be obtained by using good limiters on features and/or the predicted augmentation values.
      \newadd{red}{For instance, the augmentation value can be set to zero if it is below a certain threshold (e.g., 0.3).
      Given that such a threshold would have to be extracted using training cases and might not work for certain test cases, these limiters might be ad hoc.}
    
      \begin{figure}[h]
        \centering
        \subfigure[T3A (Flat plate length: 1.5 m)]
        {\includegraphics[width=0.45\textwidth]{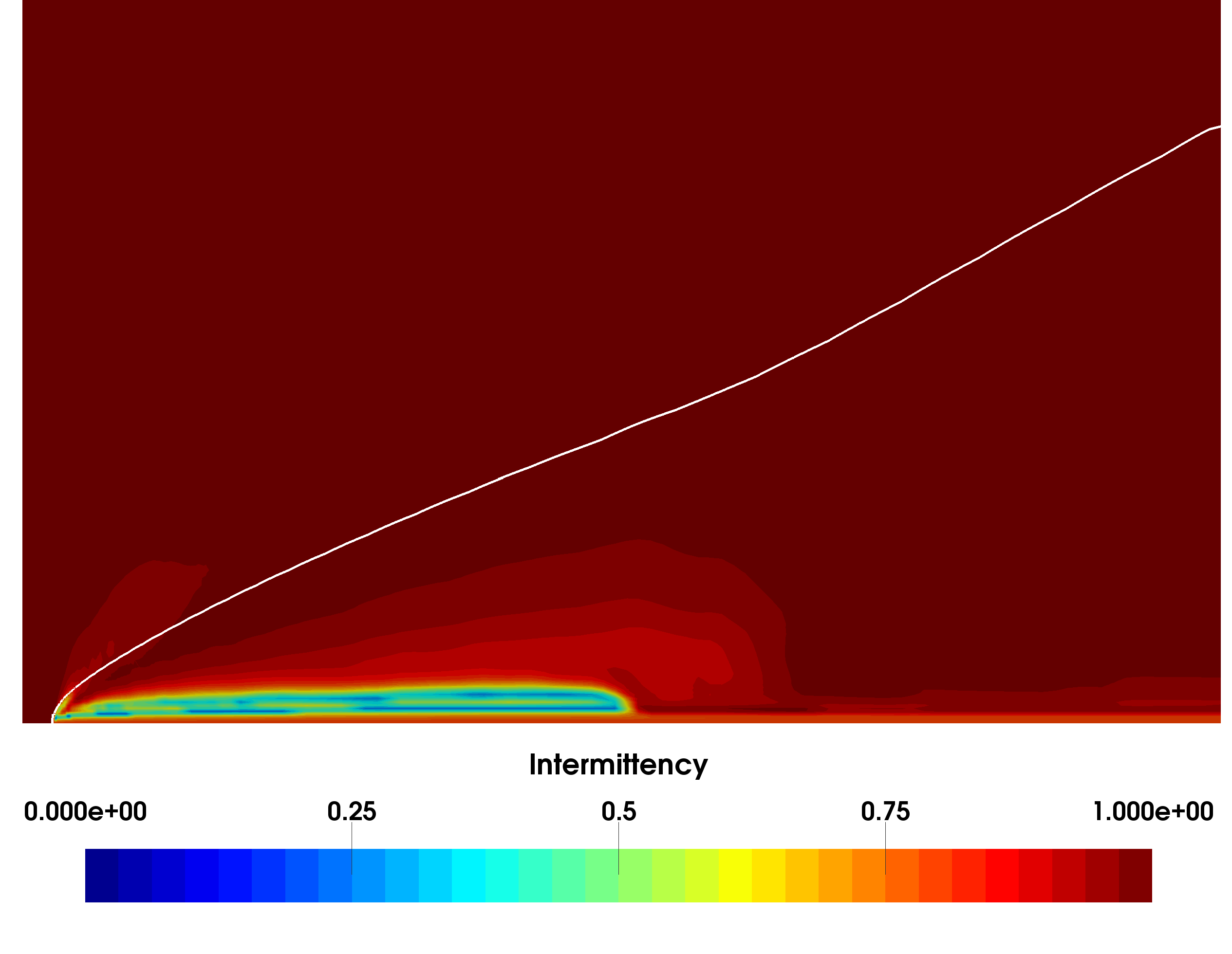} \label{fig:beta_T3A}}
        \subfigure[T3C1 (Flat plate length: 1.7 m)]
        {\includegraphics[width=0.45\textwidth]{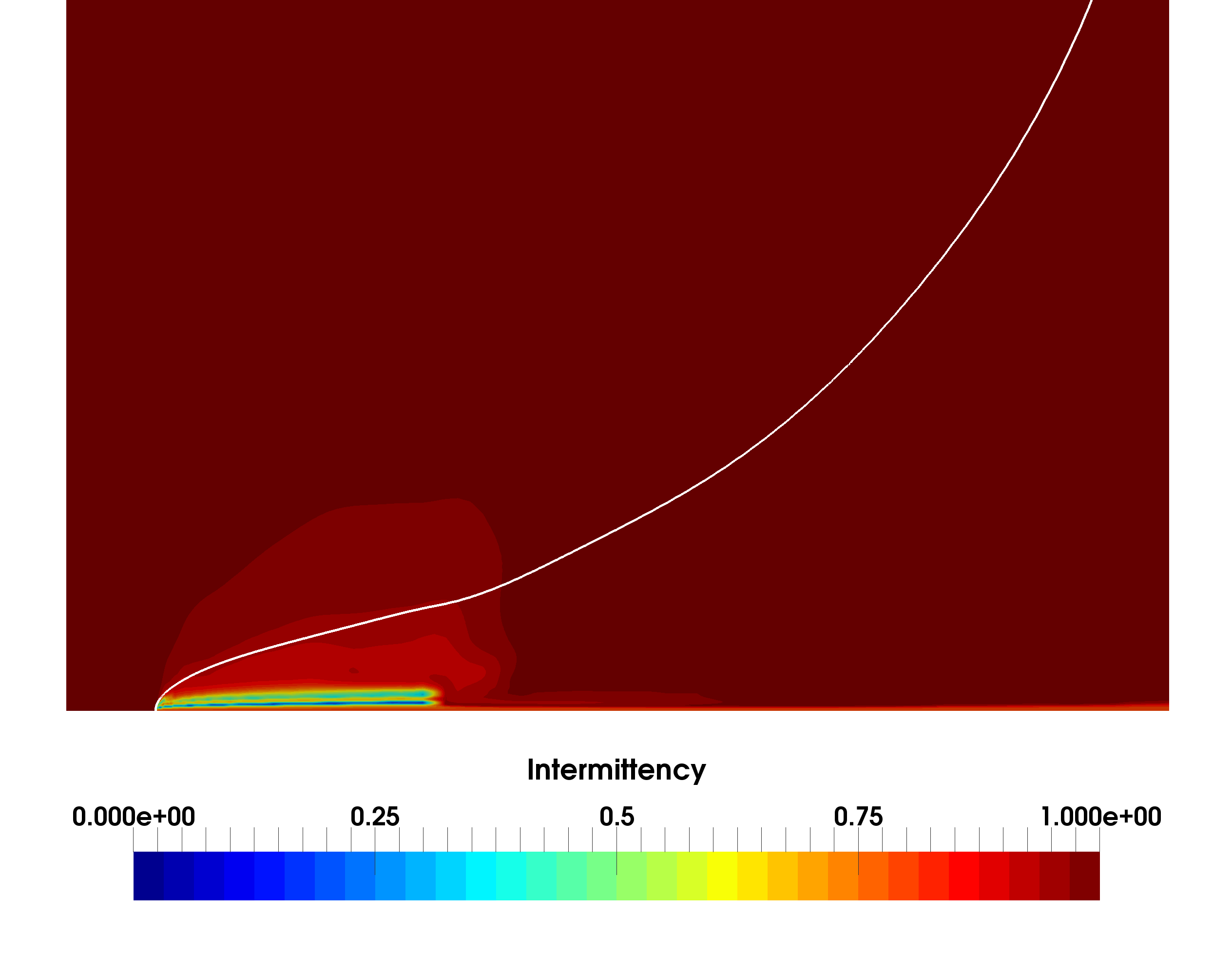} \label{fig:beta_T3C5}}
        \caption{Intermittency contours (White line marks $U=0.95U_\infty$), 40x scaling in the y-direction}
        \label{fig:beta_T3}
      \end{figure}
    
\begin{figure}[h!]
	    \centering
	    \subfigure[$\eta_1=0.05$]{\includegraphics[width=0.31\textwidth]{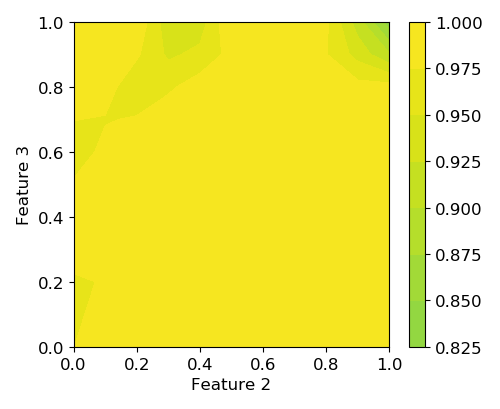}}
	    \subfigure[$\eta_1=0.15$]{\includegraphics[width=0.31\textwidth]{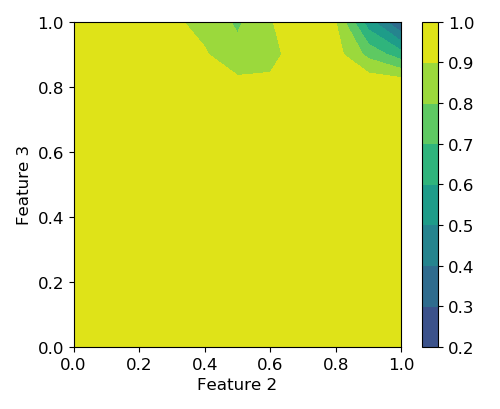}}
	    \subfigure[$\eta_1=0.25$]{\includegraphics[width=0.31\textwidth]{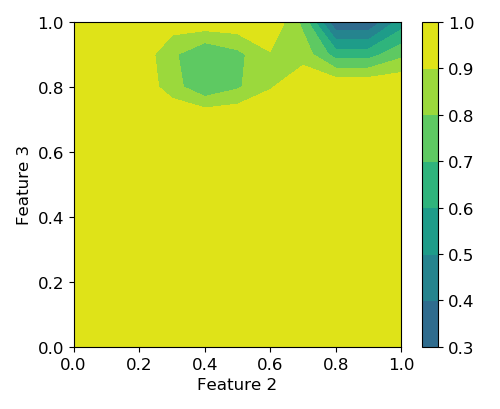}}
	    
	    \subfigure[$\eta_1=0.35$]{\includegraphics[width=0.31\textwidth]{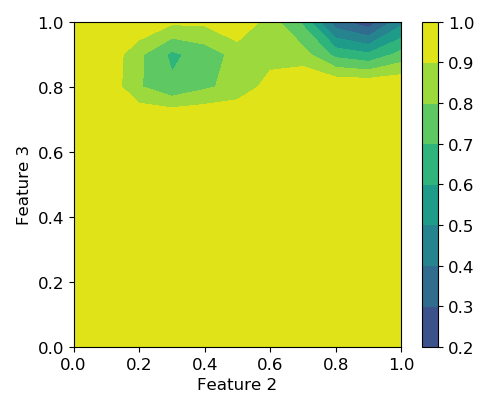}}
	    \subfigure[$\eta_1=0.45$]{\includegraphics[width=0.31\textwidth]{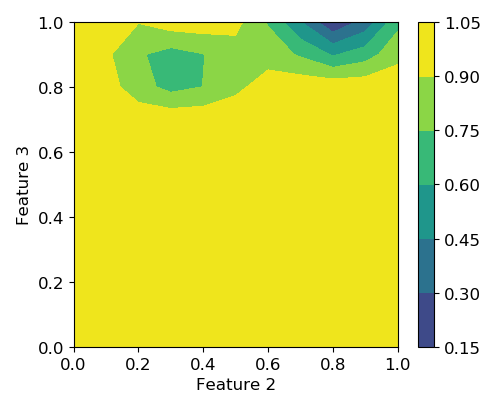}}
	    \subfigure[$\eta_1=0.55$]{\includegraphics[width=0.31\textwidth]{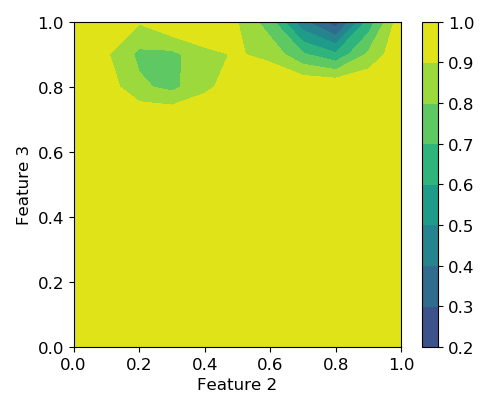}}
	    
	    \subfigure[$\eta_1=0.65$]{\includegraphics[width=0.31\textwidth]{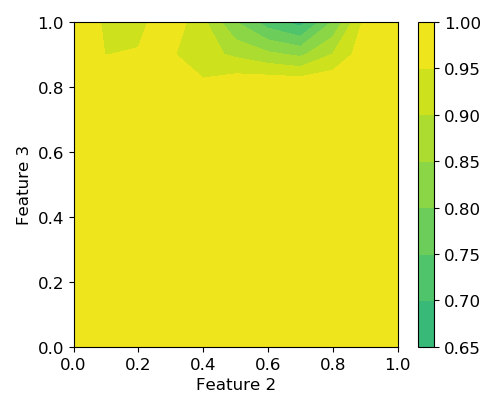}}
	    \subfigure[$\eta_1=0.75$]{\includegraphics[width=0.31\textwidth]{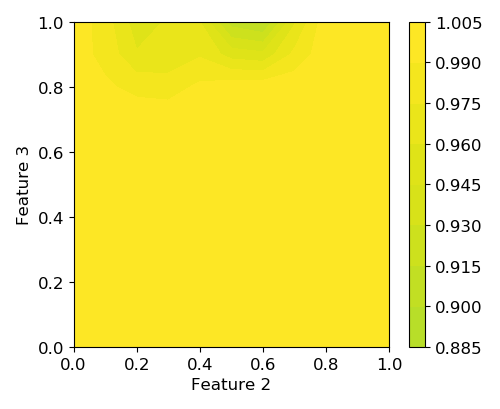}}
	    \subfigure[$\eta_1=0.85$]{\includegraphics[width=0.31\textwidth]{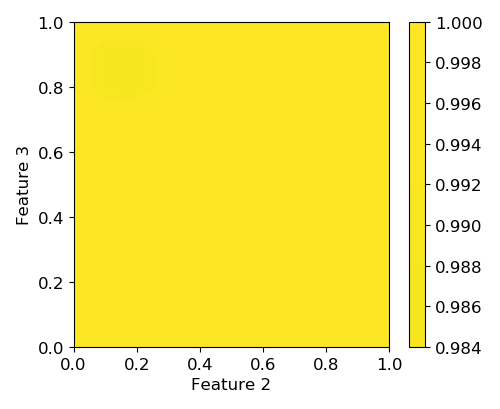}}
	    %\subfigure[$\eta_1=0.95$]{\includegraphics[width=0.19\textwidth]{figures/Ftr_09.png}}
	    \caption{Augmentation contours on feature space slices after training on T3A and T3C1 datasets}
	    \label{fig:feature_space}
      \end{figure}

      The contours of augmentation w.r.t. the feature space (slices of $\eta_1$) are shown in Figure \ref{fig:feature_space}.
      
      As can be seen in the feature space contours, the augmentation varies w.r.t. $\eta_1$ until only around $\eta_1 \approx 0.75$ above which it has a virtually constant value of 1.0 (Baseline).
      This happens because the augmentation is trained using cases involving transition occurring in regions of zero/favorable pressure gradients.
      Training the cases for the adverse pressure gradient regions would potentially populate some region with higher values of $\eta_1$.
      The augmentation assumes small values only in regions of high $\eta_3$ because low intermittency correlates to laminar regions in the flow where $\eta_3$ will be high.
      However, this is also the case in the viscous sub-layer and some parts of the buffer layer in the fully turbulent parts of the flow.
      In such a scenario, $\eta_2$ distinguishes between the laminar and turbulent parts of the flow.
      This is a clear example of how adding an extra feature can help in differentiating physical regions where the augmentation must take different values.
      
      \newadd{red}{For the purpose of comparing different learning techniques, three different set of results obtained - (1) using only the T3A case for training, (2) without using localized learning (i.e. using Neural Networks), and (3) using a finer discretization have been shown in appendices \ref{app:T3A_only}, \ref{app:without_localized_learning} and \ref{app:with_finer_grid}, respectively.
        Looking at these results, we can make the following observations:
        \begin{enumerate}
          \item As seen from the augmentation obtained by only using data from the T3A dataset for training in appendix \ref{app:T3A_only} which performs worse compared to the augmentation obtained using both the cases T3A and T3C1, adding more datasets which exhibit different physical phenomena compared to the already existing training datasets consistently improves predictive accuracy.
          \item Reducing spurious behavior using localized learning can not only make the augmentation more robust but is at times essential to infer a usable augmentation from available data. This is clearly evident when comparing the results presented in this section with the training results presented in appendix \ref{app:without_localized_learning} (which do not use localized learning and use neural networks as the functional form for the augmentation and fail miserably by predicting partially/fully laminar flow at all locations along the flat plate).
          \item During localized learning, correctly setting the range of influence that the datapoints have in the feature space is crucial in order to obtain a generalizable augmentation as mentioned in section \ref{ssec:localized_learning}. Also, as mentioned in section \ref{ssec:localized_learning}, this work does not deal with a method to optimally set the range of influence and instead focuses on why localized learning is necessary to improve generalizability and reduce spurious behavior arising from extrapolation within the bounded feature space. Nevertheless, the comparison between the test results for augmentations using different grid resolutions in the feature space (see appendix \ref{app:with_finer_grid} where a finer grid leads to the augmentation being learnt in a smaller region of the feature space which results in limited generalizability) demonstrate the importance of setting an optimal range of influence.
        \end{enumerate}
      }
    
  \subsection{Testing the model}
  
    \subsubsection{Mesh for VKI turbine cascade cases}
\begin{figure}[h]
      \centering
      \includegraphics[width=0.9\textwidth]{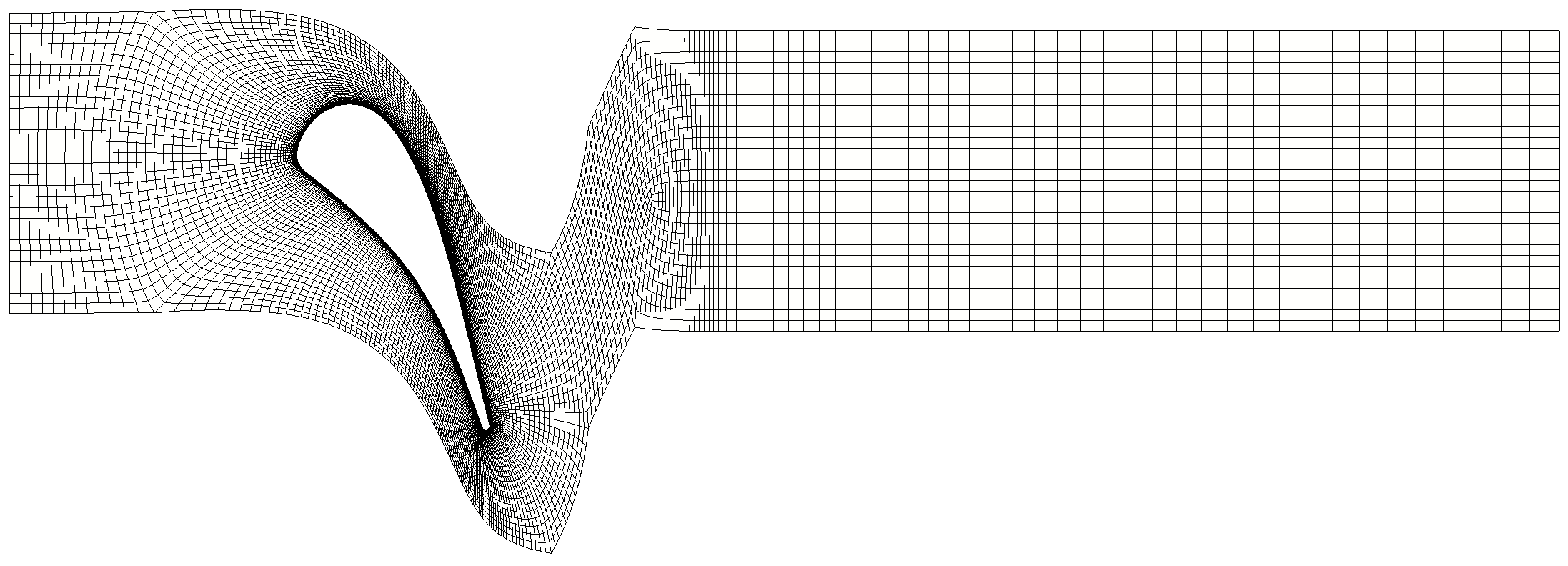}
      \caption{Mesh used for RANS simulation of VKI turbine cascade}
      \label{fig:VKI_mesh}
      \end{figure}

      The mesh used for the turbine cascade cases can be seen in Figure \ref{fig:VKI_mesh}.
      The blade chord is $0.067$ m in length and makes a $55^\circ$ angle with the streamwise direction.
      The inlet is situated $0.055$ m upstream of the leading edge, while the outlet is located $0.242$ m downstream of the leading edge.
      The mesh resolution next to the wall is on the order of $y^+\approx1$.
      
    \subsubsection{Inflow conditions for T3 test cases}
    
\begin{figure}[h!]
      \centering
      \subfigure[T3B] { \includegraphics[width=0.24\textwidth]{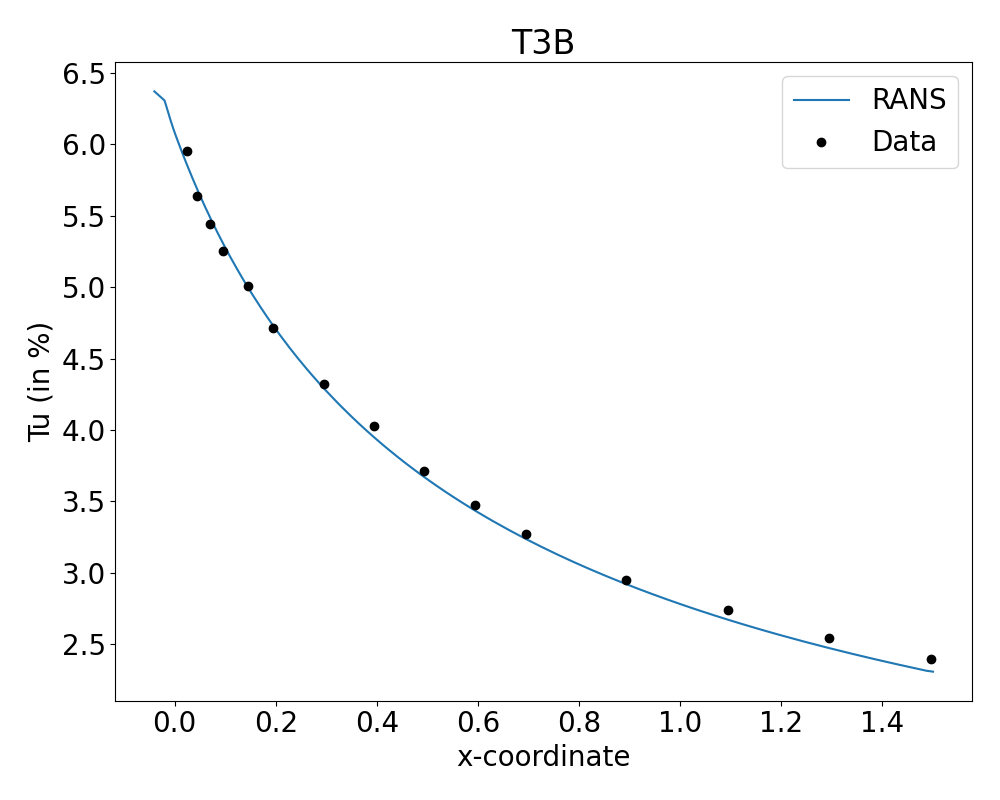} }
      \subfigure[T3C2] { \includegraphics[width=0.24\textwidth]{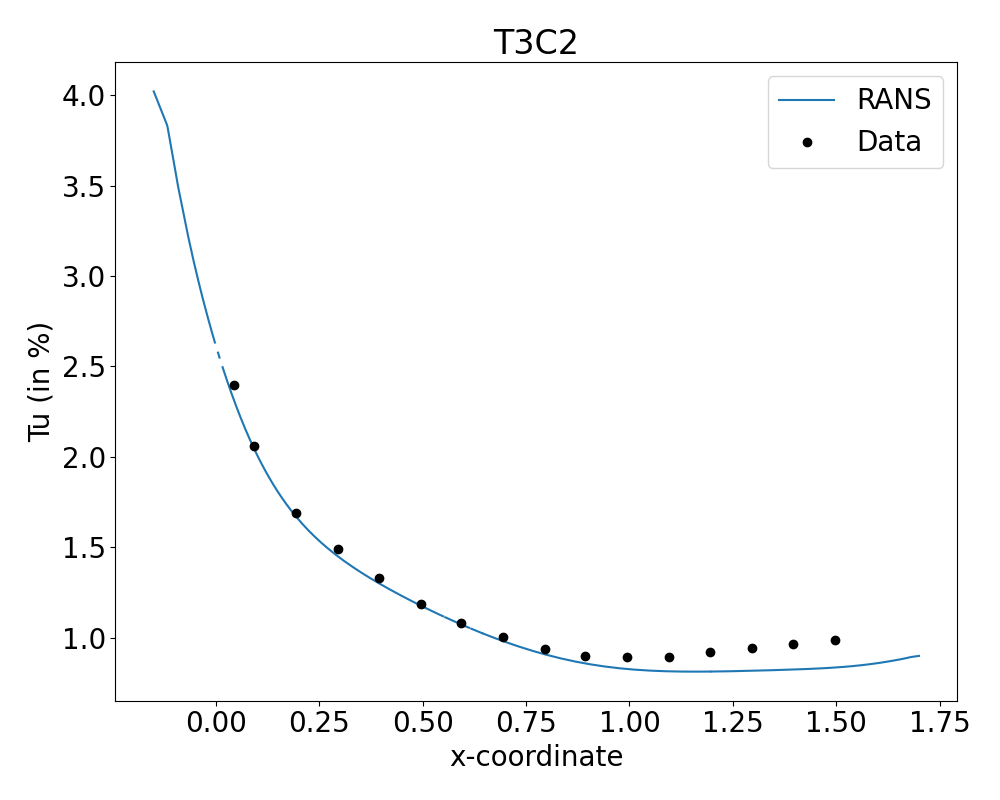} }
      \subfigure[T3C3] { \includegraphics[width=0.24\textwidth]{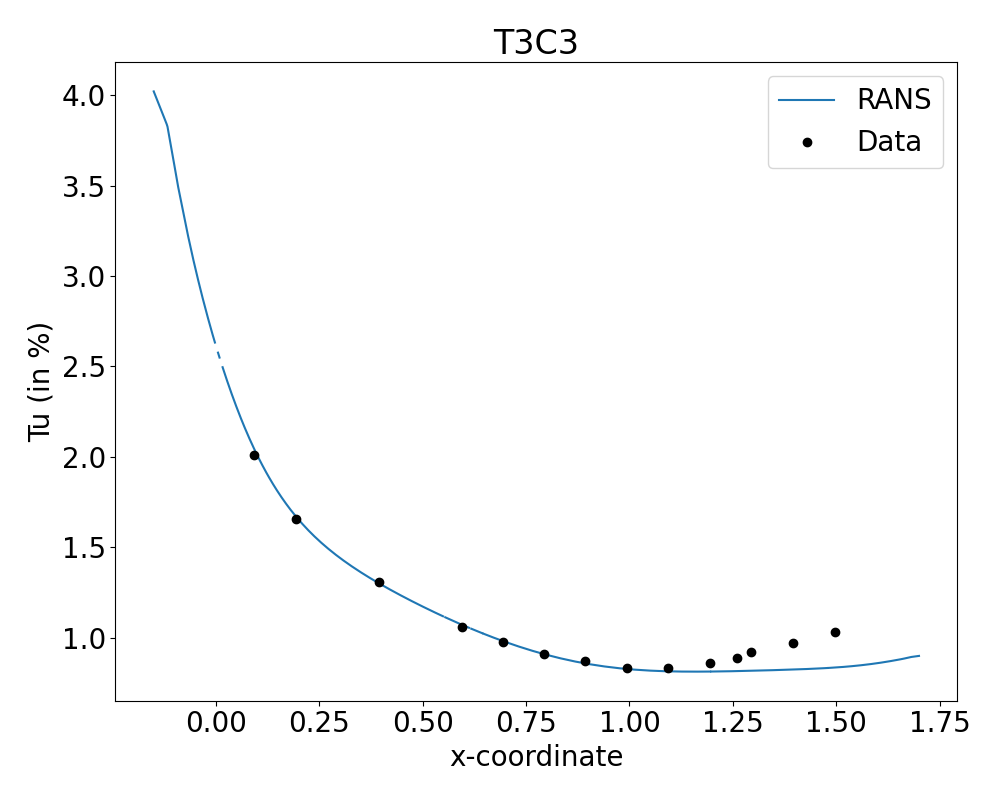} }
      \subfigure[T3C5] { \includegraphics[width=0.24\textwidth]{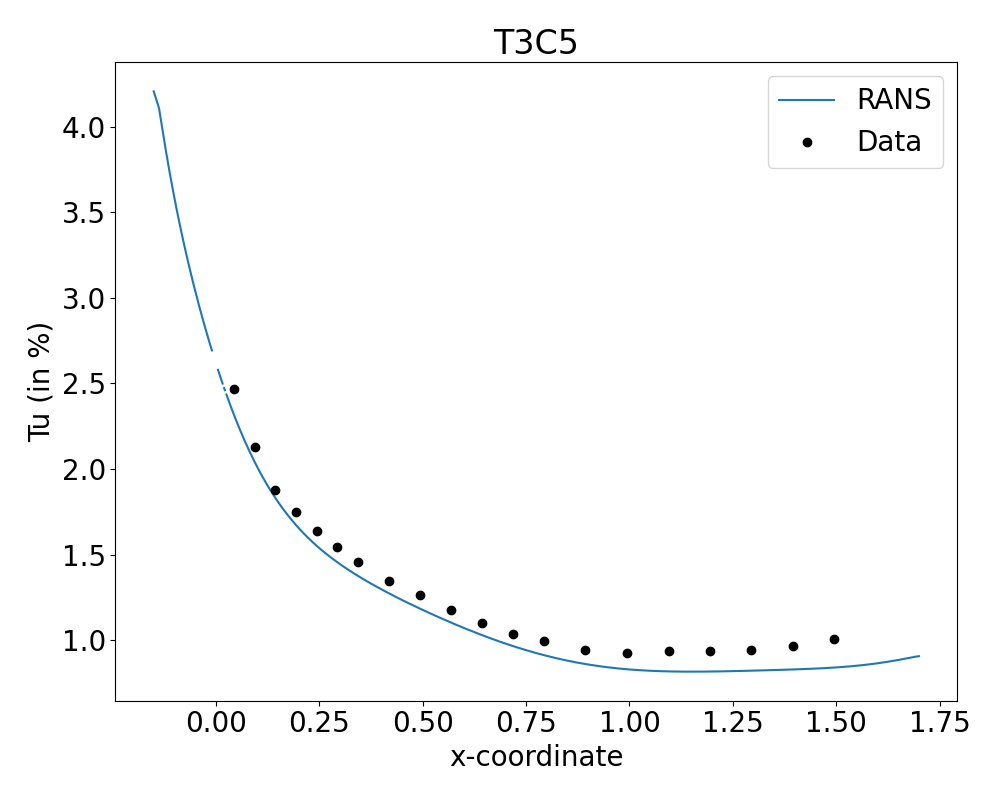} }
      \caption{Decay of freestream turbulence intensity}
      \label{fig:test_TI_decay}
      \end{figure}

      The information needed to characterize the inflow conditions for the T3 test cases has been mentioned in table \ref{table:T3_test_inflow}.
      \newadd{red}{The turbulence intensity decay plots shown in figure \ref{fig:test_TI_decay} verify the $\omega$ boundary conditions.}
    
      \begin{table}[H]
      \begin{center}
      \begin{tabular}{c|c|c|c|c}
        \hline
        \textbf{Cases} & \textbf{T3B} & \textbf{T3C2} & \textbf{T3C3} & \textbf{T3C5}  \\ 
        \hline              
        $Tu_\text{in}$                              & $0.065$   & $0.037$    & $0.034$   & $0.043$ \\
        $\nu_{t_\text{in}}/\nu$                     & $100.0$   & $12.0$     & $8.0$     & $17.0$  \\
        $L$(in $m$)                                 & $1.5$     & $1.65$     & $1.65$    & $1.65$  \\
        $Re_{L,\text{in}}$                          & $940000$  & $550000$   & $418000$  & $946000$ \\
        \newadd{cyan}{$\omega_\text{in}$(in $s^{-1}$)} & $7.943$   & $11.4083$  & $10.982$  & $18.70738$\\
        \hline
      \end{tabular}
      \caption{Inflow conditions for the T3 test cases}
      \label{table:T3_test_inflow}
      \end{center}\vspace{-1em}
      \end{table}
    \subsubsection{Inflow, wall and outflow conditions for VKI test cases}
    
      The information needed to characterize the inflow conditions for the VKI turbine cascade test cases along with pressure at outflow boundary and temperature at the isothermal wall is presented in table \ref{table:VKI_conditions}.
      \newadd{red}{
        The boundary condition for $\omega_\text{in}$ was set to ensure that the viscosity ratio $\nu_{t_\text{in}}/\nu_\text{in}$ remains between the range from $1.5$ to $50$.
      }
      
      \begin{table}[h!]
      \begin{center}
      \begin{tabular}{c|c|c|c|c}
        \hline
        \textbf{Cases} & \textbf{MUR116} & \textbf{MUR129} & \textbf{MUR224} & \textbf{MUR241} \\ \hline 
        $Tu_\text{in}$                              & $0.008$         & $0.008$         & $0.06$          & $0.06$   \\      
        $\nu_{t_\text{in}}/\nu_\text{in}$           & $3$             & $1.556$         & $43.537$        & $15.465$ \\      
        $p_{0,\text{in}}$(in bar)                   & $3.269$         & $1.849$         & $0.909$         & $3.257$  \\      
        $T_{0,\text{in}}$(in K)                     & $418.9$         & $409.2$         & $402.6$         & $416.4$  \\      
        $p_\text{out}$(in bar)                      & $1.550$         & $1.165$         & $0.522$         & $1.547$  \\      
        \newadd{red}{$T_\text{wall}$(in K)}          & $300.0$         & $300.0$         & $300.0$         & $300.0$  \\
        \newadd{cyan}{$\omega_\text{in}$(in $s^{-1}$)} & $1.5\times10^4$ & $1.5\times10^4$ & $1.5\times10^4$ & $1.5\times10^5$ \\
        \hline
      \end{tabular}
      \caption{Inflow, wall and outflow conditions for VKI test cases}
      \label{table:VKI_conditions}
      \end{center}\vspace{-1em}
      \end{table}
      
      Notice here, that MUR116 is a case with high pressure differential and low $Tu_\infty$, MUR129 is a case with low pressure differential and low $Tu_\infty$, MUR241 is a case with high pressure differential and high $Tu_\infty$, and finally MUR224 is a case with low pressure differential and high $Tu_\infty$.
      This distinction will be important later.
      
\begin{figure}[h]
        \centering
        \subfigure[T3A]
        {\includegraphics[width=0.45\textwidth]{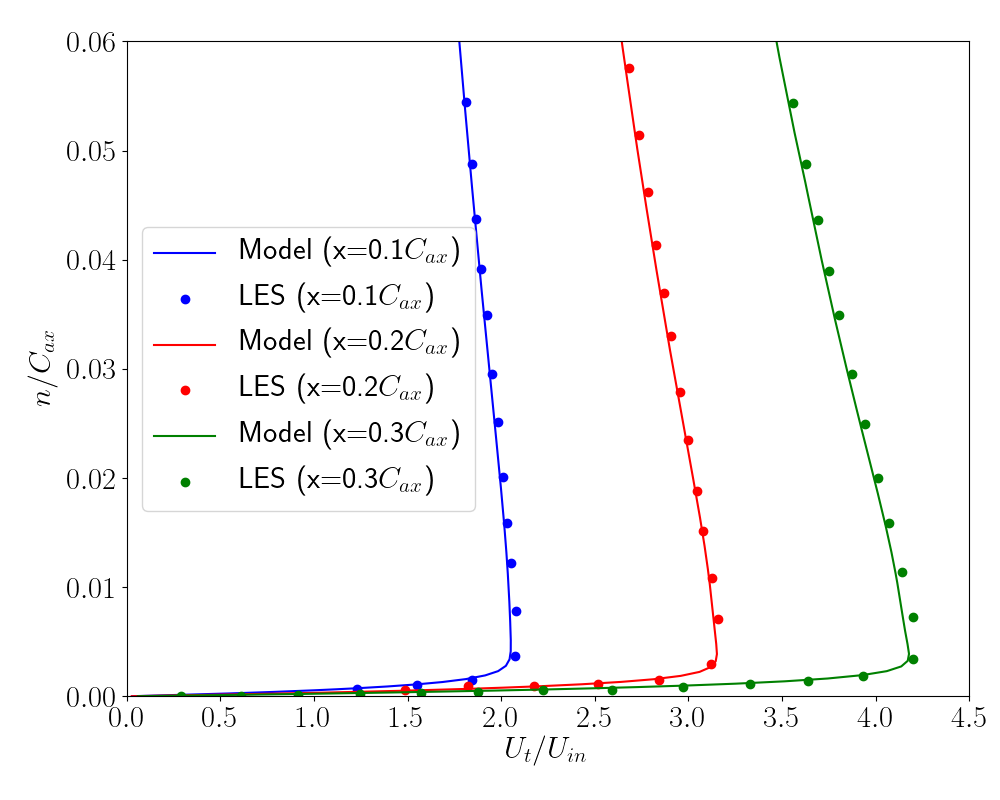} \label{fig:MUR224n_vs_U}}
        \subfigure[T3C1]
        {\includegraphics[width=0.45\textwidth]{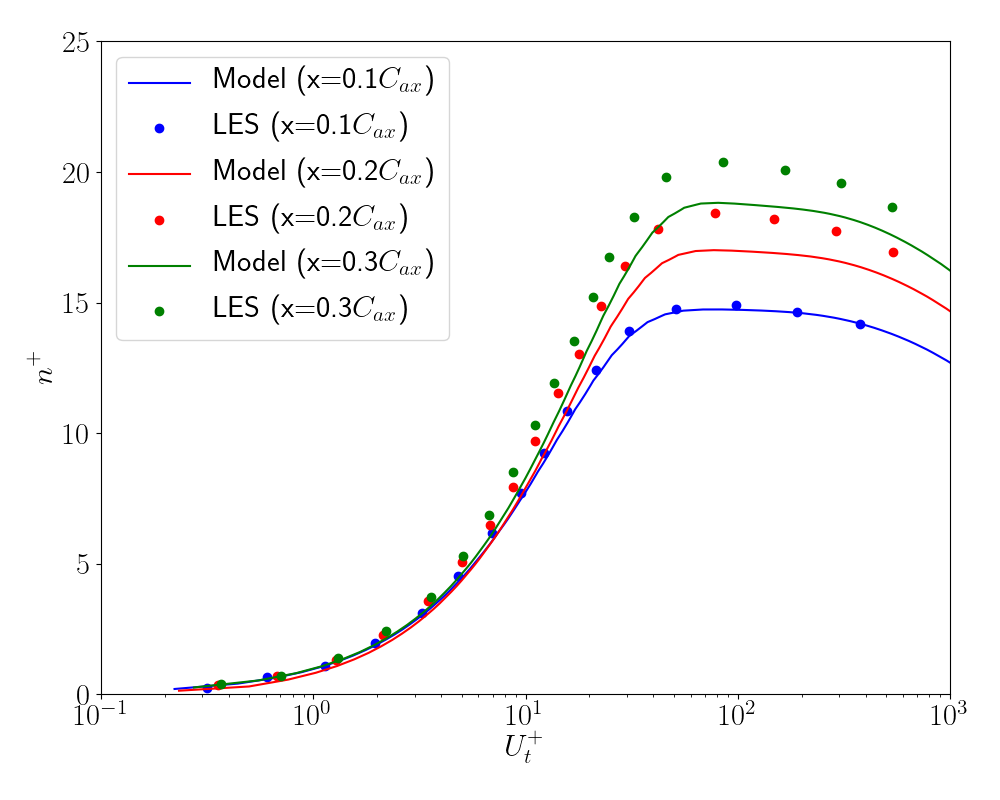} \label{fig:MUR224U_vs_logn}}
        \caption{Verification of velocity profiles for MUR224 using LES data \cite{zhao_sandberg_2020} ($C_{ax}$ refers to chord length in the x-direction)}
        \label{fig:MUR224_verification}
      \end{figure}

      The comparison of velocity profiles at three different locations on the turbine blade along the wall normal direction is shown in figure \ref{fig:MUR224_verification} for verification.
      
    \subsubsection{Predictions using the augmented model on the same geometries}
    
\begin{figure}[h]
        \centering
        \subfigure[T3B]
        {\includegraphics[width=0.4\textwidth]{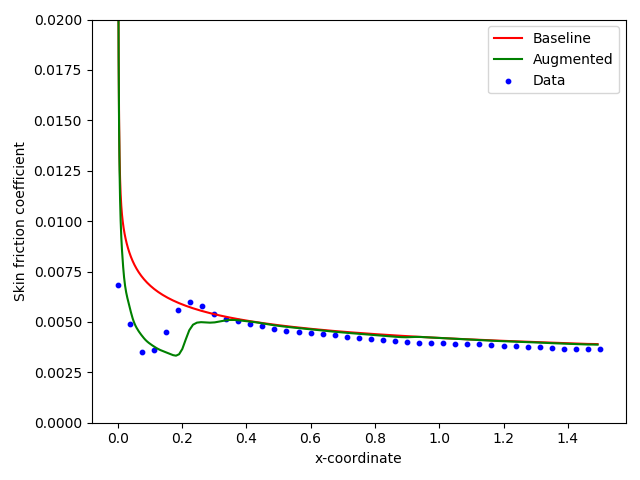} \label{fig:Cf_T3B}}
        \subfigure[T3C2]
        {\includegraphics[width=0.4\textwidth]{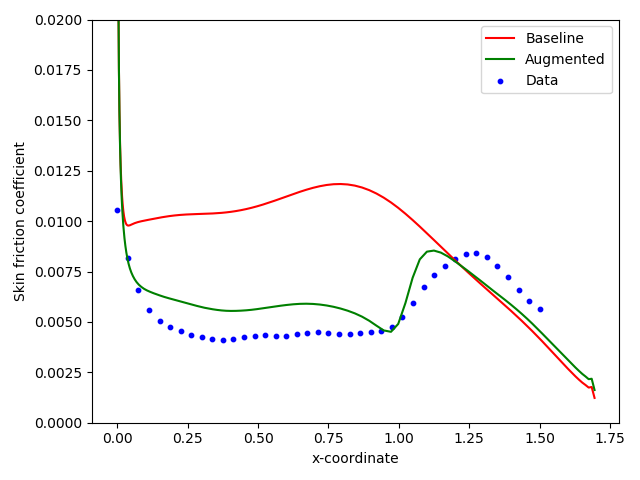} \label{fig:Cf_T3C2}}
        
        \subfigure[T3C3]
        {\includegraphics[width=0.4\textwidth]{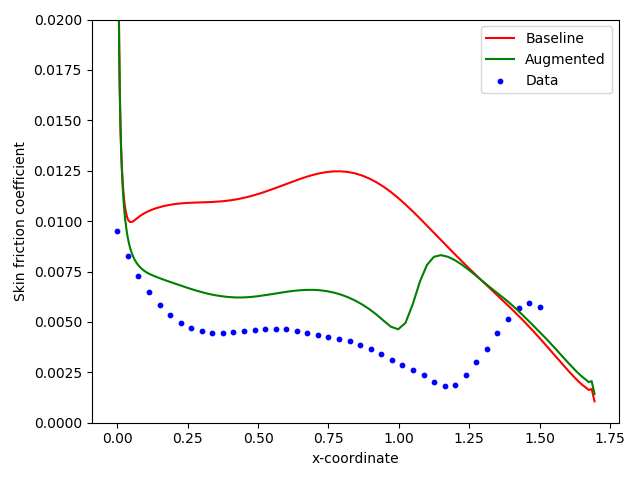} \label{fig:Cf_T3C3}}
        \subfigure[T3C5]
        {\includegraphics[width=0.4\textwidth]{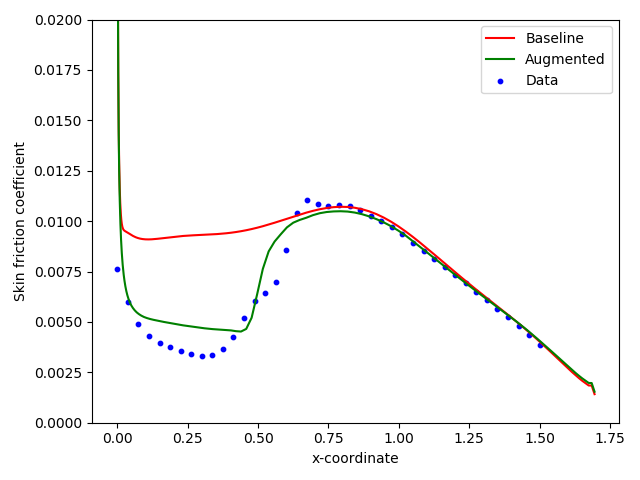} \label{fig:Cf_T3C5}}
        \caption{Predicted Skin friction coefficients for flat plate cases using the model
        trained on T3A and T3C1.}
        \label{fig:Cf_T3_test}
      \end{figure}

      Figure \ref{fig:Cf_T3_test} shows results obtained by applying the model trained on T3A and T3C1 to the T3B, T3C2, T3C3, and T3C5 cases.
      The transition locations for T3B, T3C2 and T3C5 cases are predicted reasonably well in the results shown below.
      This is because the transition occurs in all of these cases in either zero, favorable or very mild adverse pressure gradient regions.
      The model fails to predict the transition location correctly for T3C3 for this reason and instead assumes the baseline value for $\beta=1$ in the adverse pressure gradient regions, thus predicting premature transition.
      Once again, it can be noticed that the laminar part of the flow does not show fully laminar friction coefficients as can be seen when compared to data.
      This can be attributed, as explained before, to the predicted intermittency having low values in a very narrow region of the flow field.
    
    \subsubsection{Predictions using the augmented model on a different geometry}
    
\begin{figure}[h!]
        \centering
        \subfigure[MUR116]
        {\includegraphics[width=0.4\textwidth]{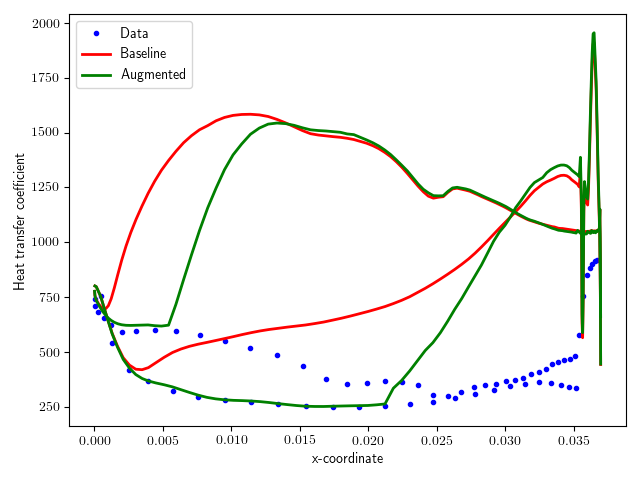} \label{fig:VKI_MUR116}}
        \subfigure[MUR129]
        {\includegraphics[width=0.4\textwidth]{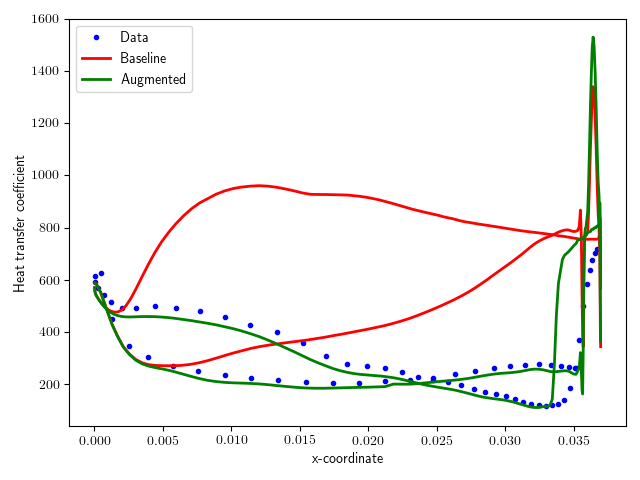} \label{fig:VKI_MUR129}}
        
        \subfigure[MUR224]
        {\includegraphics[width=0.4\textwidth]{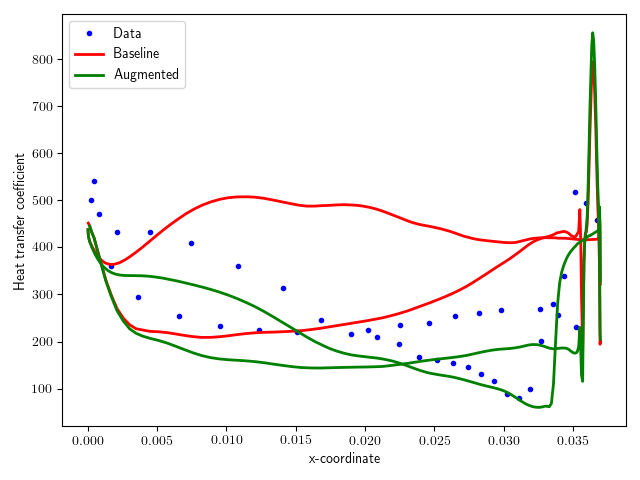} \label{fig:VKI_MUR224}}
        \subfigure[MUR241]
        {\includegraphics[width=0.4\textwidth]{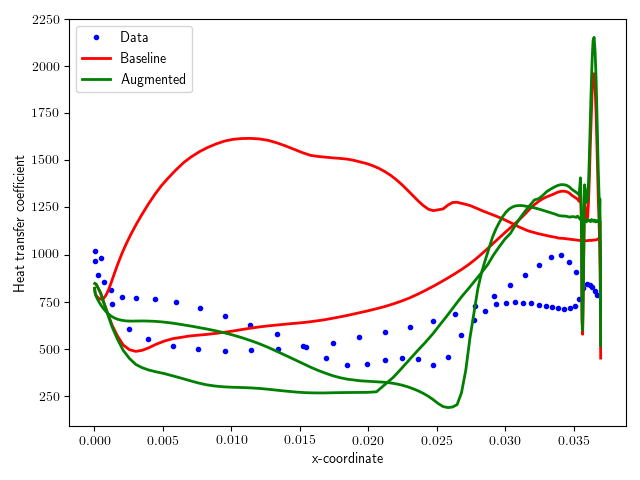} \label{fig:VKI_MUR241}}
        \caption{Predicted heat transfer coefficients for VKI turbine cases using model trained
        on T3A and T3C1.}
        \label{fig:Ch_VKI}
      \end{figure}

      Figure \ref{fig:Ch_VKI} shows the heat transfer coefficients predicted by the model trained on T3A and T3C1 on the VKI high pressure turbine cases, which are not only geometrically different from the training set, but also involve combinations of pressure gradients and freestream turbulence intensities not seen in the training data.
      Predictions with two different preset distance intervals to extract $Re_{\theta,t}$ information from the freestream are shown in Appendix \ref{app:preset} to assess the sensitivity of the solution to such modeling choices.
      For MUR224 and MUR129, the transition location is predicted fairly accurately and consistently with both the preset distance intervals.
      For MUR241, the transition location on the suction side is predicted very well.
      On the pressure side, however, the data shows a gradual transition to turbulence, while the model predicts a sharp transition at different locations within this gradual transition range when using different preset distance intervals.
      For MUR116, transition is premature for on either sides of the blade.
      
\begin{figure}[h!]
        \centering
        \subfigure[$\eta_1 = \min\left(\dfrac{Re_\Omega}{\overline{Re_{\theta,t}}},3\right)$]
        {\includegraphics[width=0.31\textwidth]{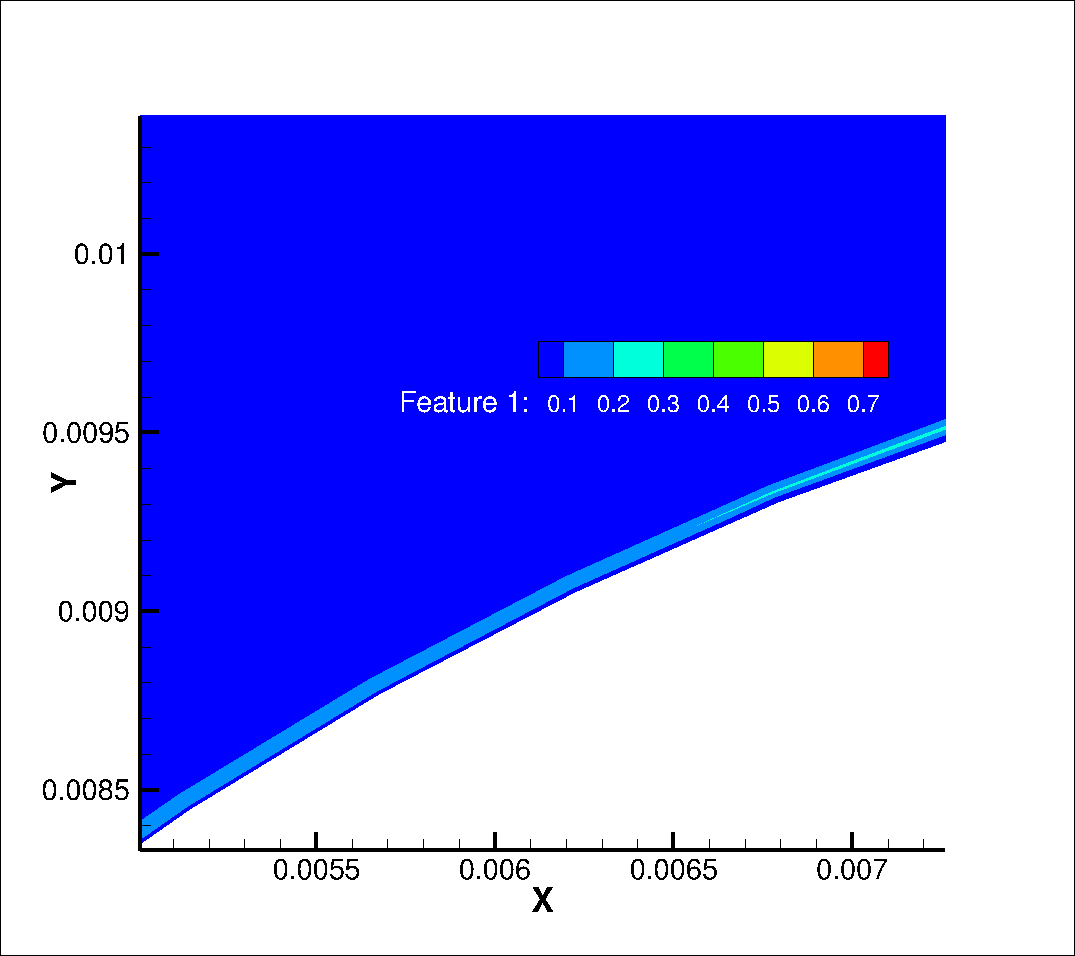}}
        \subfigure[$\eta_2 = \dfrac{\nu}{\nu+\nu_t}$]
        {\includegraphics[width=0.31\textwidth]{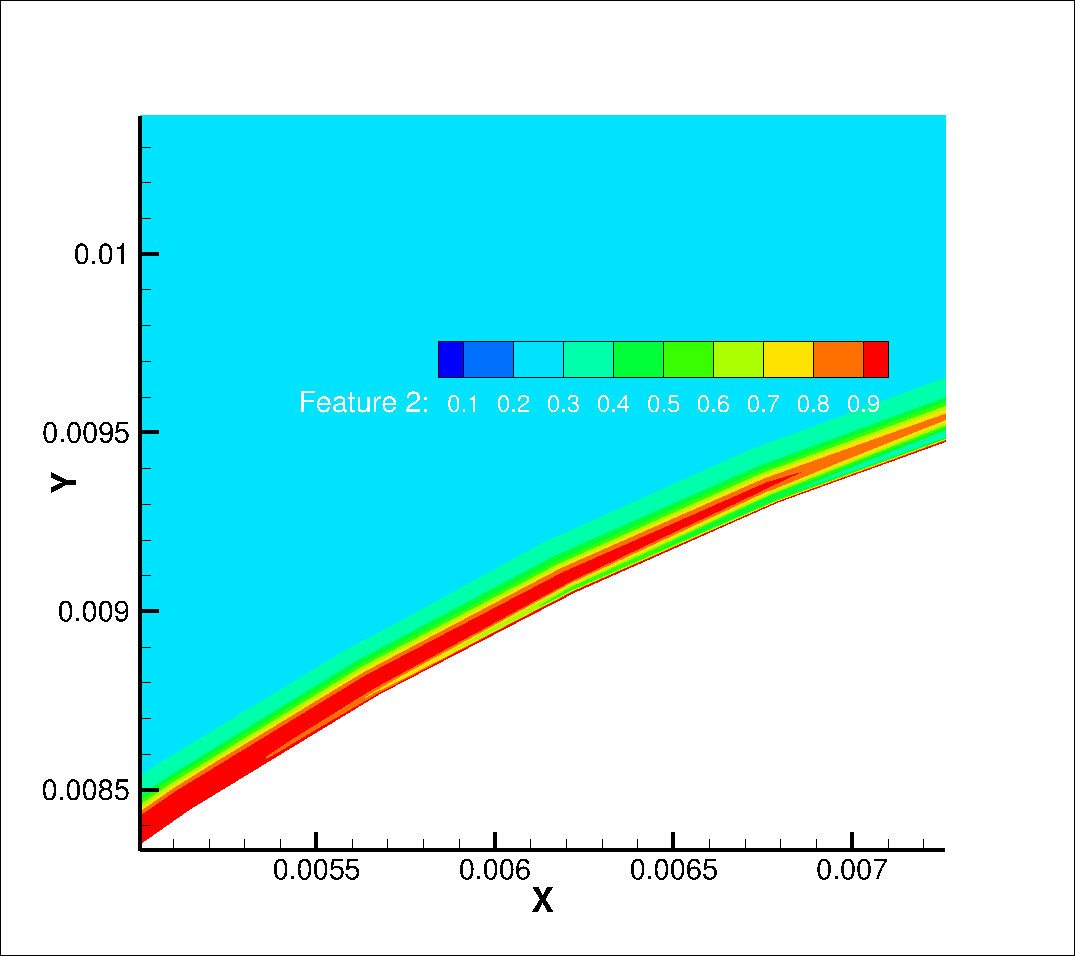}}
        \subfigure[Intermittency ($\gamma$)]
        {\includegraphics[width=0.31\textwidth]{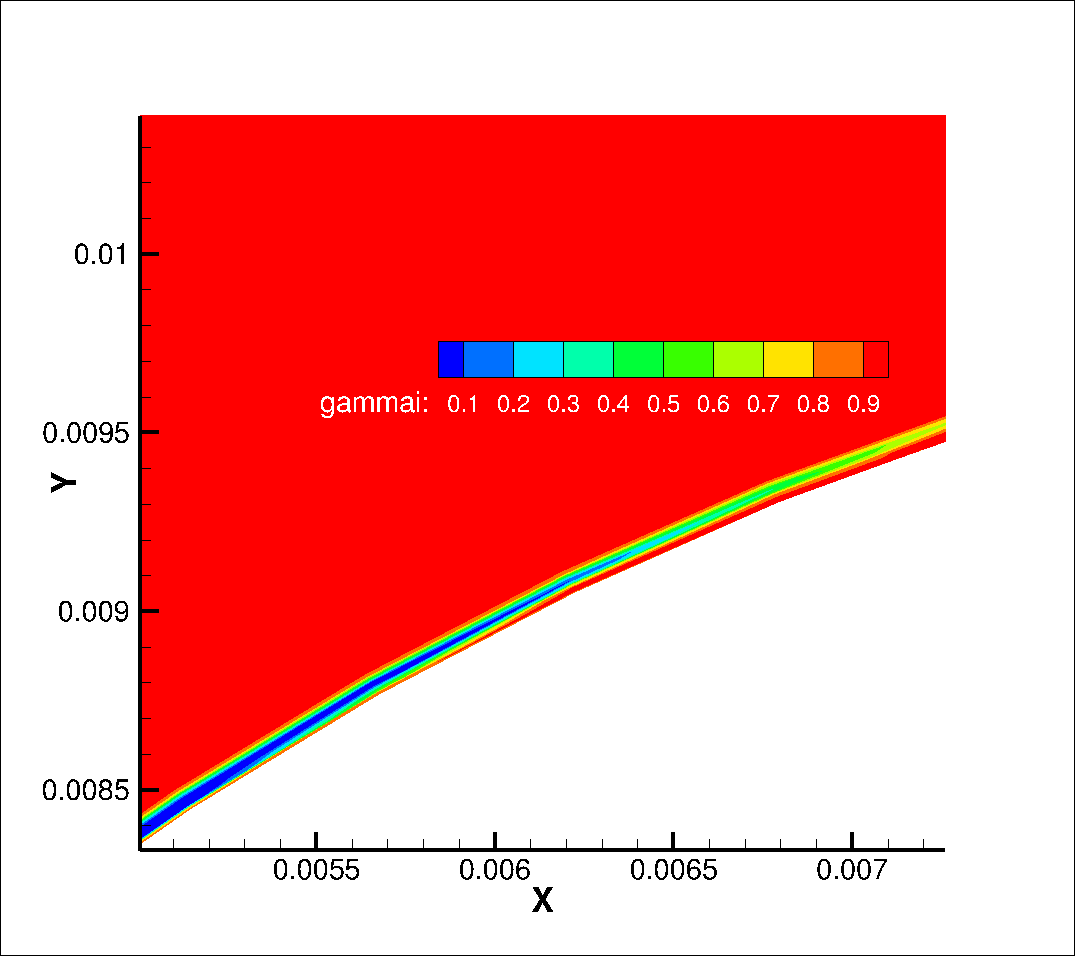}}
        \caption{Contours near the transition location on suction surface for MUR116}
        \label{fig:MUR_116_diagnostic}
      \end{figure}

\begin{figure}[h!]
        \centering
        \subfigure[$\eta_1 = \min\left(\dfrac{Re_\Omega}{\overline{Re_{\theta,t}}},3\right)$]
        {\includegraphics[width=0.31\textwidth]{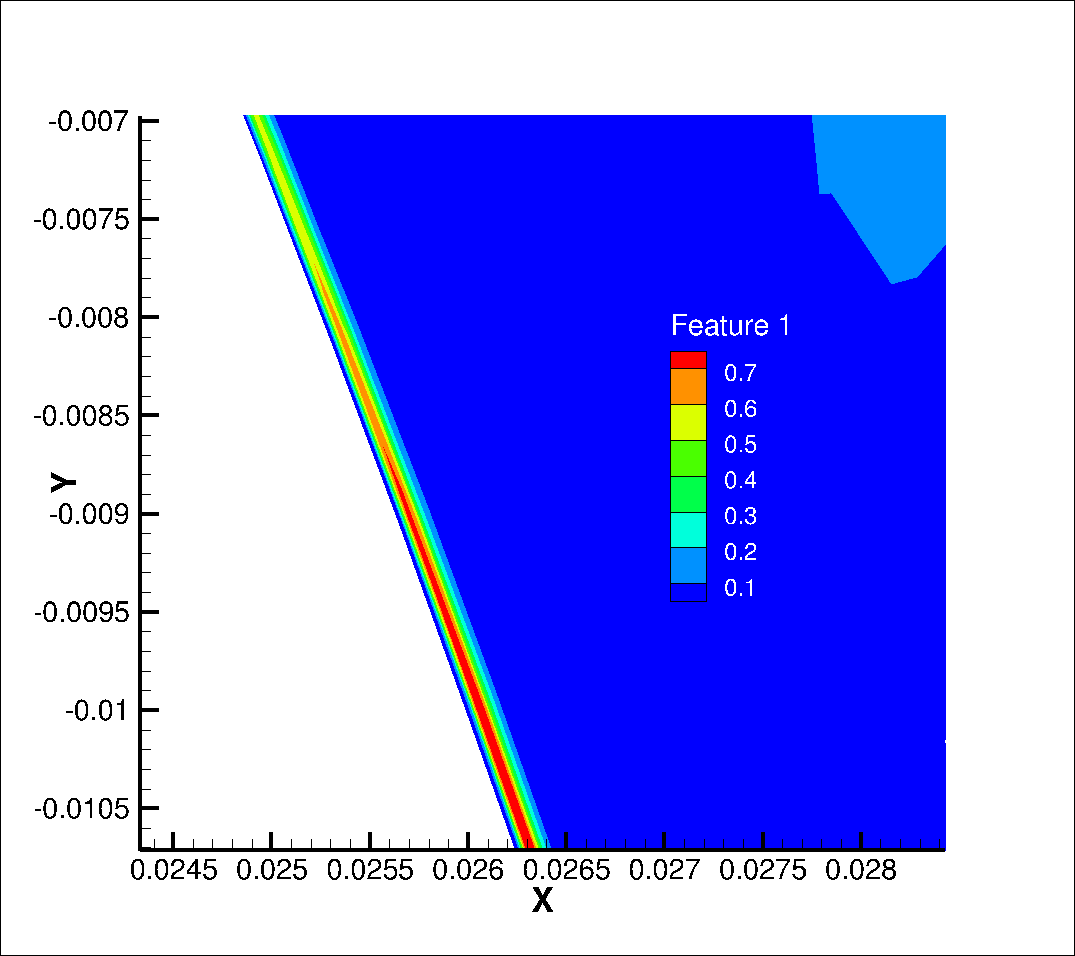}}
        \subfigure[$\eta_2 = \dfrac{\nu}{\nu+\nu_t}$]
        {\includegraphics[width=0.31\textwidth]{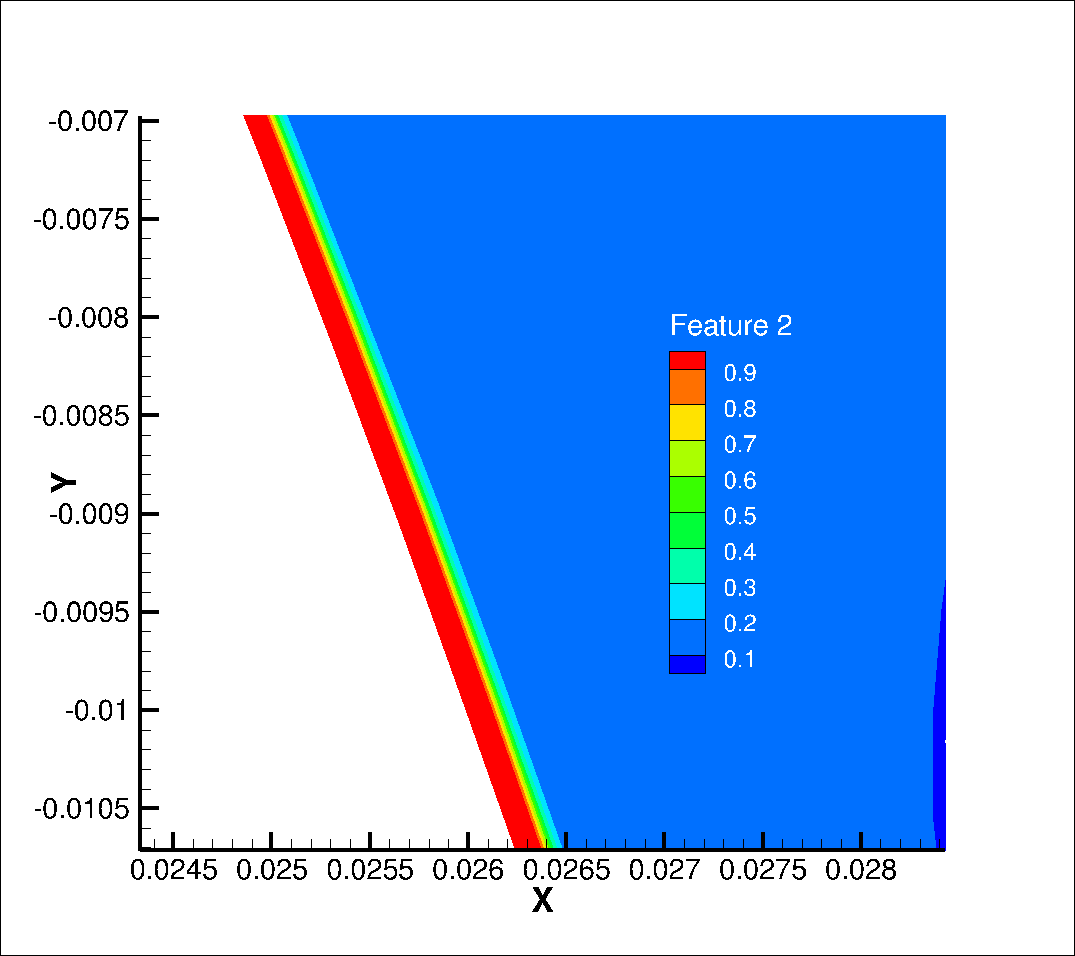}}
        \subfigure[Intermittency ($\gamma$)]
        {\includegraphics[width=0.31\textwidth]{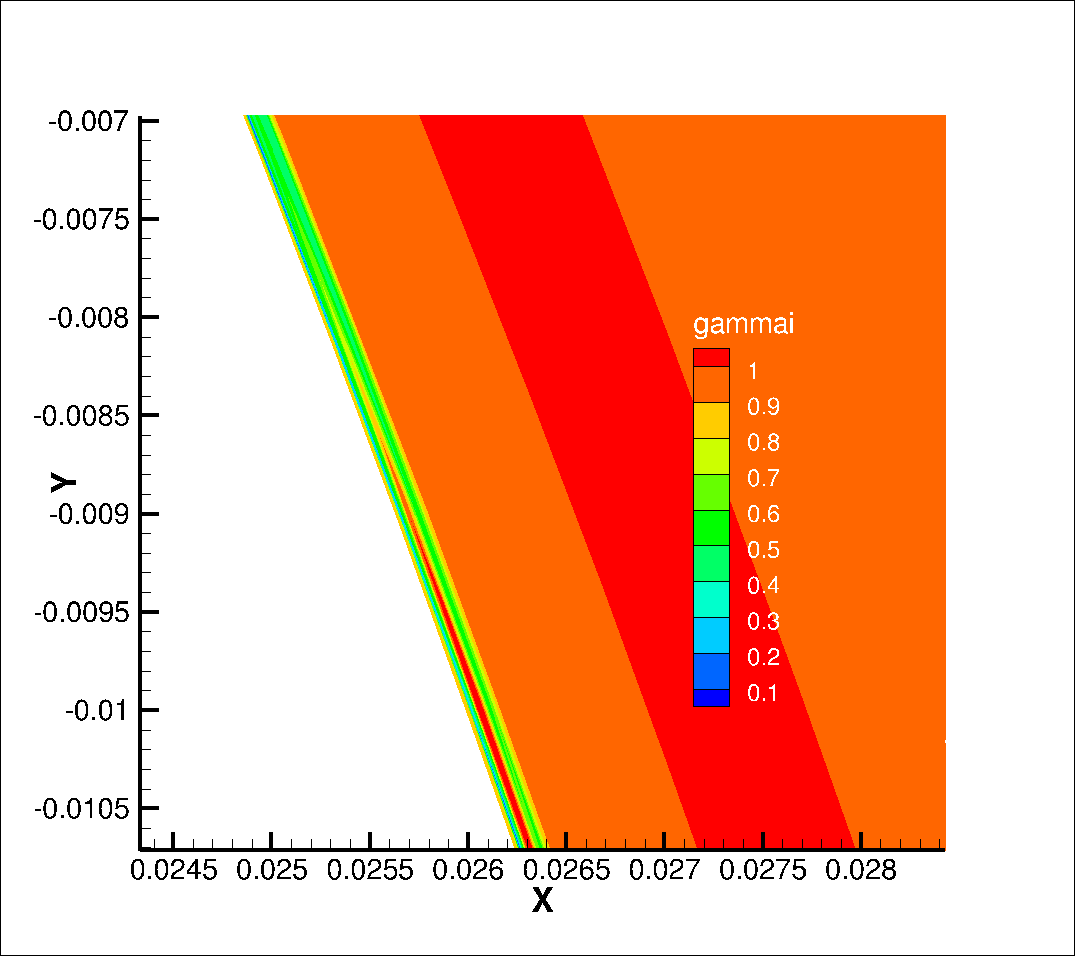}}
        \caption{Contours near the transition location on suction surface for MUR241}
        \label{fig:MUR_241_diagnostic}
      \end{figure}

      To diagnose the behavior in the MUR116 case, Figure \ref{fig:MUR_116_diagnostic} shows contours of two features and the intermittency near the transition location.
      The intermittency is seen to grow rapidly  in the region  $0.1 \leq \eta_1 \leq 0.2$ and $\eta_2 \leq 0.8$.
      This corresponds to a region in the feature space in which the augmentation does not learn significantly from the available data, leading to the augmentation reverting to its baseline behavior of $\beta=1$.
      It should be recognized however, that there is a strong feedback loop between the features, augmentation and transport processes \newdel{\sout{and thus, one should interpret causality in a more holistic sense}}.
      \newadd{red}{This means that all these quantities can influence each other in the model irrespective of the physical cause-effect relationships.}
      This behavior can be improved by bringing in more data to populate a larger part of the feature space.
      For comparison purposes, $\eta_1$, $\eta_2$ and $\gamma$ contours for MUR241 are shown in figure \ref{fig:MUR_241_diagnostic} which exhibit a physically intuitive growth of the intermittency in a region of high $\eta_1$ and $\eta_2$.
      
      \begin{figure}[h!]
        \centering
        \subfigure[MUR116]
        { \includegraphics[width=0.48\textwidth]{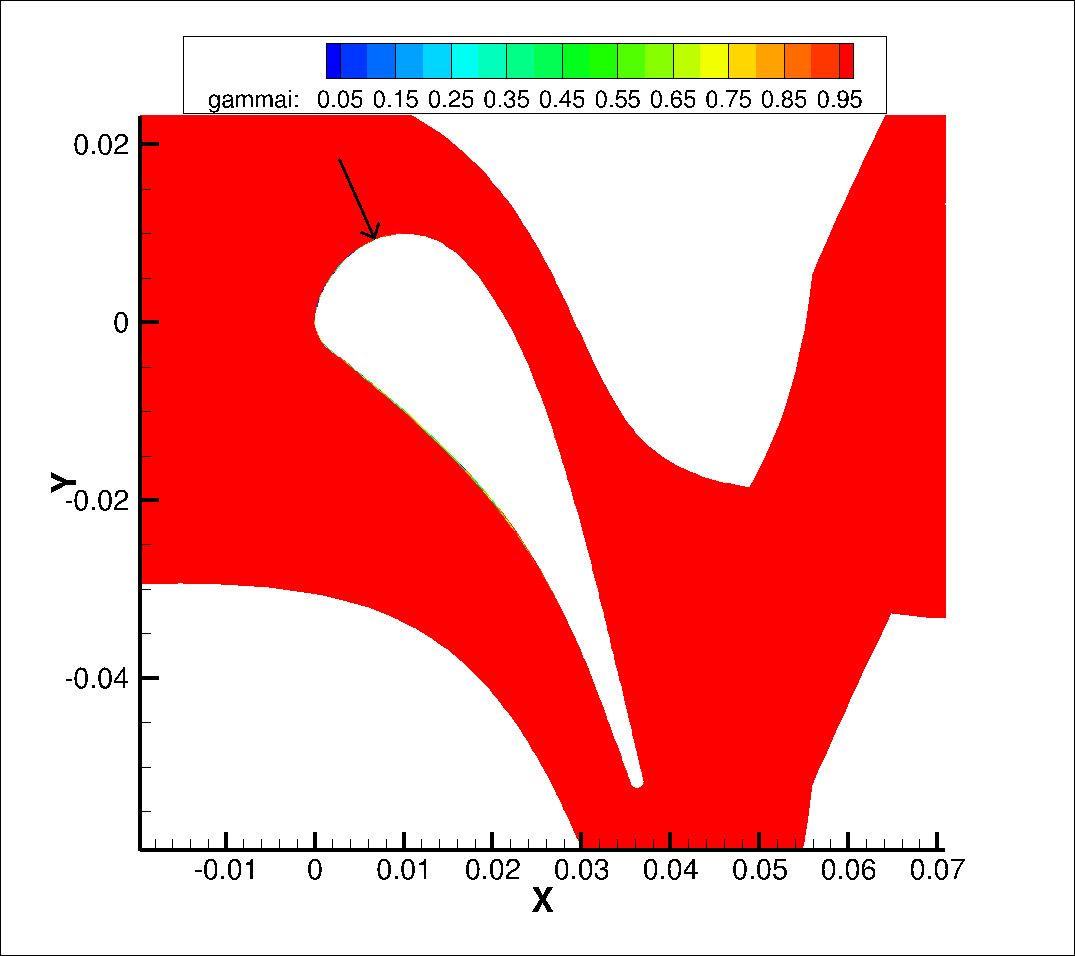} }
        \subfigure[MUR241]
        { \includegraphics[width=0.48\textwidth]{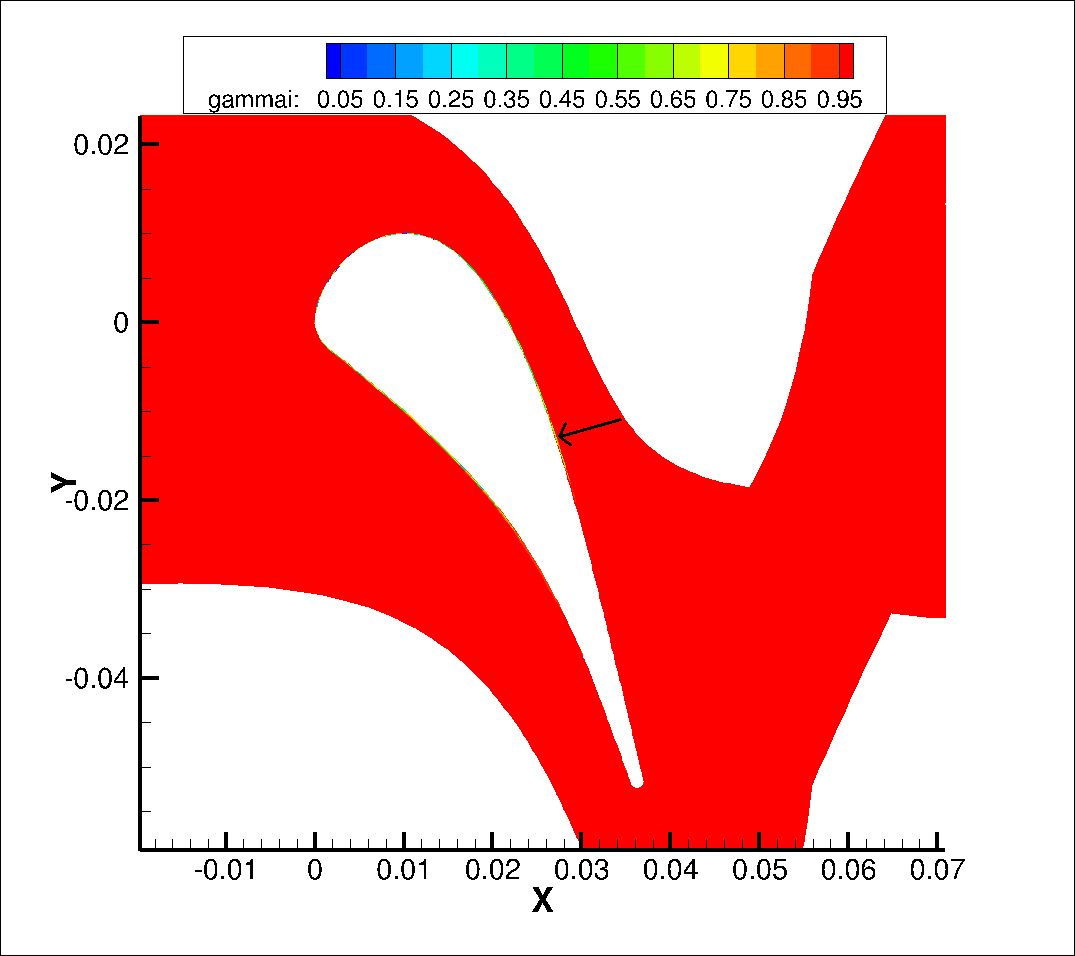} }
        \caption{Zoomed out views to indicate predicted transition locations on upper surface}
        \label{fig:MUR_Zoomed_Out}
      \end{figure}
      
      In addition, it can also be observed that - while the coefficient values are close to the data in laminar regions - there are large discrepancies between the two in the fully turbulent regions.
      This is because the turbulence model is imperfect \newdel{\sout{and should, therefore, be discounted when judging the capability of the augmentation.}}
      \newadd{red}{These discrepancies are not related to the transition phenomena and hence, are beyond the scope of the current augmentation.}
      
\section{Summary} \label{sec:Conclusions}

  This work presents a methodology that can be used to build data-driven augmentations to physics-based models.
  Building on recent work in data-driven turbulence modeling, this approach develops generalizable features-to-augmentation mapping across multiple datasets in a model-consistent manner.
  The proposed framework, referred to as ``Learning and Inference assisted by Feature-space Engineering (LIFE)'' places particular emphasis on the design of the feature space and control of the augmentation behavior in the feature space.
  Choosing features (along with appropriate physics-based non-dimensionalization) to create an ``effectively'' bounded feature space is indispensable as this boundedness ensures the tractability of the learning problem.
  In an ideal scenario, if an augmentation is learnt in this entire bounded region, an extrapolation in geometry/boundary conditions would translate to an interpolation in feature space.
  \newadd{green}{Given the availability of a limited number of datasets, human intuition and expert knowledge is essential, in the authors' opinion to formulate such features.
  Although automated feature selection techniques do exist, they are heavily (and often prohibitively) expensive to find a parsimonious combination of features that can address the augmentation, especially for complex physical problems like transition and turbulence.}
  To address the issue of the possible lack of data in certain significantly sized regions of the feature space, the notion of localized learning becomes critical to avoid spurious learning outputs.
  \newadd{green}{Localized learning refers to the modification of the augmentation function behavior at every learning step only in the vicinity of available datapoints in the feature space.}
  Additional measures, such as preventing a region in the feature space from representing different regions in the physical space where the augmentation must assume different values (i.e. ensuring a one-to-one features-to-augmentation mapping) by selecting an appropriate combination of features, also helps improve robustness.
  \newadd{green}{\textbf{The LIFE framework offers a framework along with guiding principles and techniques that are intended for use by modelers to develop data-driven augmentations for low-fidelity models using high-fidelity data.}} 
  
  To demonstrate LIFE in practice, a simple intermittency-based bypass transition model-form based on Wilcox's $k$-$\omega$ turbulence model was augmented appropriately with a careful choice of features.
  This augmentation function is inferred from two benchmark cases of flat plate transition from the T3 series of experiments.
  These two cases are characterized by transition under zero pressure gradient and favorable pressure gradient regions respectively, with an inflow freestream turbulence intensity significantly greater than $1\%$. 
  \newadd{green}{Linear interpolation on uniformly structured feature-space discretization was used as the functional form for the augmentation in order to implement a very crude variant of localized learning.
  Though more sophisticated ways to perform localized learning are possible, this implementation is showcased to demonstrate how effective and capable the notion of localized learning can be.}
  This augmented model is then used to predict transition on four other flat plate cases and four single-stage high-pressure-turbine cascade cases.
  The augmented model is shown to be able to predict the transition locations for problems in which similar physics was encountered in the dataset.
  It is noted that the focus of this work is not to develop the ultimate bypass transition model, rather to present a formalism that can be used to develop physics-constrained, data-enabled models.
  More comprehensive training datasets, especially involving higher pressure gradients and turbulence intensities can help in improving the accuracy and applicability of the model. 
  
  Currently, there are three major challenges in the LIFE framework - (1) Feature design for a given application; (2) Deciding the range of the vicinity in which a datapoint updates the augmentation; and (3) Formal uncertainty quantification.
  Designing features that can capture the behavior of an augmentation function in the simplest possible manner is a challenge that modelers have been dealing with for decades now.
  While there exist numerical diagnostics and techniques to assist and/or automate feature selection and engineering, the authors feel that domain expertise and intuition is the best way forward for problems as complex as turbulence and transition, at least given the prevailing constraints on data and computational power \newadd{red}{which render automatic feature selection infeasible.}
  As far as the range of influence that a datapoint has in the feature space is concerned, a trade-off needs to be made between over-fitting and predictive accuracy.
  If the influence is very limited, the augmented model will perform very well on the training cases, but can revert to baseline behavior for test cases which lie outside the influence of these datapoints.
  On the other hand, if the influence is too wide, the augmentation behavior might not be sufficiently resolved and hence could adversely affect predictive accuracy.
  Finally, formal uncertainty quantification techniques need to be developed for use in design of experiments to extend the capability of the LIFE framework. 
  
\section{Acknowledgements}

  Initial funding for this work came from the Office of Naval Research (ONR) under the project
  {\em Advancing Predictive Strategies for Wall-Bounded Turbulence by Fundamental Studies and
  Data-driven Modeling} (Tech. Monitor: Dr. Thomas Fu). Currently, it is being funded by Advanced
  Research Projects Agency-Energy (ARPA-E) DIFFERENTIATE program under the project {\em Multi-source
  Learning-accelerated Design of High-efficiency Multi-stage Compressor (MULTI-LEADER)} (Award
  no. DE-AR0001201) led by Raytheon Technologies Research Center (RTRC).

\bibliographystyle{unsrt}
\bibliography{References/ref}

%% APPENDIX STARTS HERE ===========================================================================================================================================================

\appendix
\section{Sensitivity evaluation using discrete adjoints} \label{app:Adjoints}

In a field inversion problem, the objective function $\mathcal{J} = \mathcal{C}(\widetilde{\bs{u}}_m) +
  \lambda_\beta\mathcal{T}_\beta(\beta)$ is a function of both $\widetilde{\bs{u}}_m$ and $\beta$, and since 
  $\widetilde{\bs{u}}_m$ is dependent on $\beta$  via $\mathscr{R}(\widetilde{\bs{u}}_m;\beta,\bs{\xi})=0$, we
  can write the following.
  $$
    \frac{d\mathcal{J}}{d\beta} = \left[\frac{\partial\mathcal{J}}{\partial\beta}\right]_{\RFvar} + \left[\frac{\partial\mathcal{J}}{\partial\RFvar}\right]_{\beta}
    \left[\frac{\partial\RFvar}{\partial\beta}\right]_{\mathscr{R}=0}
  $$
  $$
    \left[\frac{\partial\mathscr{R}}{\partial\beta}\right]_{\RFvar} +
    \left[\frac{\partial\mathscr{R}}{\partial\RFvar}\right]_{\beta}
    \left[\frac{\partial\widetilde{\bs{u}}_m}{\partial\beta}\right]_{\mathscr{R}=0} = 0.
  $$
  Now, using the above representations one can write,
  $$
    \frac{d\mathcal{J}}{d\beta} = \left[\frac{\partial\mathcal{J}}{\partial\beta}\right]_{\RFvar} - \left[\frac{\partial\mathcal{J}}{\partial\RFvar}\right]_{\beta}\left[\frac{\partial\mathscr{R}}{\partial\RFvar}\right]^{-1}_{\beta,\bs{\xi}}\left[\frac{\partial\mathscr{R}}{\partial\beta}\right]_{\RFvar,\bs{\xi}}
  $$
  Here, the sensitivities of a single objective function w.r.t. several variables need to be evaluated. In such a case, 
  it is cheaper to solve for $\dfrac{\partial\mathcal{J}}{\partial\widetilde{\bs{u}}_m}\left[\dfrac{\partial\mathscr{R}}
  {\partial\widetilde{\bs{u}}_m}\right]^{-1}$ compared to solving for $\left[\dfrac{\partial\mathscr{R}}{\partial
  \widetilde{\bs{u}}_m}\right]^{-1}\dfrac{\partial\mathscr{R}}{\partial\beta}$ and the solution to the former is 
  referred to as the vector of adjoint variables.  An algorithmic differentiation (AD) 
  package is used to evaluate the different jacobians involved in the above discrete adjoint approach. The in-house solver~\cite{duraisamy2012adjoint}
  used in this work makes use of the ADOL-C package for AD.
  
  The integrated inference and learning framework, however requires additional infrastructure for this to work. Since,
  there is an additional dependence of $\beta(\eta(\widetilde{\bs{u}}_m, \bs{\zeta}); \bs{w})$ on $\widetilde{\bs{u}}_m$
  to account for, this pathway needs to be included in the calculation of any jacobians w.r.t. states 
  $\widetilde{\bs{u}}_m$. Inclusion of the entire augmentation function in the AD calculation is unnecessary and expensive,
  however, as one only requires sensitivities of the augmentation w.r.t the features. The calculation of features 
  from states can be implemented within the AD framework relatively easily. Thus, calculating a linear approximation
  for the augmentation in feature space and using AD for this linearized function is a better and efficient alternative 
  to evaluate this Jacobian. Any other jacobians w.r.t. the augmentation can be converted to the jacobians w.r.t. 
  function parameters $\bs{w}$ using $\dfrac{\partial\beta}{\partial\bs{w}}$, which the learning framework 
  being used must be able to calculate. The sensitivity calculation for an integrated inference and learning approach
  can, thus, be calculated similar to what has been shown above for field inversion
  $$
    \frac{d\mathcal{J}}{d\bs{w}} = \left\lbrace\left[\frac{\partial\mathcal{J}}{\partial\beta}\right]_{\RFvar} - \left[\frac{\partial\mathcal{J}}{\partial\RFvar}\right]_{\bs{w}}\left[\frac{\partial\mathscr{R}}{\partial\RFvar}\right]^{-1}_{\bs{w},\bs{\xi}}\left[\frac{\partial\mathscr{R}}{\partial\beta}\right]_{\RFvar,\bs{\xi}}\right\rbrace\frac{\partial\beta}{\partial\bs{w}}.
  $$
  
\section{Results when training only with T3A} \label{app:T3A_only}
  
  The main issue with training only with T3A is that the behavior in the augmentation is learnt only on the basis of
  data from a zero pressure gradient case. The more cases are added to the mix, the stronger is the consistency of an
  optimal augmentation for different problems.
  
\begin{figure}[h!]
      \centering
      \subfigure[Optimization convergence]
      {\includegraphics[width=0.31\textwidth]{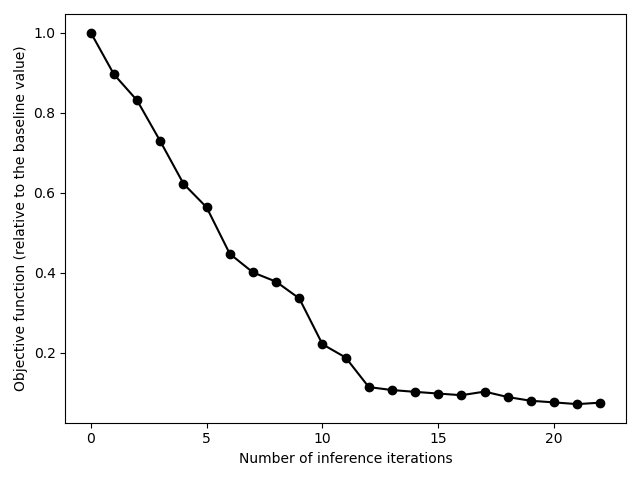} \label{appfig:optim_T3A}}
      \subfigure[Skin friction prediction]
      {\includegraphics[width=0.31\textwidth]{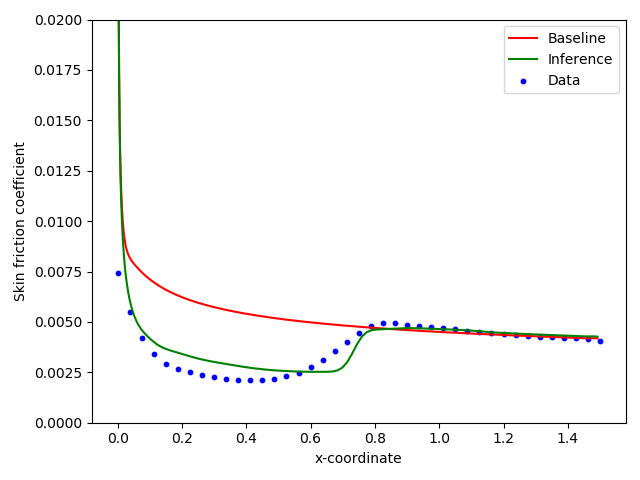} \label{appfig:Cf_T3A}}
      \caption{Augmentation training using data from T3A only}
    \end{figure}

\begin{figure}[h]
	    \centering
	    \subfigure[$\eta_1=0.05$]{\includegraphics[width=0.19\textwidth]{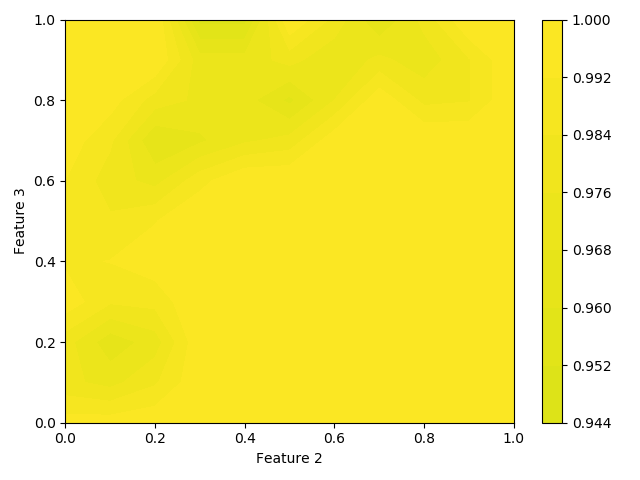}}
	    \subfigure[$\eta_1=0.15$]{\includegraphics[width=0.19\textwidth]{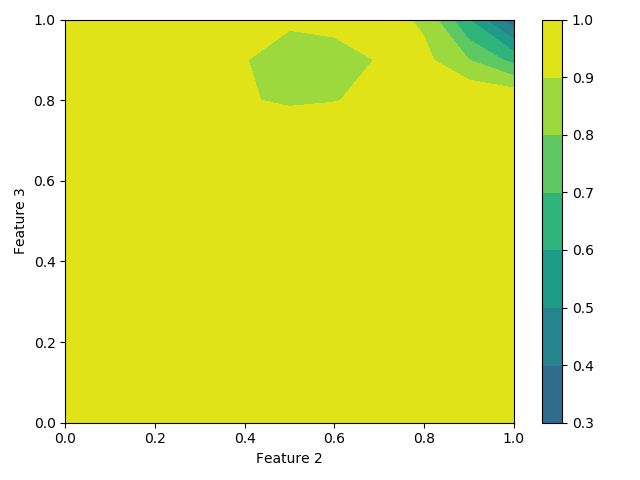}}
	    \subfigure[$\eta_1=0.25$]{\includegraphics[width=0.19\textwidth]{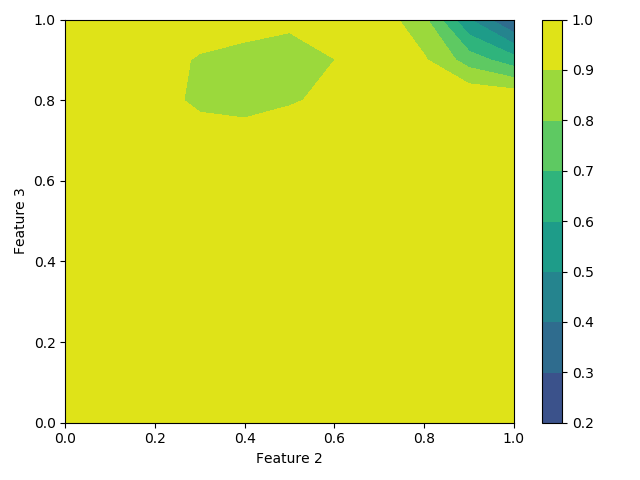}}
	    \subfigure[$\eta_1=0.35$]{\includegraphics[width=0.19\textwidth]{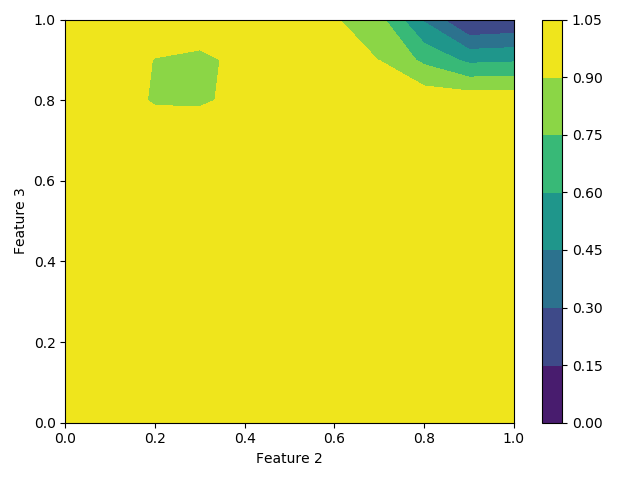}}
	    \subfigure[$\eta_1=0.45$]{\includegraphics[width=0.19\textwidth]{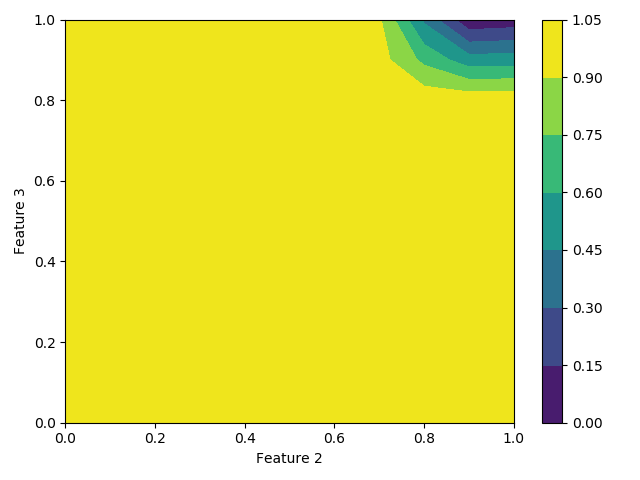}}
	    
	    \subfigure[$\eta_1=0.55$]{\includegraphics[width=0.19\textwidth]{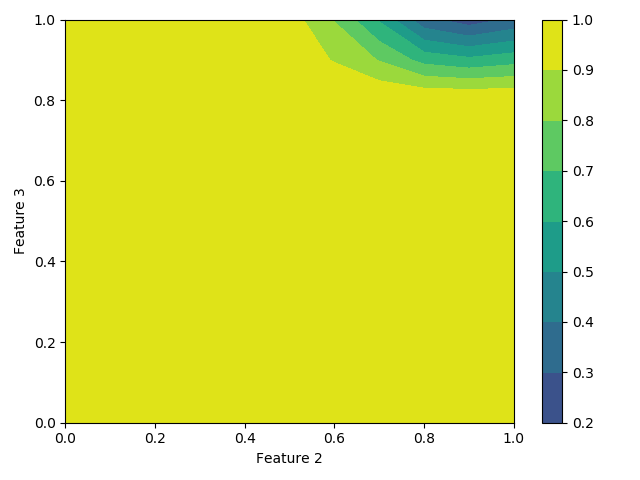}}
	    \subfigure[$\eta_1=0.65$]{\includegraphics[width=0.19\textwidth]{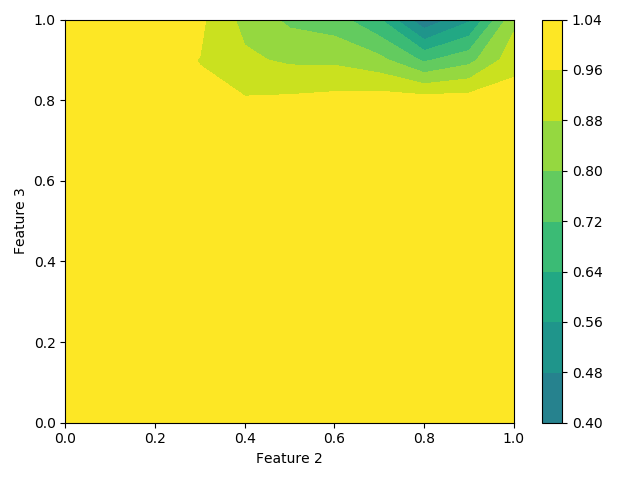}}
	    \subfigure[$\eta_1=0.75$]{\includegraphics[width=0.19\textwidth]{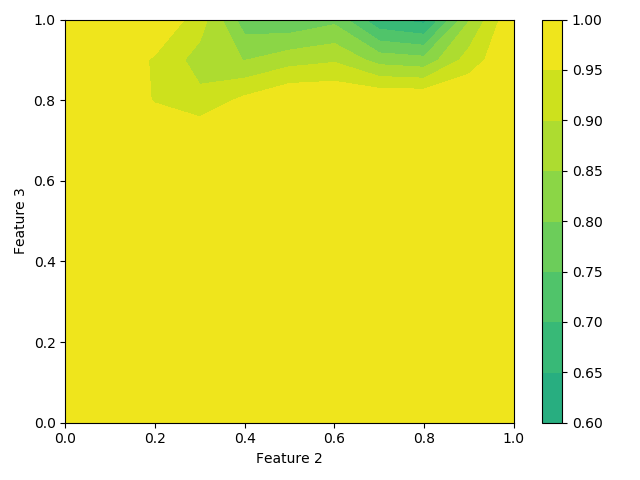}}
	    \subfigure[$\eta_1=0.85$]{\includegraphics[width=0.19\textwidth]{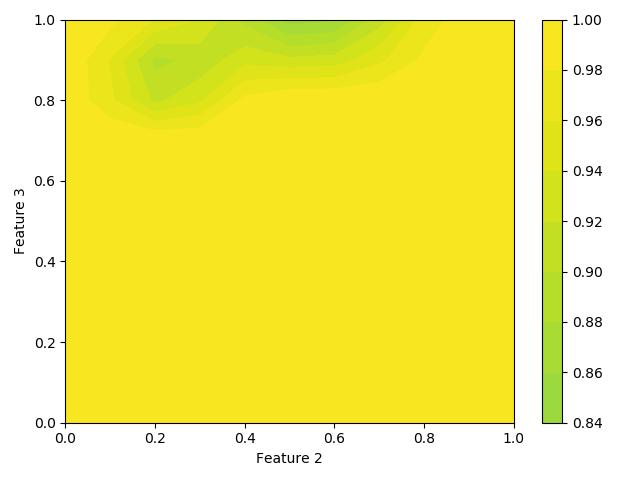}}
	    \subfigure[$\eta_1=0.95$]{\includegraphics[width=0.19\textwidth]{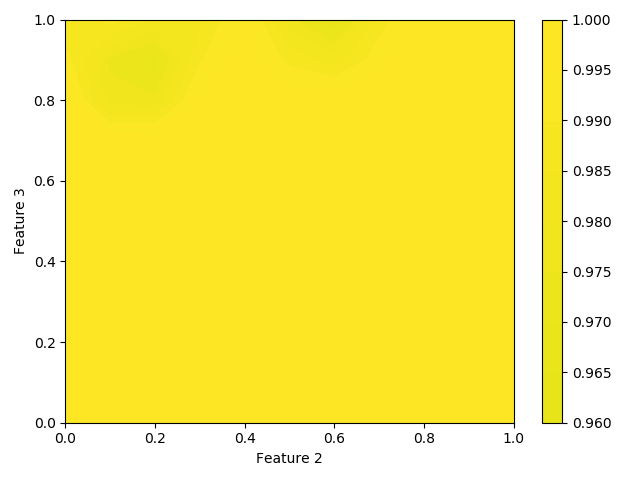}}
	    
	    \caption{Feature maps (x-axis: $\eta_2$, y-axis: $\eta_3$, uniform color-bar range [0,1] across all plots)}
	    \label{appfig:feature_space}
      \end{figure}

  {\bf Optimization convergence and feature space contours:}
    Although the convergence (Figure \ref{appfig:optim_T3A}) appears similar to that observed when both T3A and T3C1 were simultaneously used to learn
    the augmentation, the skin friction predictions (Figure \ref{appfig:Cf_T3A}) and augmentation contours in the feature space (figure \ref{appfig:feature_space}) have significant differences between the two cases.
    The effect of this difference between the two augmentations can be seen in the following sections. It is however,
    noteworthy how a single additional case with unseen physical behavior can make a considerable difference
    in the predictions. This suggests the ability of the framework to effectively extract information about the 
    features-to-augmentation functional relationship.
  
\begin{figure}[h!]
      \centering
      \subfigure[T3B]
      {\includegraphics[width=0.31\textwidth]{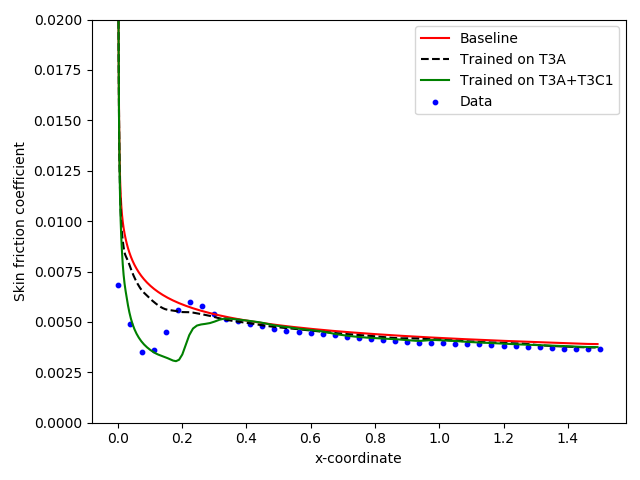} \label{appfig:Cf_T3B}}
      \subfigure[T3C1]
      {\includegraphics[width=0.31\textwidth]{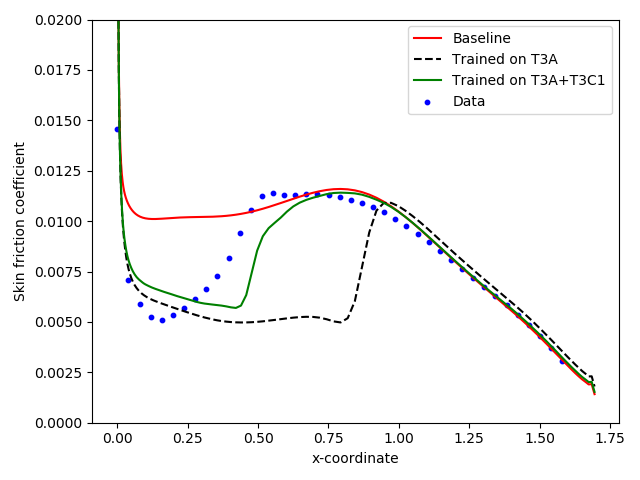} \label{appfig:Cf_T3C1}}
      \subfigure[T3C2]
      {\includegraphics[width=0.31\textwidth]{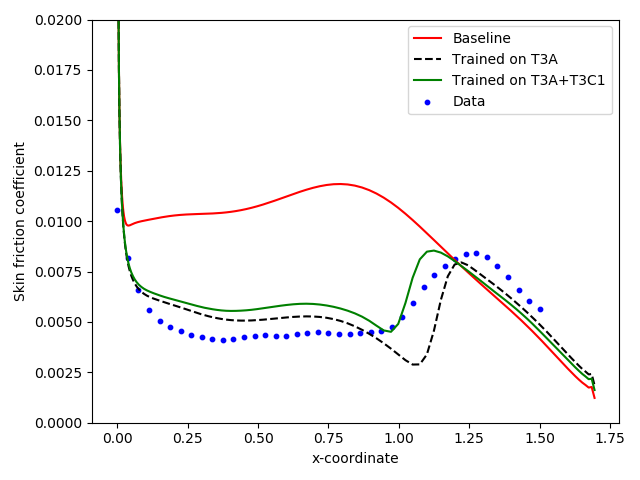} \label{appfig:Cf_T3C2}}
      \subfigure[T3C3]
      {\includegraphics[width=0.31\textwidth]{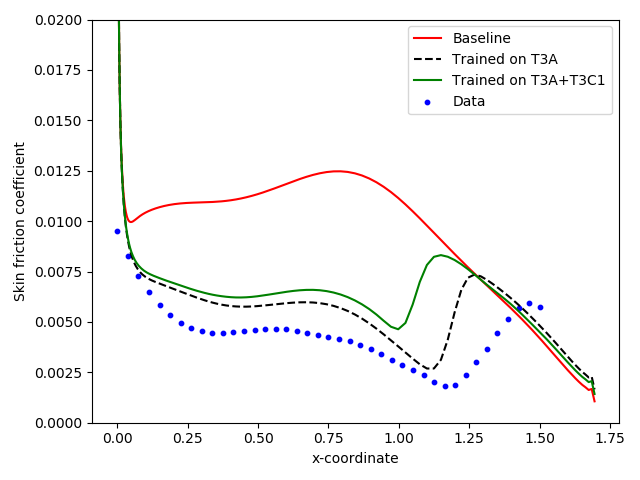} \label{appfig:Cf_T3C3}}
      \subfigure[T3C5]
      {\includegraphics[width=0.31\textwidth]{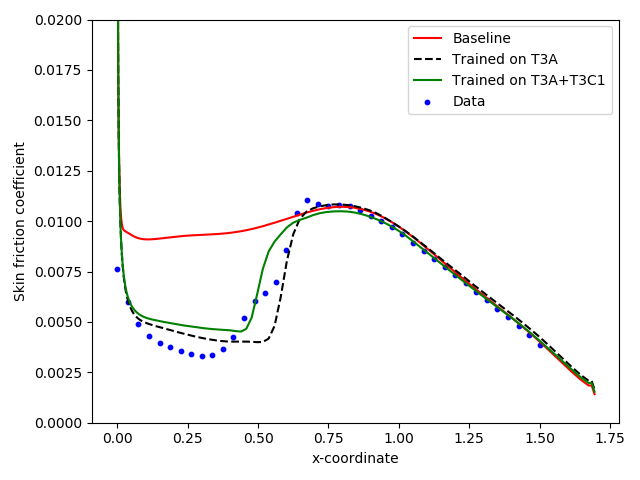} \label{appfig:Cf_T3C5}}
      \caption{Skin friction coefficients for T3C cases}
      \label{appfig:Cf_T3_test}
    \end{figure}

  {\bf Predictions on T3 cases:}
    It can be seen in Figure \ref{appfig:Cf_T3_test} that the prediction is completely wrong for T3C1 and T3B, while it
    appears reasonable for the other cases. Another important observation is that the transition locations for T3C1, T3C2 and T3C3 are over-predicted. This
    is due to the fact that the augmentation remains at a lower value for a longer distance spuriously as the 
    augmentation has no way of differentiating between how $\eta_1$ changes for different pressure gradients as the
    training is performed only for a zero pressure gradient case.
  
\begin{figure}[h]
      \centering
      \subfigure[MUR116]
      {\includegraphics[width=0.31\textwidth]{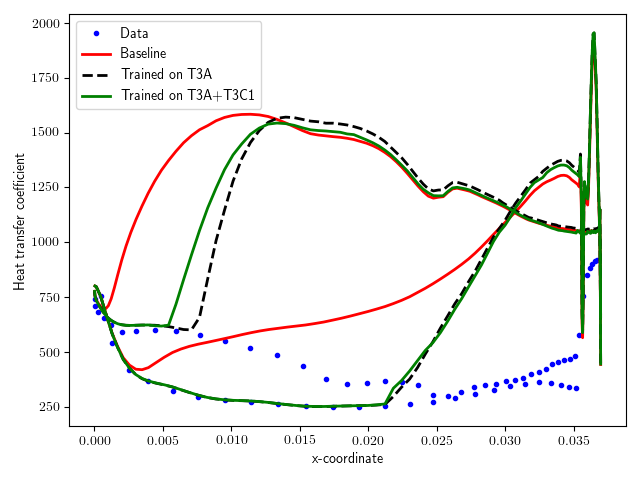} \label{appfig:VKI_MUR116}}
      \subfigure[MUR129]
      {\includegraphics[width=0.31\textwidth]{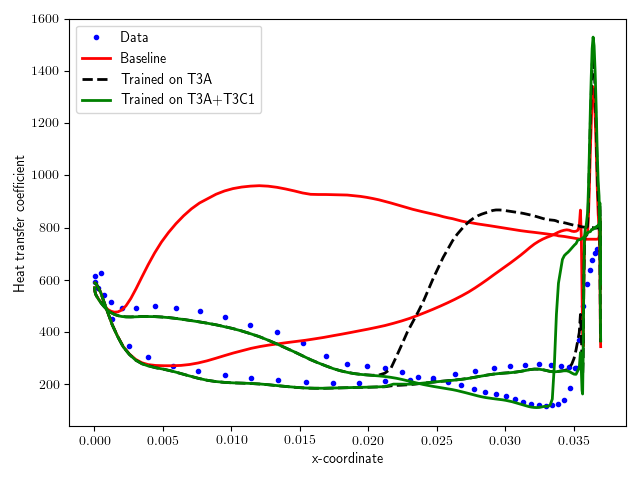} \label{appfig:VKI_MUR129}}
      
      \subfigure[MUR224]
      {\includegraphics[width=0.31\textwidth]{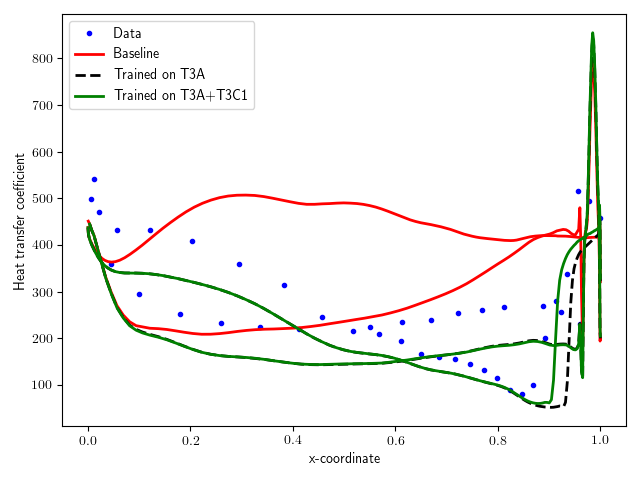} \label{appfig:VKI_MUR224}}
      \subfigure[MUR241]
      {\includegraphics[width=0.31\textwidth]{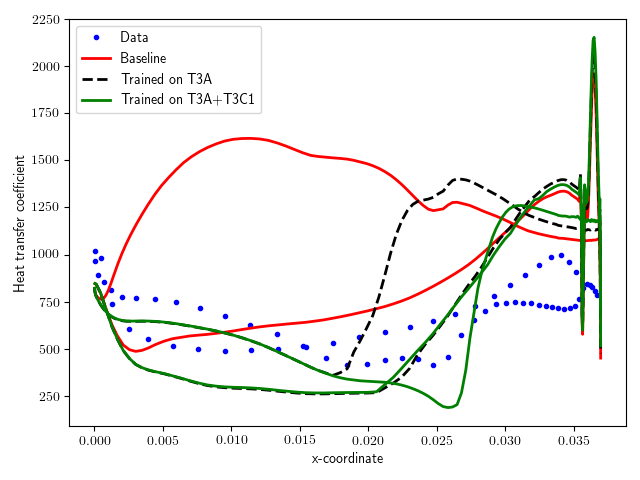} \label{appfig:VKI_MUR241}}
      \caption{Heat transfer coefficients for VKI cases}
      \label{appfig:Ch_VKI}
    \end{figure}

  {\bf Prediction on VKI cases}
    As can be seen in Figure \ref{appfig:Ch_VKI}, training a model only on the T3A data results in significantly inaccurate transition location predictions on at least one side of the blade except MUR224 when compared to the results presented in the main text where both T3A and T3C1 cases were used for training.
    This is in accordance with the explanation provided in the last section.
    Since the transition model has little information on how the features behave in the presence of non-zero pressure gradients, additional data which can highlight such behavior (in the main text as the T3C1 case) is required to extract information about the behavior of the augmentation in feature space.
  
\section{Results with varying preset distance intervals for $\overline{Re_{\theta,t}}$ extraction} \label{app:preset}

\begin{figure}[h]
      \centering
      \subfigure[MUR116]
      {\includegraphics[width=0.31\textwidth]{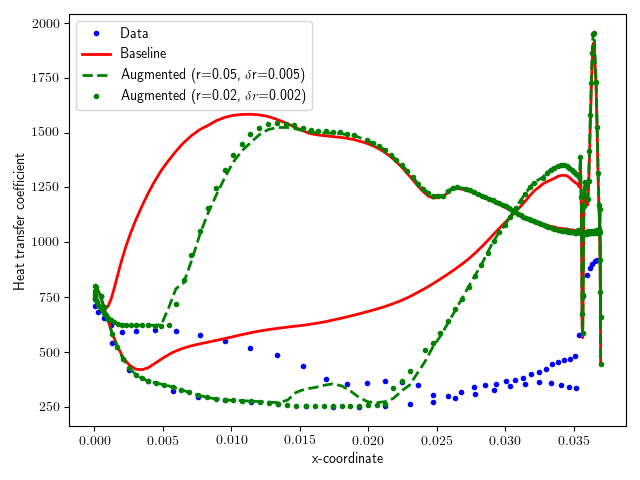} \label{appfig:VKI_MUR116pc}}
      \subfigure[MUR129]
      {\includegraphics[width=0.31\textwidth]{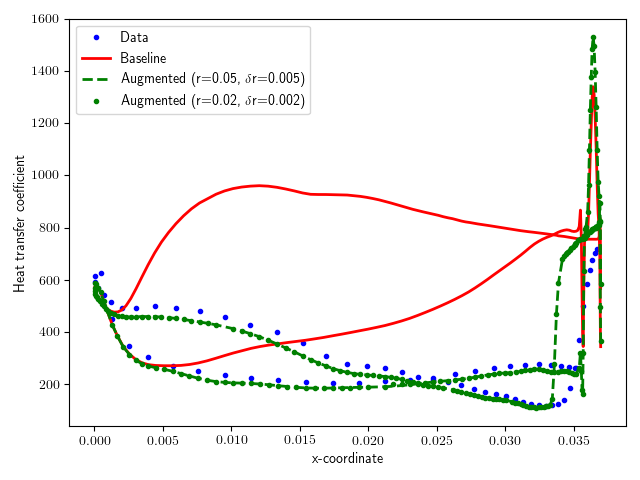} \label{appfig:VKI_MUR129pc}}
      
      \subfigure[MUR224]
      {\includegraphics[width=0.31\textwidth]{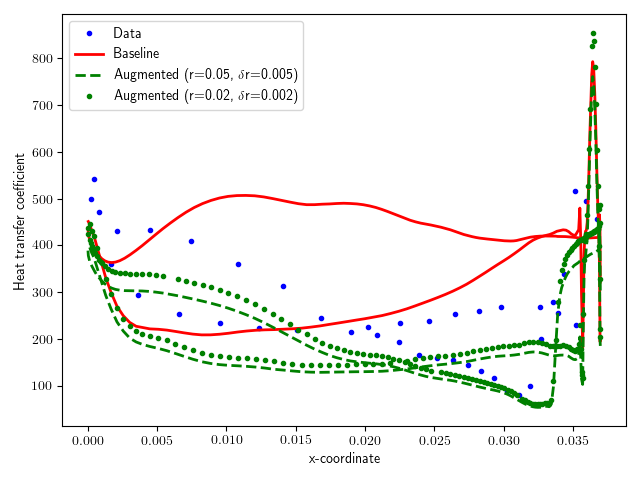} \label{appfig:VKI_MUR224pc}}
      \subfigure[MUR241]
      {\includegraphics[width=0.31\textwidth]{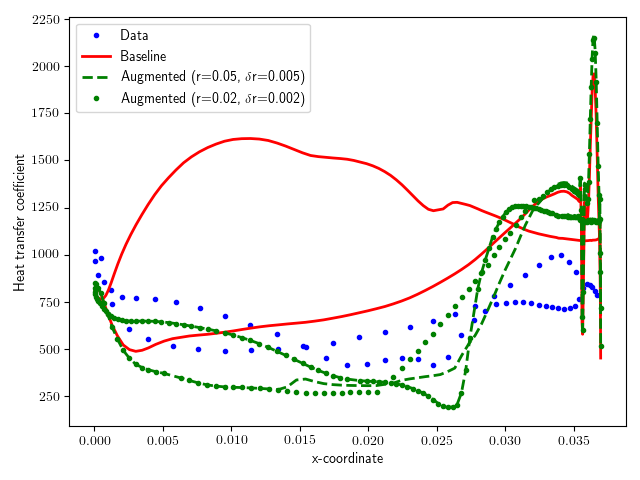} \label{appfig:VKI_MUR241pc}}
      \caption{Comparison between different preset distance intervals}
      \label{appfig:Ch_VKI_pc}
    \end{figure}

  As shown in Figure \ref{appfig:Ch_VKI_pc}, we found that varying the preset distance usually has a small effect on the predictions for the turbine blades.
  A minor discrepancy is observed in MUR116 (characterized by a small bump in the heat transfer coefficient), whereas a major discrepancy, resulting in considerable different transition behavior, is seen for MUR241.

\newadd{red}{
\section{Results without localized learning (Neural Networks)} \label{app:without_localized_learning}
  
  Three different fully connected neural network architectures were utilized as shown in table \ref{apptable:nn_arch}.
  Owing to the simple nature of augmentation behavior in feature space, a simple neural network with two hidden layers containing seven nodes each was tested.
  The results obtained during training, as shown in figure \ref{appfig:nn}, are clearly poor as they partially laminarize the entire flow rather than laminarizing the flow only before the transition location.
  Further, two other architectures with $1$ hidden layer containing $600$ nodes, and $2$ hidden layers containing $45$ and $60$ nodes respectively were tried out but yielded similarly poor results as shown in figure \ref{appfig:nn}.
  The number of parameters (weights and biases) for the latter two architectures was kept around $3000$ to match the number of parameters used in the main text during localized learning via feature-space discretization.
  \begin{table}[H]
  \begin{center}
    \begin{tabular}{c|c|c|c}
      \hline
      \textbf{Network} & \# of hidden layers & Nodes in hidden layer 1 & Nodes in hidden layer 2  \\\hline 
      Neural Network 1 & 2      & 7         & 7  \\
      Neural Network 2 & 1      & 600       & -  \\
      Neural Network 3 & 2      & 45        & 60 \\\hline
    \end{tabular}
    \caption{Description of neural network architectures}
    \label{apptable:nn_arch}
  \end{center}
  \end{table}
  The results demonstrate that localized learning is essential to both, better condition the optimization problem and also to aid the optimization trajectory in the correct direction in the augmentation space.
  The second point is especially important, as in several applications (including transition prediction), changing the augmentation behavior in regions where no data is available can in fact lead to spurious model behavior as these parts of the feature space can be accessed before the solver converges.
  This is further worsened by any feedback present between features and augmentation. It has to be mentioned that using the same software framework, neural networks have been used in the context of integrated inference and learning more successfully in Refs.~\cite{FIMLC2019a,FIMLC2019b} on a less challenging problem. The authors acknowledge that more sophisticated training methods may yield better results.
  
  \begin{figure}[h!]
    \centering
    \subfigure[Objective reduction (T3A)]
    {\includegraphics[width=0.31\textwidth]{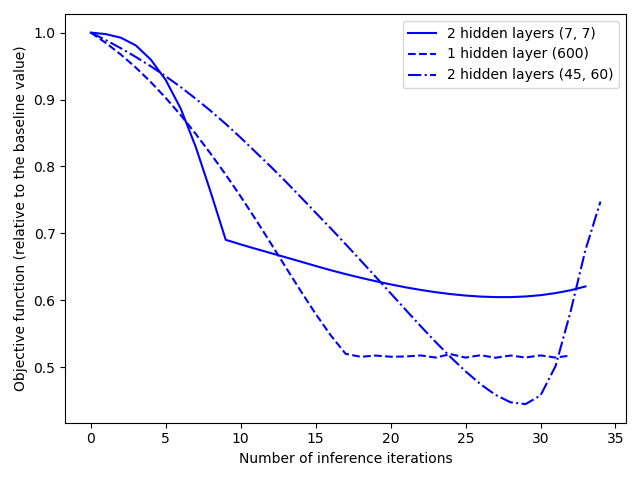}}
    \subfigure[Skin friction profile (T3A)]
    {\includegraphics[width=0.31\textwidth]{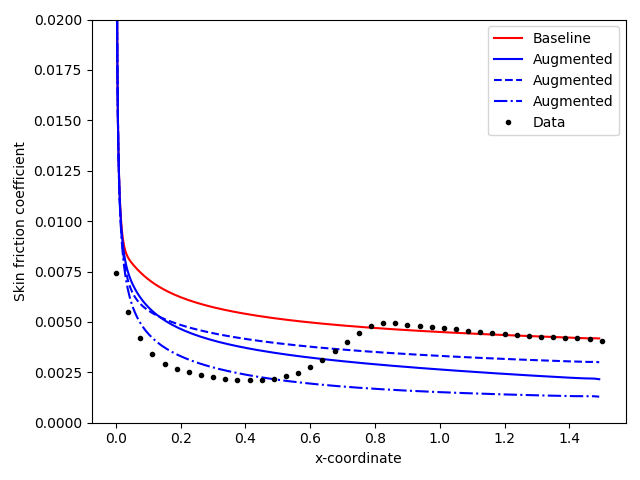}}
    
    \subfigure[Objective reduction (T3C1)]
    {\includegraphics[width=0.31\textwidth]{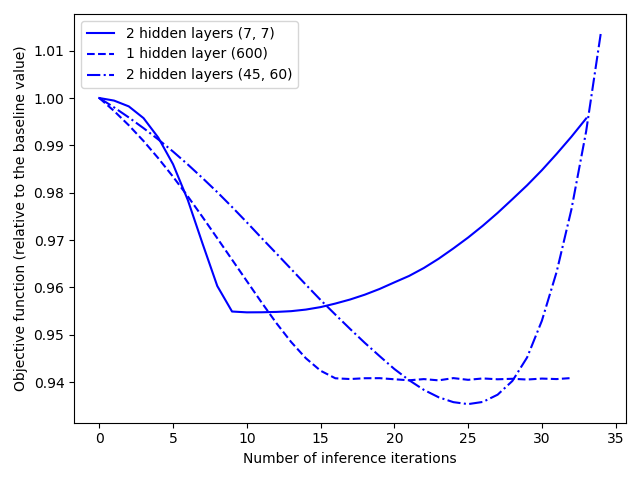}}
    \subfigure[Skin friction profile (T3C1)]
    {\includegraphics[width=0.31\textwidth]{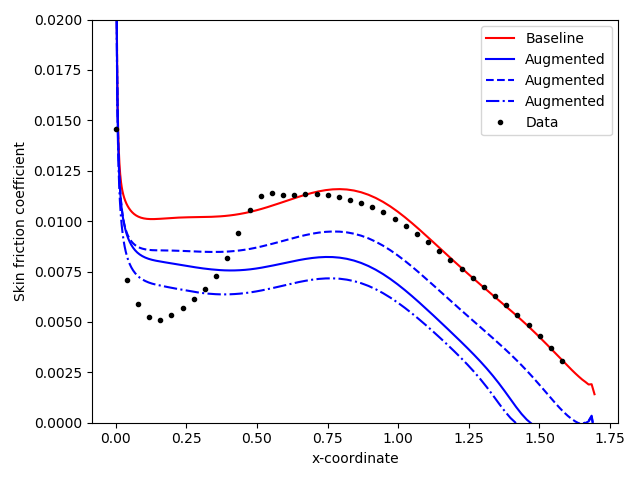}}
    \caption{Training results with Different Neural Network Architectures}
    \label{appfig:nn}
  \end{figure}

\section{Results with a finer discretization in the feature space}
\label{app:with_finer_grid}

  For  comparison purposes, a finer feature-space discretization was also used to obtain the augmentation function, the training results and augmentation contours on feature-space slices for which have been shown in figures \ref{fig:fine_training} and \ref{fig:fine_augmentation}.
  The feature space was divided into subdomains of size $1/30$ along all three feature space directions (90, 30, and 30 cells along the first, second, and third features respectively). As can be noticed in figure \ref{fig:fine_augmentation}, the influence of the changes made be the data have been restricted to smaller regions owing to the smaller cell sizes.
  Figures \ref{fig:fine_testing_T3} and \ref{fig:fine_testing_VKI} show the results from testing the augmentation on the T3B, T3C2, T3C3, T3C5, MUR116, MUR129, MUR224, and MUR241 cases.
  As can be seen from the results, while the augmentation learned on the finer grid seems to predict the transition locations for T3 cases with nearly similar accuracy (with little laminarization in the T3B case and slightly premature transition in T3C5) as its counterpart trained on the coarser grid, almost all results from the VKI cases exhibit premature transition.
  This happens because the limited region of influence that the available data has in the feature space allow some feature space locations to exhibit baseline behavior (which did not happen when using the augmentation learned on a coarser grid as the region of influence covered a larger part of the feature space).
  
  \begin{figure}[h!]
  \begin{center}
    \subfigure[Optimization history (T3A)]
    {\includegraphics[width=0.31\textwidth]{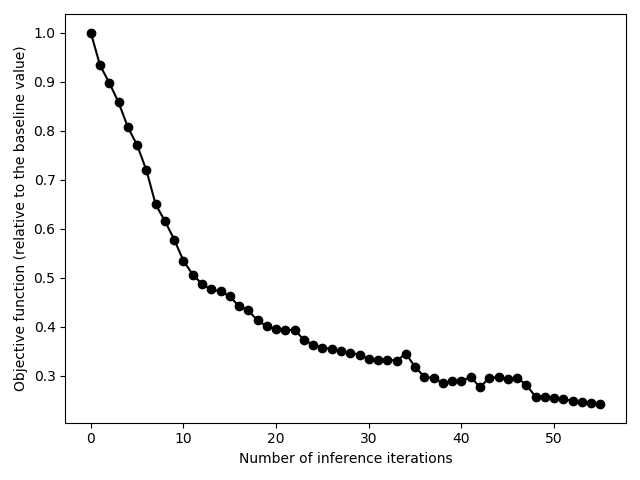}}
    \subfigure[Optimization history (T3C1)]
    {\includegraphics[width=0.31\textwidth]{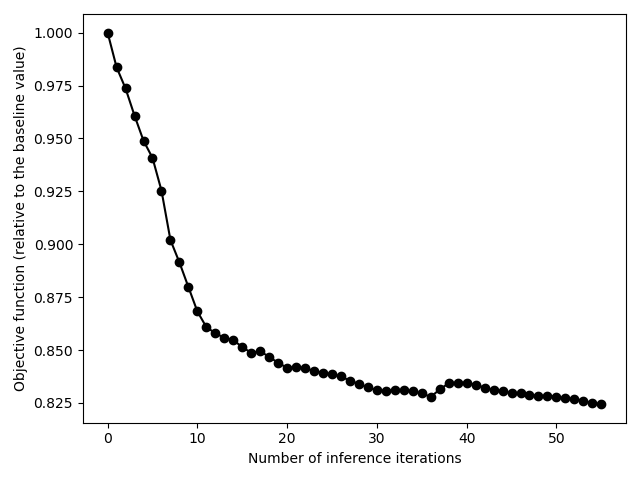}}
    
    \subfigure[Skin friction profile (T3A)]
    {\includegraphics[width=0.31\textwidth]{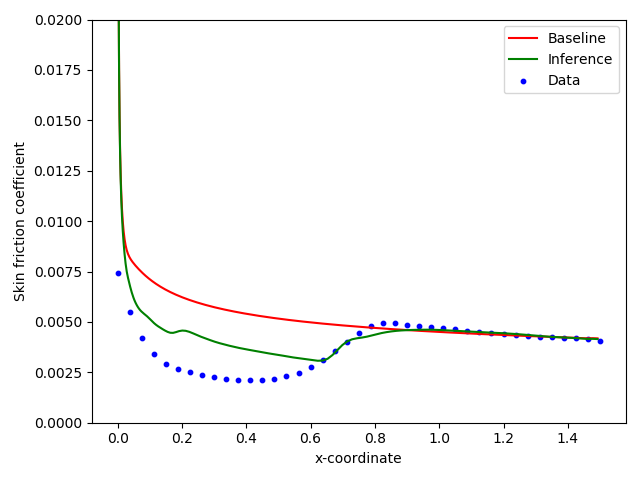}}
    \subfigure[Skin friction profile (T3C1)]
    {\includegraphics[width=0.31\textwidth]{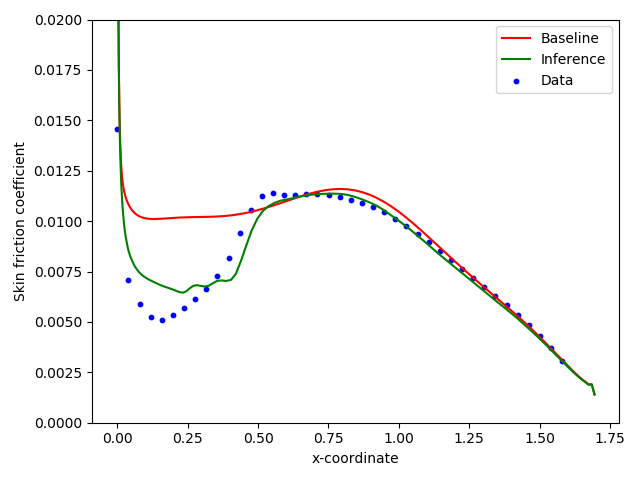}}
    \caption{Training results on a finer feature-space grid}
    \label{fig:fine_training}
  \end{center}
  \end{figure}
  
  \begin{figure}[h!]
  \begin{center}
    \subfigure[$\eta_1=0.05$]
    {\includegraphics[width=0.24\textwidth]{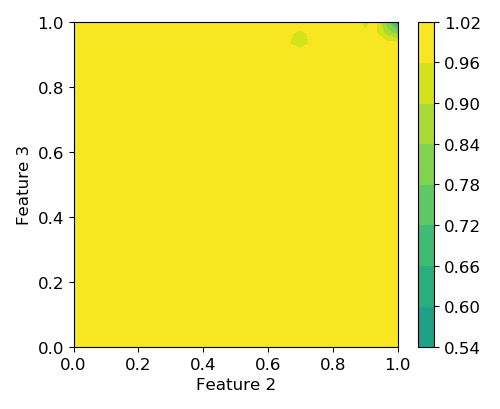}}
    \subfigure[$\eta_1=0.15$]
    {\includegraphics[width=0.24\textwidth]{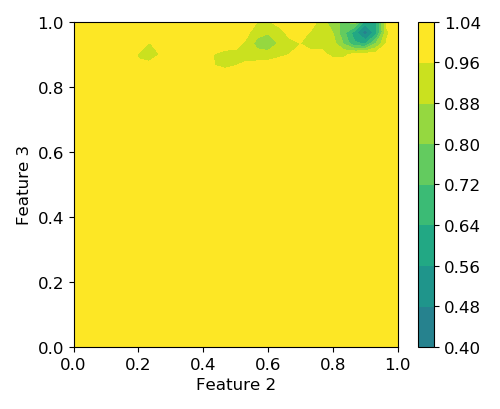}}
    \subfigure[$\eta_1=0.25$]
    {\includegraphics[width=0.24\textwidth]{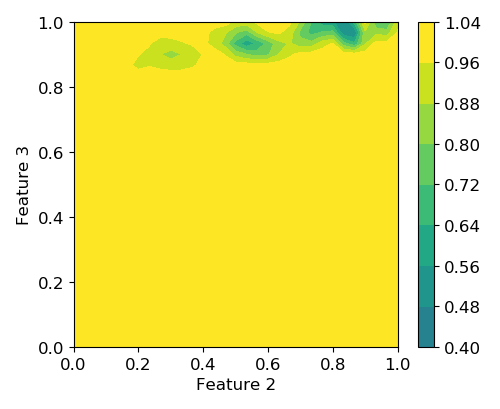}}
    \subfigure[$\eta_1=0.35$]
    {\includegraphics[width=0.24\textwidth]{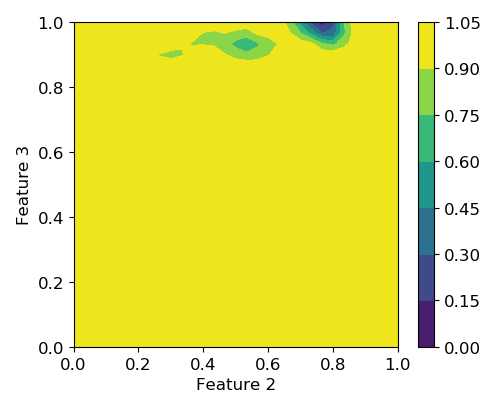}}
    \subfigure[$\eta_1=0.45$]
    {\includegraphics[width=0.24\textwidth]{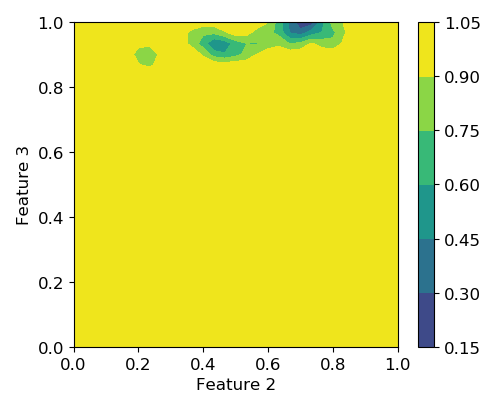}}
    \subfigure[$\eta_1=0.55$]
    {\includegraphics[width=0.24\textwidth]{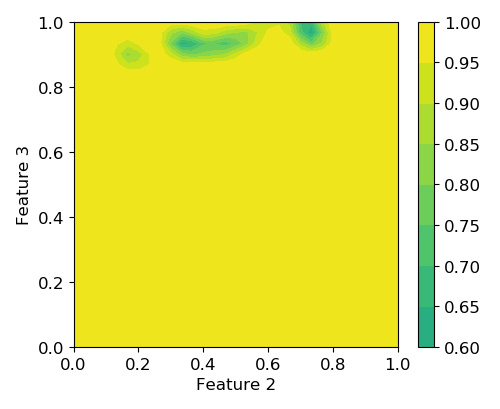}}
    \subfigure[$\eta_1=0.65$]
    {\includegraphics[width=0.24\textwidth]{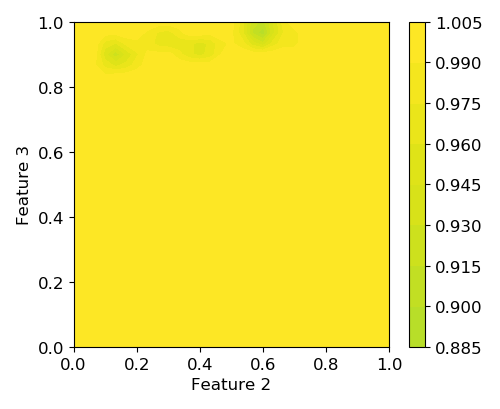}}
    \subfigure[$\eta_1=0.75$]
    {\includegraphics[width=0.24\textwidth]{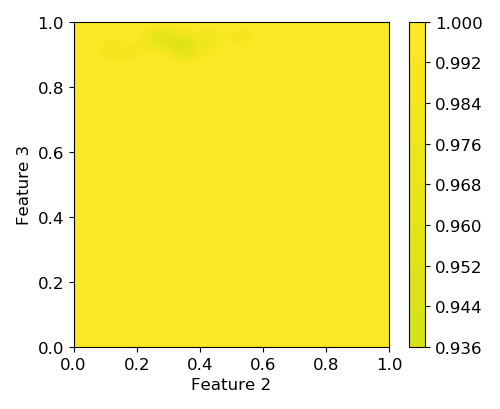}}
    %\subfigure[$\eta_1=0.85$]
   % {\includegraphics[width=0.24\textwidth]{figures/Ftr_25_fine.png}}
    \caption{Augmentation contours on feature-space slices}
    \label{fig:fine_augmentation}
  \end{center}
  \end{figure}
  
  \begin{figure}[h!]
  \begin{center}
    \subfigure[T3B]
    {\includegraphics[width=0.31\textwidth]{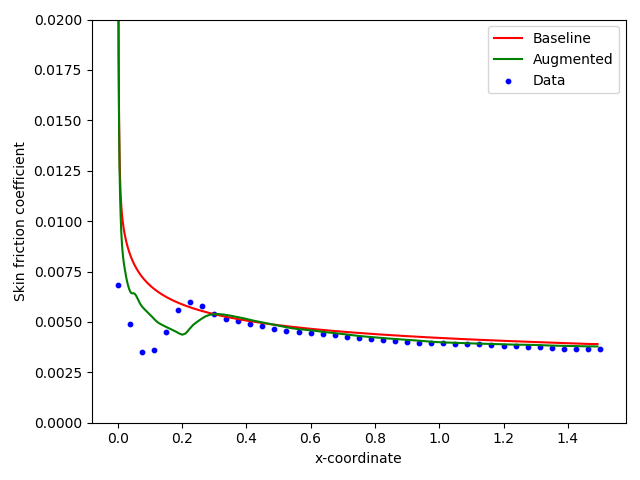}}
    \subfigure[T3C2]
    {\includegraphics[width=0.31\textwidth]{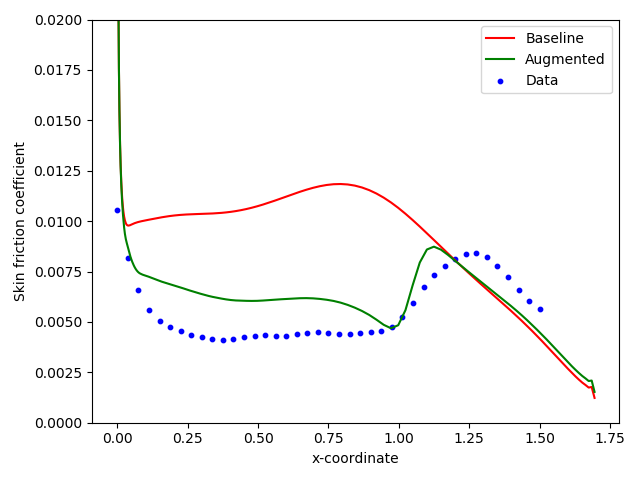}}
    
    \subfigure[T3C3]
    {\includegraphics[width=0.31\textwidth]{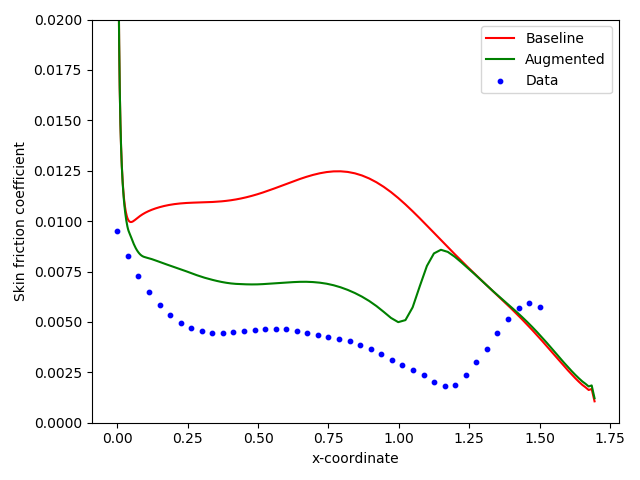}}
    \subfigure[T3C5]
    {\includegraphics[width=0.31\textwidth]{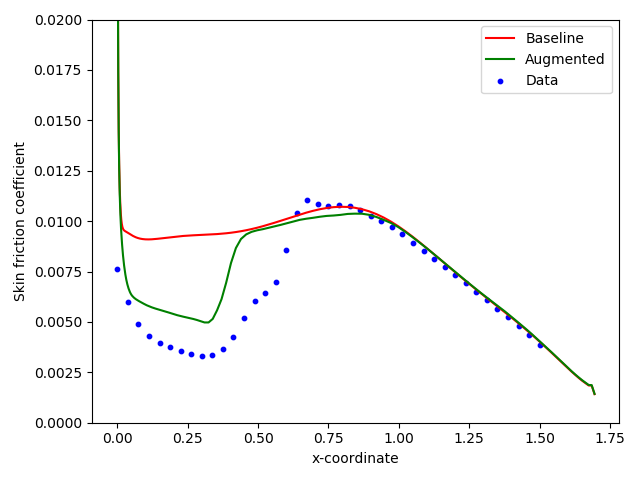}}
    \caption{Skin friction coefficient profiles for the T3 test cases}
    \label{fig:fine_testing_T3}
  \end{center}
  \end{figure}
  
  \begin{figure}[h!]
  \begin{center}
    \subfigure[MUR116]
    {\includegraphics[width=0.31\textwidth]{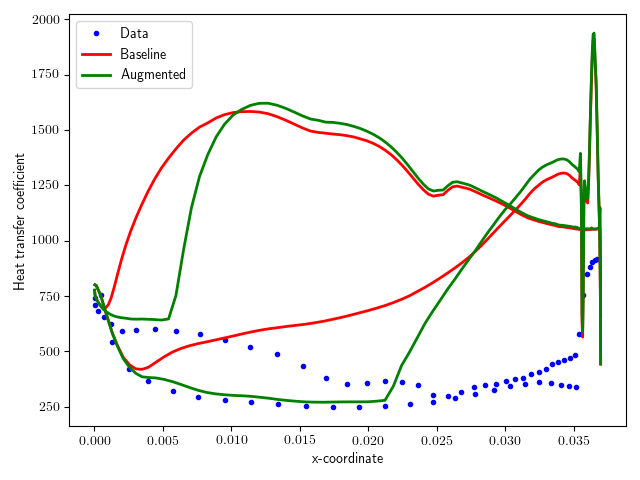}}
    \subfigure[MUR129]
    {\includegraphics[width=0.31\textwidth]{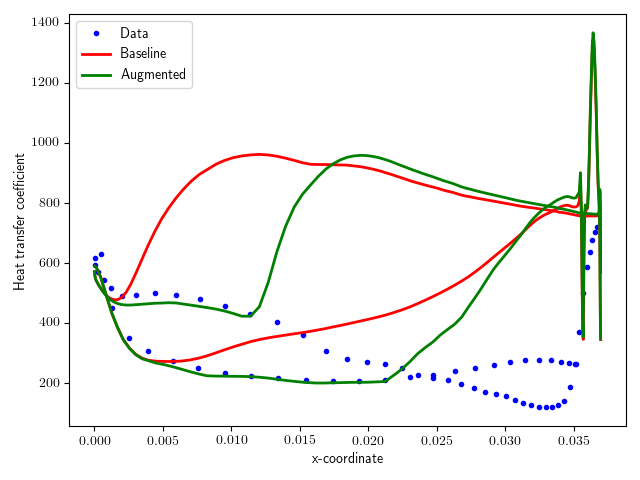}}
    
    \subfigure[MUR224]
    {\includegraphics[width=0.31\textwidth]{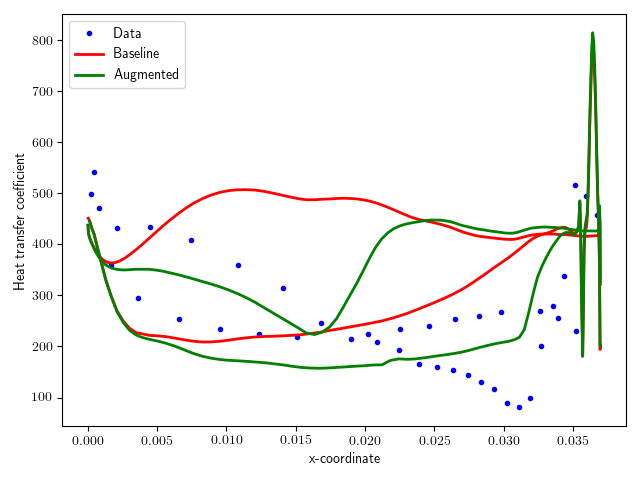}}
    \subfigure[MUR241]
    {\includegraphics[width=0.31\textwidth]{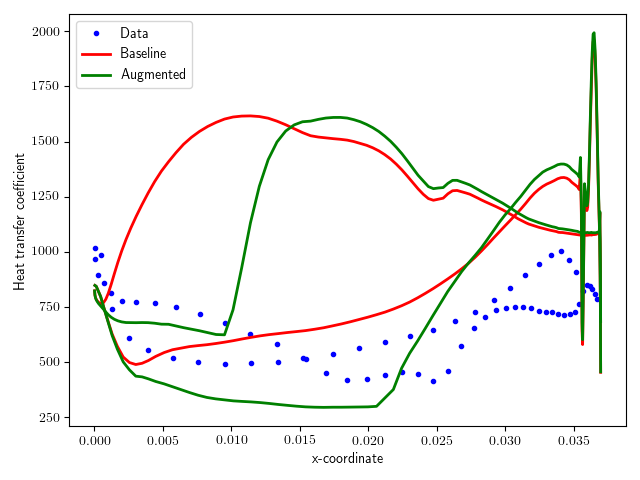}}
    \caption{Heat transfer coefficient profiles for the VKI test cases}
    \label{fig:fine_testing_VKI}
  \end{center}
  \end{figure}
}
\end{document}